\documentclass[a4paper,11pt]{article}
\usepackage{jheppub}
\usepackage{amsmath,amssymb,amsthm}
\usepackage{verbatim}
\usepackage[normalem]{ulem}
\usepackage{array}
\usepackage[table]{xcolor}
\usepackage{mathrsfs}
\usepackage{bm}
\usepackage{braket}

\usepackage{graphicx}
\newcommand{\sumint}{%
  \mathop{%
    \mathchoice
      {\raisebox{0.2ex}{\scalebox{0.88}{$\displaystyle\sum$}}\!\!\!\!\!\!\!\!\int}
      {\raisebox{0.2ex}{\scalebox{0.88}{$\textstyle\sum$}}\!\!\!\!\!\!\!\!\int}
      {\raisebox{0.15ex}{\scalebox{0.85}{$\scriptstyle\sum$}}\!\!\!\!\!\!\!\int}
      {\raisebox{0.1ex}{\scalebox{0.85}{$\scriptscriptstyle\sum$}}\!\!\!\!\!\!\!\int}
  }\displaylimits
}

\usepackage{hyperref}
\usepackage{fix-cm}
\usepackage{amsmath}

\usepackage[none]{hyphenat} 
\newcounter{fneq} 
\counterwithin{fneq}{footnote}
\renewcommand\thefneq{\thefootnote-\textsc{\alph{fneq}}}

\newcommand\dotag[1]{\refstepcounter{fneq}\label{#1}\tag{\thefneq}}

\title{Optimal paths across potentials on scalar field space}

\author[a]{Saskia Demulder}
\author[b,c]{\!\!, Dieter L\"ust}
\author[b]{\!\!, Carmine Montella}
\author[b,c]{and Thomas Raml}

\affiliation[a]{Department of Quantitative Methods, CUNEF Universidad,\\ Calle Almansa 101,
28040 Madrid, Spain}
\affiliation[b]{Max-Planck-Institut f\"ur Physik (Werner-Heisenberg-Institut),\\ Boltzmannstr. 8, 85748, Garching, Germany}
 \affiliation[c]{Arnold Sommerfeld Center for Theoretical Physics,\\
   Ludwig-Maximilian-Universit\"at, 80333 M\"unchen, Germany}

\emailAdd{saskia.demulder@cunef.edu}
\emailAdd{luest@mpp.mpg.de}
\emailAdd{montella@mpp.mpg.de}
\emailAdd{raml@mpp.mpg.de}

\abstract{Motivated by the Swampland Distance Conjecture, we study distances in field space using the framework of Optimal Transport. The associated optimisation problem naturally leads to a notion of distance in terms of a (generalised) Wasserstein distance between probability distributions over field space.
In the absence of dynamical gravity, we relate the transport problem to Hamilton-Jacobi and continuity equations arising from a WKB expansion of a Schr\"odinger equation associated with the physical configuration.
We then formulate an extension in the presence of dynamical gravity. Using the ADM formalism, we establish the corresponding transport problem through the Wheeler-DeWitt equation, giving rise to different possible choices of cost functions. The resulting notions of distances are naturally defined on the full configuration space, while an interpretation in terms of a genuine scalar field distance requires additional modifications. We further discuss several applications and examples, and indicate possible implications for different themes within the Swampland program.}

\begin{document}

\vspace*{-1.5cm}
\begin{flushright}
{\small
MPP-2026-18
  }
\end{flushright}

\vspace{1.5cm}
\maketitle
\newpage

\section{Introduction and Motivation}

It has become increasingly clear in recent years that not every effective field theory (EFT) can be consistently embedded into a theory of quantum gravity. Identifying the common properties of those EFTs that do admit an ultraviolet completion in quantum gravity is therefore essential for understanding how high-energy consistency constrains low-energy effective physics and lies at the heart of the Swampland program~\cite{Vafa:2005ui}. For pedagogical introductions to the subject, we refer to~\cite{Palti:2019pca,Grana:2021zvf,vanBeest:2021lhn,Agmon:2022thq}.

Determining which EFTs can arise from quantum gravity requires, in particular, a detailed understanding of the ways in which such theories break down. The Swampland Distance Conjecture (SDC)~\cite{Ooguri:2006in} provides a concrete diagnostic: the breakdown of the EFT description is signalled by parametrically large distances in moduli space, which are accompanied by an infinite tower of light states whose characteristic mass scale, measured in Planck units, behaves as
\begin{equation}
m_{\mathrm{tower}}(\Delta) \sim m_{\mathrm{tower}}(0)\, e^{-\alpha \Delta}\,.
\end{equation}
Here $\Delta \gg 1$ denotes the distance in moduli space, while $\alpha$ is typically an order-one constant. In string theory, and more generally in any consistent theory of quantum gravity, the appearance of such an infinite tower of light states signals the breakdown of the chosen effective description and suggests the existence of an alternative duality frame. The precise nature of the corresponding weakly coupled description is captured by the Emergent String Conjecture (ESC)~\cite{Lee:2019wij}.

While the SDC and ESC are most naturally formulated on moduli space viewed as a metric manifold, realistic string compactifications typically involve moduli stabilisation and thus scalar potentials, giving rise to more general scalar field spaces. As a consequence, physically relevant trajectories in field space are generically non-geodesic, raising the question of how to define distances once these scalar potentials are properly taken into account. In recent years, several approaches have been proposed that introduce generalised notions of distance~\cite{Lust:2019zwm,Kehagias:2019akr,Stout:2020uaf, Stout:2021ubb, Shiu:2022oti,Basile:2023rvm,Li:2023gtt,Shiu:2023bay,Palti:2024voy,Mohseni:2024njl,Debusschere:2024rmi,Demulder:2024glx,Palti:2025ydz}, as well as the corresponding interpretations of the SDC in the presence of a cosmological constant~\cite{Lust:2019zwm} or generic scalar potentials~\cite{Baume:2016psm,Valenzuela:2016yny,Blumenhagen:2017cxt,Grimm:2019ixq,Grimm:2020ouv,Klawer:2021ltm,Delgado:2022dkz,Grimm:2022sbl,Demulder:2023vlo,Grimm:2025cpq,Calderon-Infante:2026ymy}; see~\cite{Raml:2025yrb} for a recent account and summary. 

\medskip

In this work, we employ the mathematical framework of Optimal Transport (OT)~\cite{monge1781memoire,kantorovich2006translocation,villani2008optimal,ambrosio2021lectures,bernard2006monge} to define notions of distance between effective field theories in the presence of generic backgrounds, ultimately taking into account also contributions from dynamical gravity and non-vanishing scalar potentials.

\begin{figure}[h!]
    \centering
    \includegraphics[height=3.2cm]{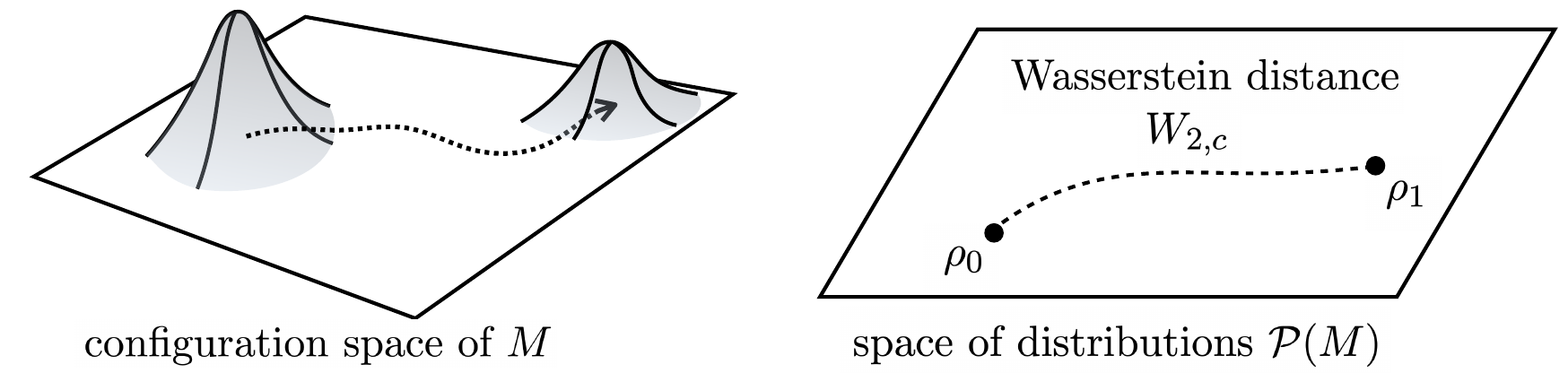}
    \caption{Points on scalar field space (or more generally the appropriate configuration space) $M$ are replaced by semiclassical distributions arising from the underlying effective theory. The associated distance in the presence of potentials, relevant also for the SDC, is calculated via the framework of Optimal Transport and the associated (generalised) Wasserstein distance $W_{2,c}$ between measures $\rho_0, \rho_1$ on $\mathcal{P}(M)$ and cost function $c(x,T(x); \,V)$.} 
    \label{fig:distributions_fieldspace}
\end{figure}

The key shift in our approach is to extend the usual pointwise description of field space to a distributional one. Instead of assigning to an EFT only a point in scalar field space (for instance, the mean value of a scalar field $\braket{\phi}$), we associate to it a semiclassical wavefunction and the corresponding probability density on field space, or more generally configuration space. The connection with the physical systems of interest is established by showing that the equations governing the critical points of the variational problem associated with the dynamical formulation of OT are closely related to the semiclassical expansion of the Schr\"odinger equation in the absence of dynamical gravity. More precisely, at orders $\hbar^0$ and $\hbar^1$ in the WKB expansion, one recovers the Hamilton-Jacobi equation and the continuity equation, which can be directly matched to the corresponding structures appearing in the dynamical (Benamou-Brenier) formulation of OT \cite{benamou2000computational,ambrosio2021lectures}. This correspondence provides the basic dictionary between semiclassical dynamics and transport on the space of probability measures. 
In particular, we employ the 2-Wasserstein distance, or suitable generalisations thereof, which can be understood as the length of minimal paths required to transport one density into another, on the space of distributions $\mathcal P(M)$. See Figure~\ref{fig:distributions_fieldspace} for a schematic illustration.
A key advantage of this approach is that the transport cost entering the Wasserstein distance can be naturally modified or extended to allow for scalar potentials entering directly into the definition of distance. We therefore study OT paths on field space in both the absence and in the presence of dynamical gravity. 

\medskip

In the standard point of view, points in scalar field space are interpreted as vacuum expectation values of scalar fields. In consistent EFTs, these values play a central role as they determine periods, gauge and Yukawa couplings, black-hole entropies, and other quantities that characterise the low-energy (effective) theory. From the perspective of the SDC, the breakdown of an EFT depends on how this entire bundle of EFT data varies across field space, illustrating why different formulations or refinements of the SDC emphasise distinct notions of distance.
However, in general, a distribution over field space contains more information than a single field expectation value, such as its mean. Indeed, specifying a distribution requires additional parameters, which may capture distinct properties. In the limit where the distribution is sharply peaked around its mean, the two descriptions are equivalent; however, away from this limit, the width of the distribution around the classical path can lead to genuinely different behaviour, motivating the distributional description on the associated metric space. Understanding how these effects on field space can furnish new tools to investigate properties of EFTs and their associated Swampland conjectures is one of the central motivations of this work.

\medskip

In the purely field-theoretic setting, the dynamical formulation of OT allows for  natural generalisations that can also incorporate scalar potentials. Furthermore, we show that for vanishing potential, and in the limit of exactly localised distributions given by Dirac delta functions, the framework directly reproduces the standard notion of moduli space distance.

In the case of dynamical gravity, the definition of the relevant cost function becomes more subtle. A qualitatively new feature is the absence of an external time variable with respect to which such dynamical notions of OT can be defined. This is directly reflected in the Wheeler-DeWitt equation, which replaces the Schr\"odinger equation as the relevant dynamical equation~\cite{DeWitt:1967yk, DeWitt:1967ub, Wheeler:1968iap}. One of the main objectives of this work is therefore to adapt the OT framework to this setting and to explore corresponding notions of transport distance between semiclassical configurations described by Wheeler-DeWitt wavefunctions, whose natural variables are fields on an (enlarged) configuration space.

To address this problem, we work within the standard ADM formalism~\cite{PhysRev.116.1322} and deparameterise the system by choosing one of the configuration-space variables as a parameter relative to the others. Since our primary interest is in extracting notions of distance associated with scalar-field configurations, we take this parameter to belong to the metric sector. We then analyse this construction in detail and introduce several candidate cost functions, potentially leading to different notions of distance. These have to be compared in view of extracting information relevant for the Swampland Distance Conjecture. In doing so, we also establish links with previously proposed notions of distance in scalar field space~\cite{Mohseni:2024njl,Debusschere:2024rmi,Li:2023gtt}, while aiming to provide a broader geometric perspective.

We then discuss several examples and applications of our OT framework. In particular, we show that this framework naturally fits within the Hamilton-Jacobi description of (fake-)supergravity, following the discussion of~\cite{Andrianopoli:2009je, Andrianopoli:2007gt, Trigiante:2012eb, Ferrara:2008hwa}. 

\medskip

While probability distributions on moduli space have appeared in various contexts, ranging from flux vacua counting to wave functions in Quantum Cosmology ~\cite{Ooguri:2005vr,Gukov:2005bg,Gukov:2005iy,Balasubramanian:2014bfa,Heckman:2013kza}, our motivation is different\footnote{For a further recent discussion of wavefunctions or probability distributions on moduli space, albeit in a different setting and motivated by distinct considerations, see~\cite{Anchordoqui:2025izb,Anchordoqui:2026nit}.}. Our goal is not to define a probability measure over the landscape. Instead, we aim to quantify how different effective field theories --- represented by semiclassical distributions over scalar field space --- are, by assigning them a suitable notion of distance motivated by OT. One of the aims of this work therefore resides in providing a further step towards a precise geometrical meaning of the notion of distance between effective field theories in quantum gravity, taking into account both non-trivial scalar potentials and contributions from dynamical gravity. 

By employing the framework of Optimal Transport, our proposed framework aims to offer a first-principles characterisation of distance with a precise geometric underpinning. A particularly appealing feature of this perspective is that --- at least in principle --- it offers a possible avenue to incorporate quantum effects, akin to tunnelling effects. More generally, this suggests that studying the full semiclassical distribution, rather than only its expectation value, may provide a useful additional tool for probing Swampland constraints beyond the strictly pointwise notion of field space distance.

\medskip

We provide a succinct and schematic summary on the following page.
The paper is organised as follows. 
Section~\ref{sec:revOT_OC} reviews the necessary elements of Optimal Transport Theory relevant for our construction, including the 2-Wasserstein distance and its extension to Tonelli and Jacobi–Maupertuis costs in the presence of scalar potentials. In Section~\ref{sec:WKB_HJprobs}, we discuss the emergence of semiclassical density distributions on scalar field space in both field-theoretic and gravitational settings. Section~\ref{sec:appl} presents some applications and examples. In Section~\ref{sec:rels}, we comment on the relation of this framework to other existing notions of distance in scalar field space and to related measures appearing in the study of moduli spaces of vacua.
We summarise our results in Section~\ref{sec:conclusions}, and conclude by outlining several interesting directions for future investigation, stressing their direct interplay with the various notions proposed within the Swampland program. Further details, together with possible connections to hyper-distinguishability and the information metric, are collected in the appendix.

\newpage
\paragraph{Schematic summary.}

Our main proposal may be schematically summarised as follows:

\begin{itemize}
\item \textbf{EFTs and semiclassical distributions.}
We use a framework in which effective field theories are represented not by points $\hat{\phi} \in \mathcal{M}_\phi$ in field space, but by probability distributions, with mean values corresponding to $\hat{\phi}$. These are obtained from the WKB expansion at orders $\hbar^0$ and $\hbar^1$ of the associated (semiclassical) wave functions. This leads to the following (schematic) replacement:
\begin{equation*}
     \textit{Point on configuration space \,\,$\hat{\phi} \in \mathcal{M}_\phi$} \,\quad  \rightsquigarrow \quad \rho[\phi] = \left|\Psi[\phi]\right|^2\,,\,\,\,\, \braket{\phi}_\rho = \hat{\phi}
\end{equation*}

\item \textbf{Distances via 2-Wasserstein Optimal Transport and Potentials.}
Given two backgrounds represented by distributions on field space $(M,\mathbf{g})$ with $\mathbf{g}$ the associated metric tensor, we define their separation using the 2-Wasserstein distance on the space of probability measures $\mathcal{P}(M)$. Scalar potentials can be incorporated through the choice of an appropriate cost function $c(x,T(x))$:
\begin{equation*}
\begin{matrix}
\textit{Distances on field space}\,\, (M,\mathbf{g})\\
\textit{with potential}\,\, V
\end{matrix} \quad \rightsquigarrow  \quad 
\begin{matrix}
\textit{Wasserstein distance}\,\, W_{2,c}\, \textit{on}\,\, \mathcal{P}(M)\\
\textit{with cost function $c(x,T(x); \,V)$}

\end{matrix}
\end{equation*}

\item \textbf{Applications \& relations to other proposals.} We apply the construction to controlled examples and study the associated WDW solutions, and using Gaussian approximations, the associated transport distances for specific choices of cost functions. Furthermore, we illustrate how our framework reproduces several previously proposed notions of scalar-field distance while clarifying the assumptions under which these notions can be treated as genuine distances versus action-type costs.
\end{itemize}

\newpage
\section{A review of Optimal Transport}\label{sec:revOT_OC}

This section aims to introduce the geometric framework underlying our notion of distance. Rather than measuring separations between points in scalar field space $M$, we will work on the space of (normalised) densities on $M$ and equip it with a genuine metric, the $2$-Wasserstein distance. 
Optimal Transport provides a canonical way to compare such densities: it measures the minimal cost required to deform one density into another, given a chosen cost on configuration space. A key advantage is that the Wasserstein distance admits both a static (Monge-Kantorovich) and a dynamical (Benamou-Brenier) formulation, the latter being directly compatible with equations of motion arising in physical settings relevant for semiclassical gravity. In particular, once one adopts a transport-based notion of distance, incorporating a scalar potential corresponds to a well-studied modification of the transport cost via Tonelli or Ma\~n\'e (equivalently Jacobi-Maupertuis) cost function, which connects naturally to Hamilton-Jacobi theory and preserves the existence of minimisers. This shift is motivated by the fact that, in a number of controlled quantum-gravity settings, one can naturally associate to a given background a semiclassical functional weight on field space. In physical examples, the width or tail behaviour of the distribution is determined by data connected with the expectation value of the field coordinates in field space plus additional data, such as charges or fluxes.
For use of OT and the Wasserstein distance in other areas of string theory and quantum gravity, see, e.g.,~\cite{DeLuca:2021mcj,DeLuca:2021ojx,Luca:2022inb,DeLuca:2025klz}; for a very recent application within holographic settings, see~\cite{Das:2026hbw,Hashimoto:2026kjy}.

\subsection{Monge-Kantorovich problem and the Wasserstein distance}\label{sec:MK-problem}\label{sec:Wasserstein}

In order to make the notion of distance between densities precise, we briefly review the Monge-Kantorovich formulation of OT and the associated Wasserstein distances. This framework provides a genuine metric on the space of normalised densities or probability distributions once a local cost on the underlying configuration space is specified. We restrict our attention to those elements that will be directly used in later sections and refer to~\cite{ambrosio2021lectures,villani2008optimal} for a more complete and pedagogical introduction to  OT.

Monge's problem~\cite{monge1781memoire} centres around the following question: ``What is the cheapest way to rearrange one distribution into the other, given a local cost function?''. Intuitively, one imagines taking a mass distribution described by an initial density $\mu_0$ and moving or transporting it into another distribution described by the density $\mu_1$, paying some cost according to how far the infinitesimal mass elements are displaced. 
In the present context, these distributions will represent semiclassical densities on scalar field space rather than literal mass profiles, which were the focus of Monge's original analysis.

Consider a smooth, connected Riemannian manifold $(M,\mathbf{g})$, together with a cost function
\begin{equation}\label{eq:cost}
    c : M\times M \to [0,\infty)\,,
\end{equation}
which, for temporary concreteness, may be taken to be the squared geodesic distance $c(x,y)=d(x,y)^2$ associated with the metric $\mathbf{g}$. Let $\mu_0,\mu_1$ be two probability measures on the manifold $M$, i.e., $\mu_0,\mu_1 \in \mathcal{P}(M)$. A map $T : M \to M$ is called a transport map from $\mu_0$ to $\mu_1$ if its push-forward satisfies
\begin{equation}\label{eq:transport}
    T_\#\mu_0 = \mu_1\,.
\end{equation}
More concretely, this means that $\mu_0\left(T^{-1}(A)\right)=\mu_1(A)$ for every Borel set $A\subset M$ . Given such a map, the total transport cost is
\begin{equation}
     C_M(T) \;=\; \int_M c\big(x,T(x)\big)\,\mathrm d\mu_0(x)\,.
\end{equation}
Different choices of the cost function correspond to different notions of distance; the quadratic cost induced by the metric distance will lead to the Wasserstein geometry used throughout this work. Mathematically, Monge's Optimal Transport problem defines the optimisation problem of minimising the following cost over all transport maps $T$
\begin{equation}\label{eq:MK_problem}
    C_M(\mu_0,\mu_1)
= \inf\left\{\,
\int_M c\left(x,T(x)\right)\,\mathrm d\mu_0(x)
\;:\; T_\#\mu_0=\mu_1
\right\}\,.
\end{equation}
This formulation is highly non-linear and, in general, may fail to admit a minimiser or admit multiple inequivalent minimising maps.

\begin{figure}[t]
    \centering
    \includegraphics[height=4.5cm]{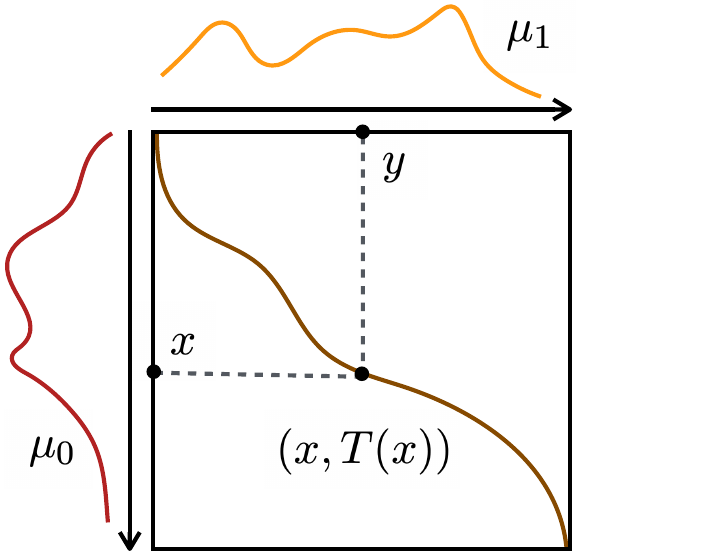}
    \caption{The Monge problem seeks an OT map $T$ between $\mu_0$ and $\mu_1$ that minimises a quadratic cost, while the Kantorovich formulation allows for general transport couplings $\pi(x,y)$ on the product space. For quadratic cost and absolutely continuous $\mu_0$, Brenier's theorem ensures that the Kantorovich minimiser is unique and that the two formulations coincide.}
    \label{fig:transportproblem}
\end{figure}

To overcome these issues, Kantorovich~\cite{kantorovich2006translocation} relaxed the problem by replacing the map $T$ with a coupling (or sometimes also called a transport plan) $\pi \in \mathcal P(M\times M)$; a probability measure on the product space. A coupling is a probability measure $\pi \in \mathcal{P}(M \times M)$ whose marginals reproduce the given measures
\begin{equation}
    (\mathrm{pr}_1)_{\#}\pi = \mu_0\,,\qquad (\mathrm{pr}_2)_{\#}\pi = \mu_1\,.
\end{equation}
where $\mathrm{pr_{1/2}}$ denote the projection onto the first and second factors of $M \times M$, respectively. The Kantorovich problem then amounts to finding a coupling $\pi$ that realises 
\begin{equation}\label{eq:optimal_transport_cost_problem}
    \inf_{\pi \in \Gamma(\mu_0,\mu_1)}
\int_{M\times M} c(x,y)\,\mathrm d\pi(x,y)\,,
\end{equation}
where $\Gamma(\mu_0,\mu_1)$ is the set of all admissible couplings with the correct marginals $\mu_0$ and $\mu_1$.

\noindent
In contrast to Monge’s formulation, this is a linear optimisation problem and always admits at least one minimiser under mild assumptions. For quadratic cost and absolutely continuous $\mu_0$, Brenier's theorem ensures that the Kantorovich minimiser is unique and concentrated on the graph of a transport map, recovering the Monge formulation~\cite{brenier1987decomposition,knott1984optimal,ambrosio2021lectures}. This result will be implicitly used in several of the examples discussed later; cf. Figure~\ref{fig:transportproblem}.

\medskip

Choosing a Finsler\footnote{A Riemannian metric or distance is a special example of a more general Finsler metric, which, in contrast to the Riemannian case, defines a possibly anisotropic and asymmetric norm on $T_xM$; cf.~\cite{bao2012introduction}.} cost like  $c(x,y)=d^p(x,y)$, the optimal value of Kantorovich's problem, determines the $p$-Wasserstein distance $W_p(\mu_0, \mu_1)$. For two probability measures $\mu_0, \mu_1 \in \mathcal{P}(M)$, the $p$-Wasserstein distance is defined via the Monge-Kantorovich formulation~\cite{Kantorovich1960}
\begin{equation}\label{eq:Wp}
	W_p(\mu_0,\mu_1)=\left(\inf_{\pi \in \Gamma(\mu_0,\mu_1)}\int_{M\times M} d(x,y)^p \, \mathrm d\pi(x,y)\right)^{1/p}\,,
\end{equation}
where $ d(x,y)$ is the distance or metric on the manifold $M$ such that the pair $(M,d)$ is a (separable and complete) metric space. We will be mostly interested in the (canonical) case of $p=2$, corresponding to the $2$-Wasserstein (or simply Wasserstein) distance
\begin{equation}\label{eq:W22}
	W_2^2(\mu_0, \mu_1) = \inf_{\pi \in \Gamma(\mu_0, \mu_1)} \int_{M \times M} d(x,y)^2\, \mathrm{d}\pi(x,y)\,.
\end{equation}
Note that in the following, we will often write the measures in the form $\rho(x)\mathrm{d}x$, with $\rho(x)$ being the density function of the probability measure. In a slight abuse of notation, we write $W_p(\rho_0,\rho_1)$, but it should be clear from the context whether $\rho$ denotes the full measure or only the density function. Furthermore, for curved manifolds, we  factor out the volume element, i.e. $\mu= \rho(x) \sqrt{g} \,\mathrm{d}^nx \equiv \rho(x)\, \mathrm{d}V_M$.

In general, the Wasserstein distance is notoriously hard to compute. However, there are two cases in which the computation of the distance drastically simplifies: in one dimension and for Gaussian distributions. The former can be intuitively expected as, in one spatial dimension, OT can never branch: mass always moves monotonically from left to right. As a result, the optimal coupling is an explicitly gradually increasing rearrangement of the two distributions.  If $F_0$ and $F_1$ denote the cumulative distribution functions\footnote{For a measure $\mu_i$, the cumulative distribution function (CDF) is defined as $F_i(x)=\int_{-\infty}^x \mu_i(t) \mathrm{d}t$.} of $\mu_0$ and $\mu_1$, then the OT map is given by $T = F_1^{-1}\circ F_0$ and the Wasserstein distance takes the closed form~\cite{villani2008optimal}
\begin{equation}
    W_2^2(\mu_0,\mu_1)
= \int_0^1 \big|F_0^{-1}(s)-F_1^{-1}(s)\big|^2 \, \mathrm ds \,.
\end{equation}
In particular, there is no need to solve any PDE or optimisation problem.

In the following subsection, we explain how to recast this static formulation into a dynamical one, which will be crucial for incorporating potentials and Hamilton-Jacobi structures. Before moving on, we review the specialisation of the Wasserstein distance to normal (Gaussian) distributions, for which it drastically simplifies.

\paragraph{Gaussian distributions.} Let $\mu=\mathcal{N}(y,\mathbf{\Sigma})$ be a $n$-dimensional multivariate Gaussian distribution with random variable vector $y$ and covariance matrix $\mathbf{\Sigma}$ with generalised variance $|\Sigma|=\det(\mathbf{\Sigma})$. Then $\mu$ is of the form 
\begin{equation}
   \mu= (2\pi)^{-n/2}|\Sigma|^{-1/2}\exp(-\tfrac{1}{2}(x-y)\cdot \mathbf{\Sigma}^{-1} \cdot (x-y))\,,
\end{equation}
and the Wasserstein distance between two such distributions $\mu_i$ over $\mathbb{R}^n$ evaluates to ~\cite{dowson1982frechet,takatsu2011wasserstein}
\begin{equation}\label{eq:W2_ndim_Guassian}
    W_2^2(\mu_0,\mu_1)=\|y_0-y_1\|^2 + \mathrm{tr}\left( \mathbf{\Sigma}_0 +\mathbf{\Sigma}_1 -2\sqrt{\sqrt{\mathbf{\Sigma}_1}\mathbf{\Sigma}_0 \sqrt{\mathbf{\Sigma}_1}} \right)\,,
\end{equation}
where $\|X\|_g^2 := g_{ij}\,X^i X^j$ denotes the squared norm of the vector $X$ with respect to the background metric $g$ on $M$.
Here $\sqrt{\mathbf{\Sigma}}$ is the unique symmetric positive-definite principal square root of $\mathbf{\Sigma}$, a positive (semi-)definite and symmetric covariance metric. It is then easy to see that in the one-dimensional case, with $\Sigma=\sigma^2$, the formula reduces to 
\begin{equation}\label{eq:wasserstein_1dgaussian}
     W_2^2(\mu_0,\mu_1)=|y_0-y_1|^2 + |\sigma_0-\sigma_1|^2\,.
\end{equation}
Finally, let us remark that for generic curved Riemannian manifolds $(M,\mathbf{g})$, the OT map is much more involved, and no closed-form analogue of \eqref{eq:W2_ndim_Guassian} is known\footnote{On curved spaces, the OT map $T$ is generally exponential, while on $\mathbb{R}^n$, it is a simple gradient of convex functions $T=\nabla \phi$; cf.~\cite{villani2008optimal}. See~\cite{gelbrich1990formula} for an extension to (separable) Hilbert spaces.}.

\subsection{Dynamical formulation of the Wasserstein distance}\label{sec:Otto}

The Monge-Kantorovich formulation reviewed above defines the Wasserstein distance as a static optimisation problem between two densities. An equivalent and conceptually useful point of view is to regard $W_2$ as the length of a shortest path in the space of normalised densities on $M$, equipped with a special metric. This interpretation relies on a simple kinematical structure, which will play a central role in later sections, where Hamilton–Jacobi systems naturally generate transport equations for semiclassical densities; we hence briefly summarise here.

A one-parameter family of densities $\rho(s,x)$ defines an admissible path in $\mathcal P(M)$ if there exists a ``time-dependent'' vector field $v(s,x)$ such that
\begin{equation}\label{eq:continuity}
    \partial_s \rho + \nabla\cdot(\rho v)=0\,.
\end{equation}
This continuity equation expresses local mass conservation and identifies the tangent vectors to $\mathcal P(M)$ with velocity fields. Within this framework, the quadratic Wasserstein metric $g_\rho^W(\cdot, \cdot)$ assigns to such a path the kinetic energy functional
\begin{equation}\label{eq:kinetic_energy}
    \|\dot\rho\|_{W}^2\equiv g_\rho^W(\delta \rho,\delta \rho) = \int_M \|v\|^2_g\, \rho\,\mathrm dV_M\,.
\end{equation}
This geometric viewpoint was formalised by Otto~\cite{otto2001geometry}, who showed that the space of probability measures can be treated, formally, as an infinite-dimensional Riemannian manifold.
In the present work, however, we will mainly use the kinematical content of this construction: the continuity equation \eqref{eq:continuity} and the associated notion of kinetic energy.
These ingredients lead directly to the dynamical formulation of OT, which we now review. Some additional details on Otto calculus, the Wasserstein metric, and the connection to gradient flows are presented in Appendix \ref{app:Otto}.

\subsubsection{The Benamou-Brenier formulation}

Benamou and Brenier showed~\cite{benamou2000computational} that the square of the $2$-Wasserstein distance $W_2^2$ can equivalently be expressed ``dynamically'' as a variational problem 
\begin{equation}\label{eq:BB_formula}
	W_2^2(\rho_0, \rho_1) = \inf_{ v(s)} \left\{ \int_0^1 \int_M \|v(s,x)\|^2_g \,\rho(s,x) \, \mathrm{d}V_M \mathrm{d}s \right\}\,,
\end{equation}
supplemented by the constraining equations
\begin{equation}
	\partial_s \rho + \nabla \cdot (\rho v) = 0\,,\qquad \text{with}\quad \rho(0,x) = \rho_0(x)\,, \quad \rho(1,x) = \rho_1(x)\,.
\end{equation} 
Minimisation then implies\footnote{This can be seen by writing the RHS of \eqref{eq:BB_formula} variationally, cf. (3.13) in~\cite{otto2005eulerian}.} that $v$ can be written in terms of a potential $\phi$
\begin{equation}\label{eq:v_potential_function}
    v(s,x)=\nabla \phi(s,x)\,,
\end{equation}
which, in turn, satisfies the Hamilton-Jacobi equation, which, with our sign convention, reads
\begin{equation}\label{eq:BB_Hamilton_Jacobi}
    \partial_s \phi - \frac{1}{2}|\nabla \phi|^2=0\,.
\end{equation}
Together, the two equations on $\partial_s \phi$ define the geodesic equations of the functional.
Along the flow, we see that the ``mass'' is conserved
\begin{equation}
	\int_M  \mathrm{d}V_M\,\rho(x)  = 1\,.
\end{equation}
The main advantage of this formulation is that $\rho(s,x)$ defines a path in $\mathcal{P}(M)$, interpolating between $\rho_0$ and $\rho_1$ for $s \in [0,1]$. The velocity describes the instantaneous movement of the probability density. The role of the emergent continuity equation is to ensure that ``mass'' is conserved along the flow in $\mathcal{P}(M)$. The variational problem selects the shortest path between $\rho_0$ and $\rho_1$ in the space of probability densities; i.e., we obtain a geodesic in $\mathcal{P}(M)$ with respect to the Wasserstein metric $\mathbf g_\rho^W$ and (minimal) length $W_2(\rho_0,\rho_1)$. This dynamical formulation will allow us to make direct contact with Hamilton–Jacobi systems that arise in semiclassical gravity.

\paragraph{Remark.} This dynamical interpretation is special to the quadratic Wasserstein case. For more general costs, one still has a well-defined static Monge-Kantorovich problem, but an equivalent Benamou-Brenier-type formulation need not exist. We will return to this point in Section \ref{sec:HJ_conteq_WdW}, when discussing distances between densities arising from WDW problems for a fixed emergent time choice.

\subsubsection{Deterministic limit}

Before introducing more general cost functions, as well as potentials, we show how the Wasserstein distance reproduces the standard notion of distance on configuration space in the deterministic limit. This limit will later arise naturally in semiclassical gravity when probability densities become sharply localised around attractor loci. It therefore plays a central conceptual role in our construction.

The deterministic limit corresponds to probability measures that are supported at a single point in configuration space, which is modelled by Dirac measures $\delta^{(n)}(x-x_0)$. In this case, the Wasserstein distance will reduce to the geodesic distance associated with the underlying metric on $M$. We now verify this statement both in the static Monge–Kantorovich formulation and in the dynamical Benamou–Brenier formulation.

We consider two Dirac measures $\rho_0=\delta^n(x-x_0)$ and $\rho_1=\delta^n(y-y_1)$ on $\mathbb{R}^n$. Since both are supported at a single point only, the set of admissible couplings $\Gamma(\rho_0,\rho_1)$ contains only one element, the trivial product measure.  
Hence, substituting into~\eqref{eq:W22}, we  obtain
\begin{equation}\label{eq:distance_Dirac_v1}
    W_2^2(\rho_0,\rho_1)= \int_{\mathbb{R}^n \times \mathbb{R}^n} d(x,y)^2\delta^{(n)}(x-x_0)\delta^{(n)}(y-y_1)\,\mathrm{d}^n x \mathrm{d}^n y=d(x_0,y_1)^2\,.
\end{equation}
Thus, the distance is simply given by the  metric distance on $\mathbb{R}^n$. This can be easily cross-checked with the previous example of the Gaussian; taking $\mathbf \Sigma_i \to 0$ and denoting $\mu_0=x_0, \mu_1=y_1$ in~\eqref{eq:W2_ndim_Guassian}, we immediately obtain the above result.

On the other hand, employing the Benamou-Brenier (BB) formula \eqref{eq:BB_formula}, there is an obvious candidate for the path $\rho(s,x)$ connecting these two distributions, given by
\begin{equation}
    \rho(s,x)=\delta(x-\gamma(s))\,,\qquad \gamma(s)=x_0+(x_1-x_0)s\,.
\end{equation}
With the distribution $\rho$ being a Dirac measure, the continuity equation fixes $v=\partial_s\gamma(s)$ (distributionally), and the Hamilton-Jacobi equation \eqref{eq:BB_Hamilton_Jacobi} is solved by $\phi(s,x)=\tfrac12|v|^2 s+v\cdot x$, thus obtaining
\begin{equation}\label{eq:phi_delta}
    \phi(s,x)=\tfrac{1}{2}(x_1-x_0)^2s+(x_1-x_0)\cdot x\,.
\end{equation}
The Benamou-Brenier formula~\eqref{eq:BB_formula} then trivially gives
\begin{equation}
	W_2^2(\rho_0,\rho_1)=\int_0^1 \int_{\mathbb{R}^n} \left(x_0-x_1\right)^2 \rho(s,x) \, \mathrm{d}^n x \,\mathrm{d}s=d\left(x_0,x_1\right)^2\,,
\end{equation}
in agreement with eq.~\eqref{eq:distance_Dirac_v1}.

The above was evaluated for the flat metric $\delta_{ij}$ on $\mathbb{R}^n$. For curved manifolds $(M,\mathbf g)$, the minimising paths $\gamma^\ast$ are the geodesics in $\mathcal{M}$, with minimal length
\begin{equation}
    d(x_0,x_1)= \inf_\gamma \left\{\int_0^1 \sqrt{g_{\gamma(s)}\left(\dot\gamma(s)\dot\gamma(s)\right)}\,\mathrm{d}s  \; \bigg| \;  \gamma(0)=x_0\,, \gamma(1)=x_1 \right\}\,.
\end{equation}
Thus, for Dirac measures $\rho_i=\delta^{(n)}(x-x_i)$, the Wasserstein distance coincides with the Riemannian geodesic distance on $M$, i.e. with the length functional minimised over curves connecting $x_0$ and $x_1$. We will return to the mechanism that leads to such a deterministic limit in the gravitational systems we will be considering in  Section \ref{sec:gaussian_collapse}.

\subsection{Optimal paths with scalar potentials}\label{sec:OTpotential}

In many physical applications, transport between semiclassical states is influenced by scalar potentials, so that the purely kinetic Wasserstein geometry does not reflect the underlying dynamics of the system. We will also encounter this problem  when considering the Swampland distance conjecture within this framework for scalar field spaces. The purpose of this subsection is to explain how such potentials can be incorporated into OT in a mathematically controlled way. We will see that OT with a  Tonelli action cost is equivalent, at the level of minimising trajectories, to standard Wasserstein transport on a Riemannian manifold equipped with a Jacobi–Maupertuis metric. This equivalence provides a practical way of computing distances in the presence of potentials.  The framework discussed so far describes OT with a purely kinetic cost, leading to geodesics of the Wasserstein metric on field space. In many situations of physical interest, however, transport is influenced by scalar potentials. Incorporating such potentials requires generalising the underlying action functional beyond the purely kinetic Benamou–Brenier form.

Starting from this general form, one can instead include a potential by choosing a so-called Tonelli Lagrangian\footnote{A Lagrangian function $L : TM \to \mathbb{R}$, for $M$ compact or non-compact manifolds, is Tonelli if:
\begin{enumerate}
    \item 	The function $L(x,v)$ is twice differentiable in both the base variables $x$ and the fibre variables $v$, with continuous derivatives: $L\in C^2(TM)$,
    \item 	$L(x,\cdot)$ is strictly convex for every $x$, i.e., $\frac{\partial^2 L}{\partial v^2}(x,v)$ is positive definite for all $(x,v)$.
    \item $L(x,v)\to +\infty$ superlinearly in $v$, i.e., $\lim_{\|v\| \to \infty} \frac{L(x,v)}{\|v\|} = +\infty$
   uniformly on compact sets of $x$.
\end{enumerate}} $L(x,v)$ and define the associated Ma\~n\'e (action) cost~\cite{bernard2006monge,fathi2010optimal}
\begin{equation}\label{eq:tonelli_cost}
	c_L(x,y)=\inf_{T>0}\ \inf_{\substack{\gamma(0)=x\\ \gamma(T)=y}} \int_0^T L(\gamma(t),\dot\gamma(t))\,\mathrm dt\,.
\end{equation}
In this subsection, we take the mechanical Tonelli Lagrangian
\begin{equation}\label{eq:tonelli_Lagr}
	L(x,v)=\frac12\|v\|_g^2 - V(x)\,,
\end{equation}
where again $\|\cdot\|_g$ is the norm induced by the metric $\mathbf g$ on $T_xM$, and $V$ is a smooth potential.  Following~\cite{bernard2006monge}, we introduce the $k$-shift
\begin{equation}\label{eq:Lk_shift}
	L_k(x,v)=L(x,v)+k\,,
\end{equation}
with $k>k_0$, where $k_0$ is the Ma\~n\'e critical value.  For $k\ge k_0$, the Ma\~n\'e cost $c_{L_k}$ is finite, satisfies the triangle inequality, and vanishes on the diagonal, $c_{L_k}(x,x)=0$, but it is generally not symmetric and hence does not define a metric.\footnote{On non-compact manifolds, analogous results hold under standard completeness/growth assumptions on the Tonelli Lagrangian and suitable decay/tightness assumptions on the measures; see, e.g.,~\cite{fathi2010optimal} for an OT treatment in non-compact settings.}

For the mechanical Lagrangian above, the Legendre transform yields the Hamiltonian
\begin{equation}
	H(x,p)=\sup_{v\in T_xM}\big(p(v)-L(x,v)\big)=\frac12\|p\|_g^2+V(x)\,.
\end{equation}
Along minimising extrema of $L_k$, one has constant energy, and in the mechanical case, this can be written as the familiar energy-shell condition
\begin{equation}\label{eq:eom_Tonelli}
	\frac12\|\dot\gamma\|^2_g + V(\gamma)=k\,.
\end{equation}
Therefore, $\|\dot\gamma\|_g=\sqrt{2(k-V)}$ whenever $k>V$ along the path.  
Equation \eqref{eq:eom_Tonelli} implies that minimising trajectories of the Tonelli action lie on a fixed energy shell. On such trajectories, the action reduces to the length functional of the Jacobi–Maupertuis metric
\begin{equation}
\mathbf g_J = 2\bigl(k-V(x)\bigr)\mathbf g\,.
\end{equation}
Thus, a dynamical transport problem with potential is equivalent, at the level of minimisers, to a geometric problem of finding geodesics in a potential-dependent Riemannian metric.

Fixing $k$ therefore fixes the Jacobi-Maupertuis (Finsler) geometry governing transport:
\begin{equation}\label{eq:Jacobim}
	\|v\|^{(F_k)}_{g_J}=\sqrt{2(k-V(x))} \|v\|_g\,,\qquad (k>\sup V)\,.
\end{equation}
Here, the supremum is taken over the relevant region, and along the energy-minimising trajectories of~\eqref{eq:eom_Tonelli} satisfying $\tfrac12\|\dot\gamma\|_g^2+V(\gamma)=k$, one has $\|\dot\gamma\|_g=\sqrt{2(k-V)}$. Hence $L(\gamma,\dot\gamma)+k= 2(k-V(\gamma))=\|\dot\gamma\|^{(F_k)}_{g_J}$ and thus
\begin{equation}
	\int_0^T\!\big(L(\gamma,\dot\gamma)+k\big)\,\mathrm dt =\int_0^T\!\|\dot\gamma\|^{(F_k)}_{g_J}\,\mathrm dt\,.
\end{equation}
In our applications, we keep $k$ fixed, viewing it as a reference energy level in configuration space that selects the Jacobi-Maupertuis geometry governing transport. Different choices of $k$ correspond to different resolutions at which the scalar potential influences the distances on field space. The resulting distances should therefore be regarded as scale-dependent cost functions rather than purely intrinsic metrics.

A central result of~\cite{bernard2006monge} states that, for $k>k_0$, the Ma\~n\'e cost and the Jacobi–Maupertuis distance differ only by a boundary term,
\begin{equation}
d_{F_k}(x,y)=c_{L_k}(x,y)+f_k(y)-f_k(x)\,,
\end{equation}
for some continuous function $f_k$. As a consequence, they induce the same optimal couplings in the Monge–Kantorovich problem. Operationally, this means that OT with Tonelli cost can be computed using the simpler geometric picture of Wasserstein transport on the Riemannian manifold $(M,\mathbf g_J)$. We will use this equivalence in the physical examples discussed in Section~\ref{sec:appl}.

Unlike the quadratic case, a simple Benamou-Brenier-type dynamical formulation for general Tonelli costs is not available in full generality; see, e.g.~\cite{elamvazhuthi2024benamou} for recent progress.

\paragraph{Wasserstein distance for Gaussians with Tonelli cost.} In many applications of interest (including the semiclassical regimes considered later), the probability measures on moduli space are sharply localised. In this situation, it is natural to approximate them by Gaussians and evaluate the Tonelli-induced transport distance locally. For the mechanical Tonelli Lagrangian
\begin{equation}
L(x,\dot x)=\tfrac12\,g(x)\dot x^2 - V(x)\,,
\end{equation}
fixing a supercritical energy $k>V(x)$ induces the Jacobi-Maupertuis metric
\begin{equation}\label{eq:Jacobi_metric}
\mathbf g_J(x)\equiv 2\bigl(k-V(x)\bigr)\,\mathbf g(x)\,.
\end{equation}
In this regime, OT with Tonelli cost reduces locally to the standard quadratic Wasserstein problem on the Riemannian manifold $(M,\mathbf g_J)$. We denote the corresponding Wasserstein distance by $W_{2,J}$.

Consider two sharply peaked one-dimensional densities approximated by Gaussians $\rho_i(x)=\mathcal N(x_{0,i},\sigma_{x,i}^2)$ with $i=1,2$, supported in a region where $g_J$ varies slowly. Introducing the locally isometric coordinate
\begin{equation}
z(x)=\int^x \sqrt{g_J(s)}\,\mathrm ds\,,
\end{equation}
the metric becomes flat, and the Gaussians map to $\rho_i(z)=\mathcal N(z_i,\sigma_{z,i}^2)$ with $z_i=z(x_{0,i})$ and  $\sigma_{z,i}^2=g_J(x_{0,i})\,\sigma_{x,i}^2$. Applying the standard one-dimensional Gaussian formula \eqref{eq:wasserstein_1dgaussian}, this time for an underlying Tonelli-cost as in \eqref{eq:tonelli_cost}, we obtain 
\begin{equation}\label{eq:Wasserstein_Gaussian_Jacobi}
W_{2,J}^2(\rho_1,\rho_2) = \left(\int_{x_{0,1}}^{x_{0,2}}\!\sqrt{g_J(x')}\,\mathrm dx'\right)^{\!2} + \left( \sqrt{g_J(x_{0,1})}\,\sigma_{x,1} - \sqrt{g_J(x_{0,2})}\,\sigma_{x,2} \right)^{\!2}\,,
\end{equation}
where the distances are measured with respect to the associated Jacobi metric controlled by the potential term appearing in the Tonelli Lagrangian in the cost-function.
In the sharp localisation limit $\sigma_{x,i}\to0$, the second term vanishes and the distance reduces to the Jacobi-Maupertuis length between the centres, providing a controlled local link between Tonelli OT and the semiclassical distance constructions used later, including the moduli-space distance discussed in Section \ref{sec:rels} and the proposal of~\cite{Mohseni:2024njl}.

\subsection*{Combining Optimal Transport and Optimal Control}\label{sec:OT_and_OC}

In the standard Monge-Kantorovich formulation reviewed in Section \ref{sec:MK-problem}, a cost function $c(x,y)$ between pairs of points $x,y\in M$ is fixed (for example, the squared Finsler distance) and the average cost
\begin{equation}\label{eq:arbitarycost_int}
    \int_{M\times M} c(x,y)\,\mathrm d\pi(x,y)
\end{equation}
is minimised over couplings $\pi$ with prescribed marginals. In this setting, the cost is specified as a function on $M\times M$, without reference to the dynamics of the paths connecting $x$ to $y$. In many physical situations, however, admissible paths are constrained by equations of motion rather than being arbitrary curves in configuration space.

This motivates the use of cost functions that are not assigned directly to pairs of points but instead arise as value functions of variational problems with dynamical constraints. Such costs naturally arise in optimal control theory; see, for instance,~\cite{Agrachev2018}. This setup corresponds to holonomic constraints\footnote{Admissible velocities span a distribution $\Delta\subset TM$. 
The constraints are holonomic if $\Delta$ is integrable, i.e., if it satisfies Frobenius' condition and defines a foliation of $M$ by submanifolds whose tangent spaces coincide with $\Delta$.  In this case, admissible paths are ordinary curves in configuration space, and the induced cost is of standard variational (Tonelli) type.
If $\Delta$ is not integrable, the constraints are non-holonomic, leading to sub-Riemannian geometry with restricted reachability and typically asymmetric costs. 
The latter case is substantially more subtle and will not be considered here; see~\cite{agrachev2009optimal}.}, and is much simpler than the genuinely non-holonomic (sub-Riemannian) case studied in~\cite{agrachev2009optimal}.
Concretely, consider a control system on a smooth manifold $M$ of the form
\begin{equation}
    \dot x(t) = F\big(x(t),u(t)\big)\,,
\end{equation}
where $u(t)$ takes values in a control set $U$ and $F:M\times U\to TM$ is a smooth map.
Given a Lagrangian $L(x,u)$, define the associated control cost between two points $x,y\in M$ as
\begin{equation}\label{eq:OC-cost}
c(x,y)=\inf_{T>0}\Big\{\;\int_0^T L\big(x(t),u(t)\big)\,\mathrm dt\;:\; x(0)=x,\ x(T)=y,\ \dot x = F(x,u) \;\Big\}\,.
\end{equation}
If the point $y$ cannot be reached from $x$ under the control system, the cost is taken to be $c(x,y)=+\infty$. Given such a cost, one may then define a Monge-Kantorovich problem exactly as before, by minimising $\int c\,\mathrm d\pi$ over couplings with fixed marginals.

The Tonelli-Ma\~n\'e costs of Section \ref{sec:OTpotential} arise as special cases of this construction.
Indeed, taking
\begin{equation}
    F(x,u)=u\,,\quad L(x,u)=\tfrac12\|u\|_g^2 - V(x)\,,
\end{equation}
optionally, with the supercritical shift $L\mapsto L+k$, the cost in eq.~\eqref{eq:OC-cost} reduces precisely to the Ma\~n\'e potential associated with the Tonelli Lagrangian. In this sense, OT with Tonelli cost can be viewed as an OT problem whose cost is generated by an underlying optimal control problem.

This observation will be used later to relate our transport-based distances to previously proposed distance measures that incorporate gravitational or dynamical constraints.

\section{Distributions on configuration space and cost functions}\label{sec:WKB_HJprobs}

In the semiclassical regime, the wave function on configuration space admits a WKB (Wentzel–Kramers–Brillouin) form whose leading phase, determined by the principal function, solves a Hamilton-Jacobi equation, while the leading amplitude induces a continuity equation for a density $\rho$ transported along the associated classical flow. This provides a canonical way to associate a given effective field theory background not only with a point in scalar field space but also with a non-trivial distribution on it.

Before explaining in detail how these Hamilton-Jacobi systems can give rise to the sought-after distributions, we review some basic facts about WKB approximations and the Hamilton-Jacobi formulation in general. The one-dimensional examples discussed thereafter serve as first (non-gravitational) realisations to illustrate the approach. We then discuss in Section~\ref{sec:ADM_WdW} and Appendix~\ref{sec:HJ_conteq_WdW} how Hamilton-Jacobi systems likewise arise in the context of the ADM formalism and the Wheeler-DeWitt equation.

\subsection{A brief WKB refresher}\label{sec:WKB_refresher}

In one dimension, the Schr\"odinger equation for a particle of mass $m$ moving in a potential $V(x)$ is given by
\begin{equation}
\left(-\frac{\hbar^2}{2m} \frac{\partial^2 }{\partial x^2} + V(x) \right) \Psi(x,t) = i \hbar \frac{\partial}{\partial t} \Psi(x,t)\,.
\end{equation}
Then any solution $\Psi(x,t)$ induces a continuity equation of the form 
\begin{equation}
    \partial_t \rho + \partial_x j=0\,, 
\end{equation}
where $\rho$ is the probability density $\rho = |\Psi(x,t)|^2$, and $j$ the probability current is given as $j = \frac{\hbar}{m} \mathfrak{Im}\left(\Psi^\ast \partial_x \Psi \right)$. For $\hbar \ll 1$, one can employ the (semiclassical) approximation, in which the (Lorentzian\footnote{We will discuss the Euclidean case, which will play a role in some gravitational settings, later.}) WKB approximation yields solutions of the form (see, e.g.,~\cite{sakurai2020modern,griffiths2018introduction})
\begin{equation}\label{eq:WKB_ansatz}
\Psi(x,t) = A(x,t) e^{\frac{i}{\hbar} S(x,t)},
\end{equation}
where $S(x)$ is given by an expansion in $\hbar$, i.e., $S(x)=S_0(x,t) + \hbar S_1(x,t)+ \hbar^2 S_2(x,t)+\dots$, and $A(x,t)$ is the (real) amplitude. Substituting this ansatz into the Schr\"odinger equation, one can solve the resulting system order by order. In particular, at order $\hbar^0$, one finds that $S_0$ is the classical action satisfying the eikonal equation
\begin{equation}\label{eq:eikonal}
\left(\partial_x S_0(x,t)\right)^2 = 2m\left(-\partial_t S_0(x,t) - V(x)\right)\,.
\end{equation}
At order $\hbar$, on the other hand, one obtains
\begin{equation}
   \partial_t A^2 + \partial_x \left(\frac{A^2 \partial_x S_0}{m}\right)=0\,,
\end{equation}
which, upon identifying $\rho_0=A^2$ as a (classical) leading order probability amplitude, is nothing else than a  continuity equation with probability current $j_0= \tfrac{A^2 \partial_x S_0}{m}$.

\paragraph{Stationary solutions.} For a time-independent problem with conserved energy $E$, we can write
\begin{equation}\label{eq:principal_function_WKB}
    S_0(x,t) = W(x) - E t\,, \qquad A(x,t)= A(x)\,,
\end{equation}
such that \eqref{eq:eikonal} reads
\begin{equation}\label{eq:dx_W}
\left(\frac{\mathrm{d} W(x)}{\mathrm{d}x}\right)^2 =\left(\frac{\mathrm{d} S_0(x)}{\mathrm{d}x}\right)^2 = 2m(E - V(x))\,.
\end{equation}
Note that in this case, $S_0=W(x)-E t$ is exactly the classical Hamilton principal function \eqref{eq:H_principle_function} which we will encounter in the next section, while $W(x)$ is Hamilton's characteristic function \eqref{eq:H_characteristic_function}. Defining the classical momentum as $p(x,E) = \sqrt{2m(E - V(x))}$, the WKB solution then takes on the form\footnote{The associated probability density is
\begin{equation}\dotag{fneq:aa_3}
    \rho_0 = |\psi_0(x)|^2= \frac{|C_+|^2+|C_-|^2}{p(x,E)} + \frac{2}{p(x,E)}\mathfrak{Re}\left(C_+ C_-^\ast e^{ \frac{2i}{\hbar} \int p(x,E) \mathrm d x} \right)\,.
\end{equation}.}
\begin{equation}\label{eq:WKB_solution}
\psi_0(x) = \tfrac{1}{\sqrt{p(x,E)}} \left( C_+ e^{ \frac{i}{\hbar} \int p(x,E) \mathrm d x } + C_- e^{ -\frac{i}{\hbar} \int p(x,E) \mathrm d x} \right) \,,
\end{equation}
giving an oscillatory behaviour in classically allowed regions where $E > V(x)$ and $C_\pm$ are (generically complex) constants that are fixed by boundary conditions and normalisation.
When a particle encounters a potential barrier where $E < V(x)$, it enters a classically forbidden region where the momentum becomes imaginary
\begin{equation}
p(x,E)= i\bar p(x,E) = i \sqrt{2m(V(x) - E)}\,.
\end{equation}
In this region, the WKB solution is
\begin{equation}
\psi_0(x) = \tfrac{1}{\sqrt{\bar p(x,E)}} \left( \bar C_+ e^{ \frac{1}{\hbar} \int \bar p(x,E) \mathrm d x  } + \bar C_- e^{ -\frac{1}{\hbar} \int \bar p(x,E) \mathrm d x} \right) \,,
\end{equation}
which describes exponential decay, i.e., limited transmission through the potential barrier.
Note that at the turning points, the approximated solutions become singular; hence, these points require a separate, and more careful treatment when working with the WKB approximation; cf.~\cite{griffiths2018introduction,bender1999advanced}. Full (approximate) solutions can then be found, at least under certain conditions, by appropriately patching solutions across the turning points.

More generally, for a generic $S(x,t)$, the WKB approximation reads
\begin{equation}\label{eq:WKB_solution_t}
	\Psi_0(q,t) = \sumint \mathrm{d} \alpha  \frac{C(\alpha)}{\sqrt{|J(q,t;\alpha)|}} \exp\left[\frac{i}{\hbar} S_0(q,t;\alpha)\right]\,,
\end{equation}
where $S_0(x,t;\alpha)$ is the leading order classical Solution, $|J(q,\tau;\alpha)|$ is the Jacobian (Van Vleck) determinant for the associated flow of the ``branch'' $\alpha$, and we sum/integrate over all possible branches determined by the initial conditions of the system\footnote{For example, in the time-independent case, $\alpha=\{+,-\}$ corresponds to the two possible signs for $p(x,E)$.}, cf.~\cite{Visser:1992pz} and references therein. The probability density is analogously given by $\rho(q,t) = |\Psi_0(q,t)|^2$.

\paragraph{Remark.}\label{remark:KG_WdW} It is important to recall that the Schr\"odinger formalism extends naturally to Quantum Field Theory. Indeed, as originally discussed in \cite{Luscher:1985iu,Symanzik:1981wd}, the Schr\"odinger picture of QFT is obtained by promoting the canonical momentum to the functional operator $\pi(x)\ \rightarrow\ \hat{\pi}(x) = -i \frac{\delta}{\delta \phi(x)}\,,$
so that the Hamiltonian is represented as 
\begin{equation}
    \hat{H}\left[\phi(x), -i \frac{\delta}{\delta \phi(x)}\right]
=
\int_{\Sigma} d\Sigma  \; \hat{\mathcal{H}}\left[\phi(x), -i \frac{\delta}{\delta \phi(x)}\right]\,,
\end{equation}
with $\Sigma$ the spatial hypersurface orthogonal to the time coordinate $t$, and $\hat{\mathcal{H}}$ the Hamiltonian density. Accordingly, the Schr\"odinger equation is to be interpreted as a functional differential equation,
\begin{equation}
i \partial_t \Psi[\phi(x),t] = \hat{H}\,\Psi[\phi(x),t]\,.
\end{equation}
The definition of the cost function can then be easily modified once the changes are implemented. This will be different for dynamical gravity\footnote{Accordingly, the Hamiltonian contains both a quadratic momentum contribution, proportional to $(\pi(x))^2$, and a gradient term of the form $(\nabla \phi(x))^2$, which acts as an additional effective potential associated with the spatial variation of the scalar field. In the following, we will mostly focus on symmetric configurations, for which this term vanishes. We refer the reader to~\cite{Luscher:1985iu,Symanzik:1981wd,Weinberg:1982id} for more complete discussions.} as we will discuss in Section~\ref{sec:ADM_WdW}.

Additionally, when we consider alternative wave equations, for example the WDW equation, the WKB ansatz requires additional care in defining a probability density. The Wheeler-DeWitt equation is of Klein-Gordon type on minisuperspace, with an associated conserved current but no positive-definite probability density in general. In particular, generic WKB solutions contain both ``ingoing'' and ``outgoing'' (i.e., positive and negative frequency) components, and the current is not sign-definite. A consistent probabilistic interpretation therefore requires selecting a suitable branch of the solution (e.g., outgoing modes or a choice of intrinsic time slicing) so that the flux defines a positive measure on the space of classical trajectories~\cite{DeWitt:1967yk,Hartle:1983ai,Vilenkin:1987kf}. See, e.g.,~\cite{Isham:1992ms,Halliwell:1989myn} for reviews.

\subsection{From Hamilton-Jacobi to distributions}\label{sec:associatedWKB}

In Section \ref{sec:revOT_OC}, we explained how distances between density measures are defined once a cost on field space is chosen. In the following, we illustrate how, by identifying Hamilton-Jacobi systems arising in semiclassical gravity, one can associate to a given theory a WKB measure on scalar field space. This is the object whose Wasserstein distance we will compute in Section \ref{sec:appl} in several different explicit settings.

\subsubsection{Basics of Hamilton-Jacobi theory}\label{sec:basics_HJ}

Let $H(q,p,t)$ be the Hamiltonian of a classical system on a $n$-dimensional manifold $M$ with configuration space variables $(q^1,\dots,q^n;p_1,\dots,p_n)$. The equation 
\begin{equation}\label{eq:Hamilton_Jacobi}
    \frac{\partial}{\partial t} S + H(q,\nabla S,t)=0
\end{equation}
is known as the Hamilton-Jacobi equation and provides a defining equation for $S(q,t)$ the classical (on-shell) action. The solution $S_0(q,t,q_i,t_i)$ to a given initial configuration $(q_i,t_i)$ is then referred to as Hamilton's principal function and is nothing else than the action functional $S[q(t)]$ evaluated on the classical solution
\begin{equation}\label{eq:H_principle_function}
    S_0(q,t,q_i,t_i)=S[q_{\mathrm{classical}}(s)]\bigr|_{t_i}^{t}\,.
\end{equation}
In solving the system defined by this equation, one usually makes use of canonical transformations from the pair $(q,p)$ to new canonical coordinates $(Q,P)$, and $S_0$ can be viewed as the generating function for this canonical transformation.

In case the Hamiltonian $H$ is time-independent, i.e. $H(q,p,t)=E$, the Hamilton-Jacobi equation takes the simple form
\begin{equation}
    -  \frac{\partial}{\partial t} S_0  = H(q,p,t)=E\,,
\end{equation}
such that
\begin{equation}\label{eq:H_characteristic_function}
    S_0(q,t) = W(q)-E\,t\,.
\end{equation}
Here, $W(q)$ is Hamilton's characteristic function, which satisfies
\begin{equation}
    p_i =  \frac{\partial S_0}{\partial q^i}  =  \frac{\partial W}{\partial q^i}\,.
\end{equation}
For further details, we refer to the standard literature~\cite{landau1960mechanics, arnol2013mathematical,goldstein1950classical}.

\subsubsection{Obtaining the distribution}\label{sec:obt_distr}

Consider a classical Hamilton-Jacobi system on a (for now arbitrary) configuration space with coordinates $q^i$ and metric $g_{ij}$, governed by a principal function $S_0(q,t)$. Then the semiclassical WKB approximation yields a wavefunction of the form \eqref{eq:WKB_solution_t}, with $S_0$ solving the Hamilton-Jacobi equation~\eqref{eq:Hamilton_Jacobi}. The associated density $\rho=|\Psi_0|^2$ obeys a continuity equation
\begin{equation}\label{eq:continuity_equ_HJ}
	\partial_\tau \rho+\nabla_i(\rho\,v^i)=0\,,	\quad	v^i=-\nabla^i S_0=-g^{ij}\partial_j S_0\,,
\end{equation}
which describes transport along the classical flow generated by $S_0$. Working with a (classical) Hamiltonian with a generic potential $V(q,t)$ of the standard form 
\begin{equation}
    H=\frac{1}{2}g_{ij}\dot q^i \dot q^j +
V(q,t)\,,
\end{equation}
the flow generated by the continuity equation and $S_0$ will be geodesic with respect to $g_{ij}$ in the case of $V=0$, while in the generic case $V\neq 0$, the flow is geodesic with respect to the Jacobi metric $\mathbf g_J=2(E-V)\mathbf g$.

\medskip

Hamilton-Jacobi equations of this type naturally govern the effective one-dimensional dynamics of scalar fields in gravitational backgrounds, such as those arising in attractor flows, domain walls, and fake supergravity descriptions. Given a Hamilton-Jacobi principal function $S_0(q,t)$ --- or equivalently the characteristic function $W_0(q)$ in stationary situations as defined in eq. \eqref{eq:H_characteristic_function}  --- the associated classical trajectories are obtained from its gradient. The corresponding semiclassical density is then transported along this flow according to a continuity equation of the form \eqref{eq:continuity_equ_HJ}.

This structure will be illustrated in several examples analysed in Section \ref{sec:fakeSUGRA_examples}. As will be discussed there, the pair $( W_0,\rho)$ can be reinterpreted directly in the language of OT: $W_0$ plays the role of a velocity potential, and $\rho$ is the transported density entering the continuity equation. Allowing for scalar potentials then leads\footnote{At least in the field-theoretic (non-gravitational) cases.} to a natural Jacobi-type deformation of the transport cost, as analysed in Section \ref{sec:OTpotential}.

\paragraph{Comments on the Euclidean vs. Lorentzian picture.} Anticipating the applications, we now focus on effective one-dimensional reductions with coordinate $\phi$, metric $g(\phi)$, and evolution parameter $\tau$. We work strictly at leading order in the semiclassical expansion and restrict to a single WKB branch, so that the wavefunction takes the form
\begin{equation}\label{eq:wavefunction_LE}
	\Psi(\phi,\tau)=A(\phi,\tau)\exp\left(\frac{i}{\hbar}S_0(\phi,\tau)\right)\,,
\end{equation}
So far, we have implicitly taken the phase $S_0$ to be real (Lorentzian branch). However, we will also be interested in the Euclidean case, which will become relevant in certain semiclassical gravitational settings and, in particular, in the setting of fake supergravity in Section~\ref{sec:fake_SUSY_HJ}. In the latter case (Euclidean branch), one has to Wick rotate such that $S_0=i S_E$ and, therefore\footnote{For the moment we again define the densities via the canonical assignment $\rho=|\Psi|^2$; see, however, the remark on p.\,\pageref{remark:KG_WdW}.}
\begin{align}
	\Psi_L(\phi) &=A(\phi)\exp\left(\frac{i}{\hbar}S_L(\phi,\tau)\right)\,, & 
	\rho_L(\phi) &=A^2(\phi,\tau)\,,\label{eq:Lor_dens} \\ 
	\Psi_E(\phi) &=A(\phi,\tau)\exp\left(-\frac{1}{\hbar}S_E(\phi,\tau)\right)\,, &
	\rho_E(\phi) &=A^2(\phi,\tau)\exp\left(-\frac{2}{\hbar}S_E(\phi,\tau)\right)\,.\label{eq:Eucl_dens}
\end{align}
In the physical applications that we consider here, the relevant branch depends on the framework under consideration. In the examples below, we hence proceed pragmatically and choose the branch depending on the system under consideration. The WKB densities obey the continuity equation~\eqref{eq:continuity_equ_HJ} and will provide the canonical semiclassical density measures on scalar field space, whose Wasserstein distances we will study in the following sections. We note that the normalisability of the WKB density is not automatic and depends on the stabilising properties orthogonal to the classical trajectory of the Hamilton-Jacobi flow. Since on Euclidean branches the phase becomes purely imaginary, $S_0=iS_E$, the resulting density $\rho_E\propto e^{-2S_E/\hbar}$ is exponentially weighted and hence generically better behaved than the Lorentzian counterpart and, in particular, normalisable under appropriate conditions.

\subsubsection{Deterministic limits}\label{sec:det}

We analyse the local structure of semiclassical densities arising from the Hamilton-Jacobi and WKB frameworks and clarify how sharply localised limits relate to the deterministic distances discussed in Section \ref{sec:Wasserstein}.

\paragraph{Gaussian approximation near a minimum.}\label{sec:gaussian_limit} Semiclassical densities derived from Hamilton-Jacobi flows are complicated functions at generic points in moduli or field space. However, in certain situations, for example, in the presence of a stable critical point, the distribution is sharply peaked near a critical point of the principal function $S_0$ (or the characteristic function $W_0$, respectively). In a neighbourhood of such a point, the distribution can be approximated by a particular class of Gaussian distributions.

Working in Riemann normal coordinates $\xi^i$ centred around the point of interest $\phi_*$ (e.g., the attractor point), such that $G_{ij}(\phi_*)=\delta_{ij}$, we expand  the principal function as
\begin{equation}
    S_0(\phi) = S_{0,*} + \tfrac12\,\xi^i H_{ij}\,\xi^j + O(\xi^3)\,,
\end{equation}
where $H_{ij}=\nabla_i\nabla_j  S|_{\varphi_*}$ is the Hessian $\mathbf{H}$ at $\varphi_*$. The Euclidean WKB density obtained from the Hamilton-Jacobi system \eqref{eq:Eucl_dens} then takes the local form
\begin{equation}\label{eq:approx_gaussian}
    \rho_E(\xi)\propto\exp\left[-\frac{2}{\hbar}\left( S_{0,*} + \tfrac12\,\xi^T \mathbf{H}\,\xi\right)\right]\propto \exp\left[-\frac{1}{\hbar}\,\xi^T \mathbf{H}\,\xi\right]\,.
\end{equation}
This is a centred Gaussian with covariance matrix $\mathbf \Sigma = \frac{\hbar}{2}\,\mathbf{H}^{-1}$. In the one-modulus case, with coordinates $\varphi$ and $\kappa=\partial_\varphi^2 S_0|_{\varphi_*}$, this reduces to
\begin{equation}\label{eq:gaussapprox_var_clean}
    \rho_E(\varphi)\simeq \exp\Big[- \frac{\kappa}{\hbar}(\varphi-\varphi_*)^2\Big]\,, \qquad \sigma_\varphi = \sqrt{\frac{\hbar}{2 \partial_\varphi^2 S_0}}\Bigg|_{\phi_*}\,.
\end{equation}
This approximation is valid only locally, around the attractor point. We will return to this point in Section \ref{sec:fakeSUGRA_examples}.

\paragraph{Collapse to a Dirac measure.}\label{sec:gaussian_collapse} Using the Gaussian approximation, we can precisely define in what sense semiclassical densities can collapse to sharply localised configurations or points in the pertinent space. Within the Gaussian approximation \eqref{eq:approx_gaussian}, the covariance $\mathbf \Sigma$ controls the width of the distribution, and the limit $\mathbf \Sigma \to 0$ leads to a deterministic collapse $\rho_E \rightarrow \delta_{\phi_{cl}}$. This collapse can arise in different physical regimes or limits. One distinguished case is the semiclassical limit $\hbar\to0$ at fixed Hessian, for which $\mathbf \Sigma = \hbar\,\mathbf H^{-1} \to 0$, directly leading to localisation of the density. Yet another possibility arises through the relation between entropy and the Hessian. In large-entropy attractor regimes, the Hessian $\mathbf H$ typically grows with the charge or entropy scale, so that $\mathbf \Sigma \sim \hbar\,\mathbf H^{-1}$ shrinks as $S_{\rm BH}$ increases.

Once the semiclassical density collapses to a Dirac measure, distances between such configurations can be meaningfully compared to point-particle distances on moduli space. As shown in Section \ref{sec:Wasserstein}, when both measures are Dirac deltas and the transport cost is fixed, the Wasserstein distance reduces to the corresponding geodesic distance associated with that cost (metric or Jacobi, depending on the setup). We will see all these cases exemplified in the later Section \ref{sec:appl}.

\subsubsection{An illustrative (non-gravitational) example}

Before turning to explicit realisations of the Hamilton-Jacobi system within gravitational theories, we present a simple class of examples that illustrate two key points. First, we demonstrate how the WKB prefactor naturally induces a probability density on configuration space. Second, we show how the inclusion of a potential naturally leads to a modified cost of Tonelli- or Jacobi-Maupertuis-type in the associated OT problem. We consider one-dimensional semiclassical systems with simple potentials $V(x)$.

\paragraph{Symmetric double well.} Let $V$ be the symmetric double-well potential
\begin{equation}
	V(x) = \lambda (x^2 -a^2)^2\,.
\end{equation}
The potential has local minima at $x=\pm a$ with $V(\pm a)=0$ and a local maximum at $x=0$ with $V(0)=\lambda a^4$. For a given fixed energy $E>\lambda a^4$, the potential has (outer) turning points at 
\begin{equation}
    x^\circ = \pm\sqrt{a^2 + \epsilon}\,,
\end{equation}
with $\epsilon=\sqrt{E/\lambda}$, while for $E<\lambda a^4$, there are additional (inner) turning points at 
\begin{equation}
    x^\ast = \pm\sqrt{a^2 - \epsilon}\,.
\end{equation}
Hence, there is a forbidden region $(-x^\ast, x^\ast)$ through which no classical path connecting the two vacua exists. Using the WKB approximation, we can associate wave functions with the two vacua at $\pm a$, around which we can approximate the potential near $\pm a $ as harmonic
\begin{equation}
    V(x) \approx \frac{1}{2}\omega^2(x \mp a)^2\,, \qquad \omega =a\sqrt{8\lambda}\,.
\end{equation}
Therefore, the (ground state) wave function for the local minimum at $a$ reads 
\begin{equation}
    \psi_{\pm a}(x) = \left(\tfrac{1}{\pi \sigma^2}\right)^{1/4} e^{-(x\mp a)^2/(2\sigma^2)}\,, \quad \sigma^2=\frac{\hbar}{\omega}
\end{equation}
which is nothing else than the local harmonic oscillator approximation.
\begin{figure}[t]
    \centering
    \includegraphics[height=4.35cm]{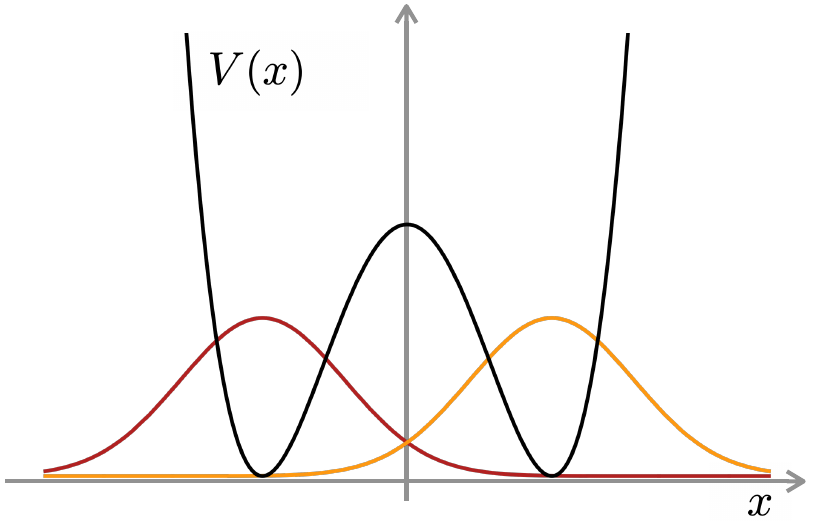}
    \caption{The well-known (symmetric) double well potential can be viewed as a toy model for calculating the distance between two local minima or vacua in a non-gravitational setting. The minima can be approximated by a harmonic potential and hence naturally give rise to Gaussian wave functions and distributions. The distance between the two vacua in our framework is then measured by the (modified) Wasserstein distance with Tonelli cost between the two distributions.}
    \label{fig:placeholder1}
\end{figure}
Following our proposed method, we can calculate the distance between the two vacua by calculating the associated Wasserstein distance (with Tonelli cost) for the two normalised densities
\begin{equation}
    \rho_a(x)= | \psi_a(x)|^2\,, \quad \rho_{-a}(x)= | \psi_{-a}(x)|^2\,,
\end{equation}
associated with the minima. We have to distinguish between the case $E> V(0)$ where there is a classical trajectory between the vacua and the case $E<V(0)$, in which the vacua are separated by a potential barrier, and any transition necessarily proceeds via tunnelling.

\paragraph{i) $\bm{E>V(0)}$:} In the classically allowed regime, we may directly apply formula~\eqref{eq:Wasserstein_Gaussian_Jacobi} to obtain
\begin{equation}
    W_{2,J}^2(\rho_a, \rho_{-a}) \simeq\left(\sqrt{2E}\int_{-a}^{a} \left(1-\frac{V(s)}{2E}\right)\mathrm{d}s \right)^2\simeq 2E \left( 2a - \frac{1}{2E}\frac{16 a^5\lambda}{15}\right)^2\,,
\end{equation}
where we used the fact that the variances and vacuum expectation values of both distributions agree and thus cancel, so that we obtain
\begin{equation}
    W_{2,J} \simeq \sqrt{2E}2a - \frac{16 a^5 \lambda}{15 
    \sqrt{2E}}\,.
\end{equation}
For large $E\gg 1$, the formula reduces to
\begin{equation}\label{eq:modulidistance1}
    W_{2,J} \simeq \sqrt{2E}2a = \sqrt{2E}\Delta x\,,
\end{equation}
which is just the field-space distance\footnote{Since we need to have $E>\lambda a^4$ to stay in the regime in which the above analysis is valid, the first factor dominates and $W_{2,J}$ is always positive.} weighted by the energy $\sqrt{2E}$, which --- after normalising by $\sqrt{2E}$ ---  agrees with the distance of~\cite{Mohseni:2024njl}.

\paragraph{ii) $\bm{E<V(0)}$:} When the energy lies below the top of the barrier at $x=0$, the two wells are separated by a classically forbidden region, and the only way to transition between the vacua is via tunnelling. In this case, the transport problem must incorporate forbidden segments, requiring an appropriate modification of the cost functional. A natural choice is to employ the absolute-value Jacobi cost\footnote{In particular, this does not define a Riemannian metric globally.} (cf. \eqref{eq:Jacobi_metric})
\begin{equation}
    g_J^\ast(x) = 2 |E-V(x)|g(x)\,,
\end{equation}
which remains real and well-defined across both allowed and forbidden regions.
Then, using \eqref{eq:Wasserstein_Gaussian_Jacobi} and the fact that the standard deviations agree, we obtain
\begin{align}
    W_{2,J}^2(\rho_a, \rho_{-a}) &=\left(\int_{-a}^{-x^\ast} \sqrt{g_J^\ast(s)}\mathrm{d}s +\int_{-x^\ast}^{x^\ast} \sqrt{g_J^\ast(s)}\mathrm{d}s +\int_{x^\ast}^{a} \sqrt{g_J^\ast(s)}\mathrm{d}s \right)^2\nonumber\\
    &=\left(2\int_{-a}^{x^\ast} \sqrt{2(E-V(s))}\mathrm{d}s +2\int_{0}^{x^\ast} \sqrt{2(V(s)-E}\mathrm{d}s \right)^2\,,
\end{align}
where we used the symmetry of the potential. Although the above can be solved analytically, the regime of primary interest for us is the limiting case of zero $E$, corresponding to pure tunnelling behaviour. In this limit of $E \to 0$, we have $x^\ast \to a$, and therefore the above formula reduces to
\begin{equation}\label{eq:S_int}
   W_{2,J}= 2\int_{0}^{x^\ast} \sqrt{2(V(s)-E)}\mathrm{d}s
   \simeq 2\int_{0}^{a} \sqrt{2V(s)}\mathrm{d}s=\frac{4}{3}\sqrt{2\lambda}a^3\,,
\end{equation}
which is nothing else than the Euclidean instanton action $S_{\mathrm{inst}}$ for the double well potential. The distance between the vacua grows with the height of the barrier in the middle --- parametrised by $\lambda$ --- and diverges in the limit $\lambda \to \infty$.

\paragraph{Asymmetric double well.} One can easily generalise the above analysis to the case of an asymmetric potential, which can be seen as a naive field theoretic toy model for vacuum decay. To avoid a repetitive discussion, we only give a brief qualitative illustration of the main properties. 

Consider a potential with a local and a global minimum, located at $x=x_1$ and $x=x_2$, respectively. We can approximate the potential around each minimum as harmonic
\begin{equation}
    V(x) \simeq V(x_{0,i}) + \frac{\omega_i^2}{2}(x-x_{0,i})^2\,,
\end{equation}
with $\omega_i = \sqrt{V''(x_{0,i})}$ and associate to it Gaussian distributions $\mathcal{N}(x_i,\sigma_i)$ with $\sigma_i^2=\frac{\hbar}{\omega_i}$. We again have to distinguish between the classically allowed and forbidden cases:

\paragraph{i) $\bm{E>V(0)}$:} Similarly to the symmetric case, we obtain
\begin{equation}
    W_{2,J}^2 \simeq 2E\left(\Delta x - \frac{1}{\sqrt{2E}}\int_{x_1}^{x_2}V(s)\mathrm{d}s \right)^2 + \left(\sqrt{2(E-V(x_1))}\sigma_1 -\sqrt{2(E-V(x_2)}\sigma_2 \right)^2\,.
\end{equation}
In the limit of $E \to \infty$, this gives
\begin{equation}
    W_{2,J}^2 \simeq 2E \left(|x_2-x_1|^2 +|\sigma_2-\sigma_1|^2 \right)\,.
\end{equation}

\begin{figure}[t]
    \centering
    \includegraphics[height=4.3cm]{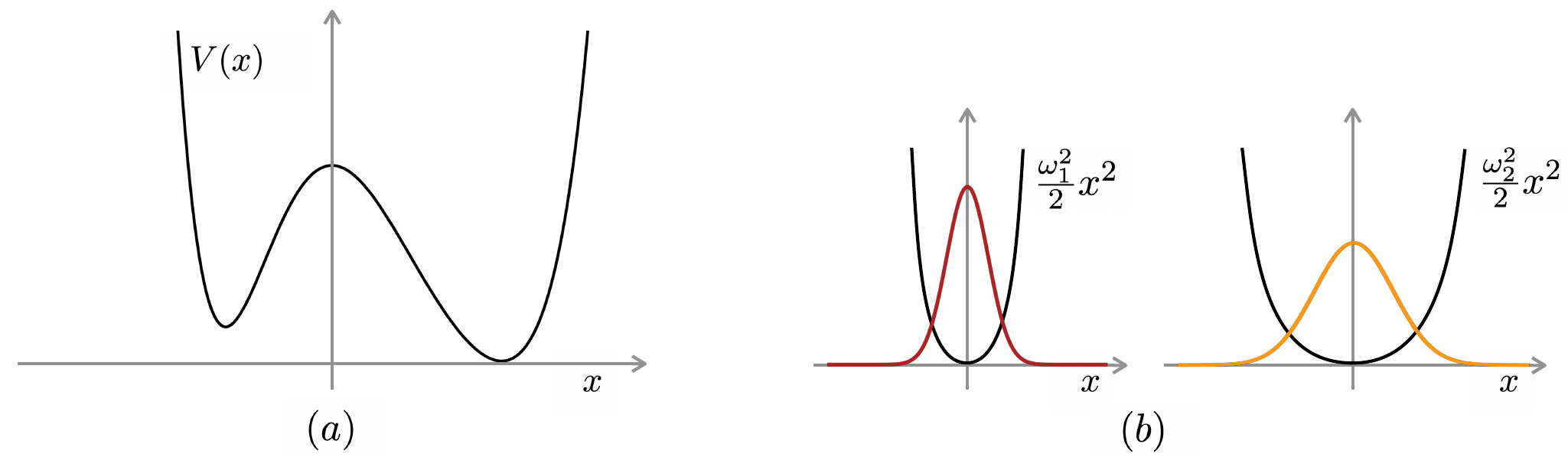}
    \caption{In case the double well potential is asymmetric --- although the minima can be approximated as harmonic --- the resulting curvature is different hence giving rise to wavefunctions with different variances and generically different vacuum expectation value. In this case  the Wasserstein distance (with Jacobi metric) explicitly depends also on the vacuum expectation values and variances.}
    \label{fig:placeholder2}
\end{figure}

\paragraph{ii) $\bm{E<V(0)}$:} Since the minima now lie at different vacuum energies, we cannot simply send $E \to 0$.  Instead one has to do a more careful analysis and split the path into allowed and disallowed regions. However, qualitatively --- if the difference in vacuum energy is small --- the resulting distance will be similar to \eqref{eq:S_int}, however now with an additional contribution from the standard deviations of the form 
\begin{equation}
   \left(\sqrt{2V(x_1)}\sigma_1-\sqrt{2V(x_2)}\sigma_2 \right)^2\,.
\end{equation}

\medskip

\noindent In both cases i) and ii), we see that not only does $\Delta x \to \infty$ create infinite distances, but also divergent variances $\sigma_i \to \infty$ or a infinitely high potential barrier $V(0) \to \infty$ can lead to an infinite Wasserstein distance.

\medskip

These examples illustrate a generic feature of the distance --- or more precisely cost --- between vacua or local minima of the potentials, as measured in the Gaussian approximation. At least qualitatively, it can be read off from the curvature of the potential (determining the variance of the distribution) as well as the separation of the minima (mean of the distribution) and the vacuum expectation value\footnote{Furthermore, the resulting cost, also depends on the available energy $E$.}. In principle, this naturally provides a cost for potentials of the form presented in~\cite{Shiu:2023bay}. However, as we will see in the following, in the presence of gravity, the situation generically becomes more involved.

\paragraph{Remark.} Note that although the vacua in the above examples corresponded to a common configuration space, namely a common space of probability measures, we treated them as separate distributions, normalised independently. This allowed us to calculate the distance between the two configurations where we have a Gaussian distribution at one vacuum to the one where the distribution is centred at the other vacuum. One might be tempted to define an overall distribution that takes into account relative probabilities of the vacuum configurations; however, it is then not immediately clear what the relevant transport problem is through which the distance in probability space is defined.

Furthermore, we want to stress that the above example is meant to serve as a toy example. In particular, a full analysis would require also a careful analysis of the behaviour at turning points, where the approximated solutions become singular in general; cf. Section~\ref{sec:WKB_refresher} above.

\subsection{ADM formalism and the Wheeler-DeWitt equation}\label{sec:ADM_WdW}

We will briefly review the ADM (Arnowitt-Deser-Misner) formalism and its relation with the Wheeler-DeWitt equation. For a more exhaustive and rigorous formulation, see~\cite{PhysRev.116.1322, Bergmann:1972ud, Hanson:1976cn, DeWitt:1967yk, Isham:1992ms, Hartle:1972ya, Halliwell:1988wc}.  
Consider a globally hyperbolic spacetime endowed with a metric $(M,g_{\mu\nu})$ foliated by spacelike hypersurfaces $\Sigma_t$, where $t$ parameterise the different foliations along $M$. 
The ADM coordinates define a line element $ds^2$
\begin{equation}\label{eq:ADM_metric}
	\mathrm ds^2 = -N^2 \mathrm dt^2 + h_{ij}\big(\mathrm dx^i + N^i \mathrm dt\big)\big(\mathrm dx^j + N^j \mathrm dt\big)\,,
\end{equation}
where $h_{ij}$ denotes the components of the induced metric $\mathbf h$ on $\Sigma_t$, $N$ is the lapse function, and $N^i$ the shift vector. Intuitively, the lapse function $N$ gives information about the time lapse between the events in two different foliations, while the shift vector represents how much two points on the same hypersurface are displaced. 
Up to boundary terms with respect to the action
\begin{equation}
    S_{EH} = \int d^dx \sqrt{-g} R\,,
\end{equation}
in the coordinate system \eqref{eq:ADM_metric}, the momentum conjugate to $h_{ij}$ is
\begin{equation}
	\pi^{ij}(x) = \frac{\delta \mathcal{L}}{\delta \dot{h}_{ij}} = \sqrt{h}\,\big(K^{ij}-h^{ij}K\big)\,,
\end{equation}
where $K=h^{ij}K_{ij}$ is the extrinsic curvature defined by
\begin{equation}
	K_{ij} = \frac{1}{2N}\big(\dot{h}_{ij}-D_iN_j-D_jN_i\big)\,.
\end{equation}
The ADM action takes the Hamiltonian form
\begin{equation}
	S_{\mathrm{ADM}} = \int \mathrm dt \int_{\Sigma}\mathrm d^{d-1}x\, \sqrt{h}
	\Big(\pi^{ij}\dot h_{ij}-N\mathcal H-N^i\mathcal H_i\Big)\,,
\end{equation}
with the (Dirac) constraints 
\begin{equation}\label{eq:WdW_quantum_constr}
\mathcal{H} = \frac{1}{\sqrt{h}}\Big(\pi^{ij}\pi_{ij}-\frac{1}{d-2}\pi^2\Big)-\sqrt{h}\,R=0\,,\qquad
\mathcal{H}_i=-2D_j\pi^j{}_i=0\,.
\end{equation}
Canonical quantisation promotes $h_{ij}(x)\mapsto \hat h_{ij}(x)$ and $\pi^{ij}(x)\mapsto -i\hbar\,\frac{\delta}{\delta h_{ij}(x)}$, so that the Dirac constraints \eqref{eq:WdW_quantum_constr} return
\begin{equation}
	\hat{\mathcal{H}}\,\Psi[\mathbf h] \approx 0\,,\qquad \hat{\mathcal{H}}_i\,\Psi[\mathbf h] \approx0\,,
\end{equation}
where $\approx$ denotes weak (on-shell) equivalence. From now on, we will again use the equality symbol, except in special cases.
For the Hamiltonian constraint, one often writes schematically the so-called Wheeler-DeWitt (WDW) equation
\begin{equation}\label{eq:WdW}
\left[-\hbar^2\,G_{ijkl}\frac{\delta^2}{\delta h_{ij}\,\delta h_{kl}}+\sqrt{h}\,R\right]\Psi[\mathbf h]=0\,,
\end{equation}
where the DeWitt supermetric is
\begin{equation}
	G_{ijkl}=\frac{1}{2\sqrt{h}}\Big(h_{ik}h_{jl}+h_{il}h_{jk}-\frac{2}{d-2}h_{ij}h_{kl}\Big)\,.
\end{equation}
The DeWitt metric equips the configuration space or superspace with an indefinite (Lorentzian) metric structure; the momentum constraint then enforces invariance under spatial diffeomorphisms.

In the presence of matter fields, for instance scalar fields $\phi^a$ with metric $G_{ab}(\phi)$ and potential $V(\phi)$ one obtains the schematic extension
\begin{equation}
\left[-\,G_{ijkl}\frac{\delta^2}{\delta h_{ij}\,\delta h_{kl}}
-\,\frac{1}{\sqrt{h}}\,G^{ab}(\phi)\frac{\delta^2}{\delta\phi^a\,\delta\phi^b}
+\sqrt{h}\,R+\sqrt{h}\,V(\phi)\right]\Psi[\mathbf h,\phi]=0\,,
\end{equation}
up to factor ordering ambiguities, which we will choose such as to be compatible with the invariance under diffeomorphism~\eqref{laplace-beltrami}. This is the ``frozen formalism'': no external time variable is present, and semiclassical evolution must be extracted relationally along WKB branches.

In the oscillatory WKB regime, the WDW equation admits two local semiclassical branches corresponding to opposite orientations of the classical flow in minisuperspace. Restricting to a single branch and evaluating the state on a hypersurface transverse to this flow yields a positive, conserved measure that can be interpreted as a probability distribution~\cite{Halliwell:1989myn}. In what follows, we will work directly with the induced semiclassical density defined by such systems, which is sufficient for defining Wasserstein distances between different configurations.

A characterising feature of Wheeler-DeWitt systems is the absence of a preferred external time parameter. In contrast to the Hamiltonian systems discussed above, where the Hamilton-Jacobi equation directly generates a continuity flow for the associated semiclassical density, the WDW equation is intrinsically timeless. However, in the semiclassical regime, one may restrict to a given WKB branch and choose a suitable internal clock, for instance a scalar field or the scale factor, provided it is monotonic along the corresponding classical trajectories. Once such an emergent time variable is fixed, the timeless WDW constraint induces a reduced Hamilton-Jacobi problem together with an associated continuity equation for the semiclassical density, as reviewed in Section  \ref{sec:obt_distr}. In this sense, the corresponding probability distribution may again be viewed as ``evolving'' along a well-defined flow, now defined relationally with respect to the chosen clock. The derivation is given in Appendix \ref{sec:HJ_conteq_WdW}.

\subsection{Cost functions in the presence of gravity}\label{sec:cost_and_gravity}

Contrary to the fake supergravity systems  we will discuss in Section \ref{sec:fake_SUSY_HJ}, in Wheeler-DeWitt problems, there is no unique a priori preferred transport cost. Once an internal clock has been chosen, the WKB approximation produces a continuity equation for semiclassical densities --- see Appendix \ref{sec:HJ_conteq_WdW} --- but the reduced Hamiltonian is generically not quadratic in the momenta. As a result, the transport problem naturally associated with the reduced dynamics is not, in general, the quadratic Wasserstein problem. Instead, one is led to a static Monge-Kantorovich problem with a non-quadratic cost. From the mathematical point of view this is not exotic: transport problems with action-type, relativistic, and Lorentzian costs have been studied extensively in the OT literature; see, e.g.,~\cite{brenier2004extended,bertrand2013optimal,suhr2018t,mccann2023d}.

\paragraph{Monge-Kantorovich problem with DBI-type cost.} It is important to note that the optimisation problem for Wheeler-DeWitt systems differs from the standard quadratic Wasserstein case reviewed in Section \ref{sec:Otto}. There, the static Monge-Kantorovich problem admits an equivalent dynamical Benamou-Brenier formulation, in which transport is governed by a continuity equation together with a quadratic kinetic action. This equivalence is special and does not hold for a generic cost.

In the Wheeler-DeWitt setting, once an internal clock $T$ has been chosen, the WKB approximation produces a continuity equation of the form \eqref{eq:continuity_equ_HJ} for semiclassical densities on configuration space. Concretely, after splitting the configuration-space variables as $q^i=(T,q^\alpha)$, one solves the Hamiltonian constraint for the momentum conjugate to $T$,
\begin{equation}
p_T+h_T(T,q^\alpha,p_\alpha)=0\,,
\end{equation}
thereby defining a reduced Hamiltonian $h_T$ which governs the flow with respect to the chosen ``emergent" time $T$. In the situations relevant for us, $h_T$ is generically of square-root form. The induced semiclassical flow is therefore well defined, but it does not imply that the associated OT problem is of Benamou-Brenier type. Instead, we are naturally led to a static Monge-Kantorovich problem with a cost determined by the reduced dynamics encoded in $h_T$. Importantly, the optimal coupling does not necessarily coincide with the semiclassical flow generated by $h_T$.

The reduced Hamiltonian $h_T$ determines the cost through a reduced Lagrangian obtained by a Legendre transform from $h_T$,
\begin{equation}\label{eq:reduced_Lagrangian}
L_\ast(T,q^\alpha,\dot q^\alpha)=\sup_{p_\alpha}\bigl(p_\alpha \dot q^\alpha - h_T(T,q^\alpha,p_\alpha)\bigr)= -\sqrt{2V}\,\sqrt{1-G_{\alpha\beta}\dot q^\alpha \dot q^\beta}\,,
\end{equation}
where in the last step, we used the square-root form of $h_T$ obtained from the Hamiltonian constraint. Unlike the Tonelli case discussed in Section~\ref{sec:OTpotential}, this Lagrangian is not itself a metric distance on configuration space. Rather, it should be regarded as an action density defining an endpoint cost, in the same spirit as relativistic and Lorentzian transport costs\footnote{Note that due to the generically indefinite signature of $G_{\alpha \beta}$, the Lagrangian/cost might not be monotonic or even be degenerate along the (on-shell) transport process. Even if $G_{\alpha \beta}$ is positive definite, which, for example, can be the case for reductions of 4D static maximally symmetric black hole solutions, the square root type still forbids a direct Benamou-Brenier formulation analogue.} studied in~\cite{brenier2004extended,suhr2018t,mccann2023d}. 

Given two endpoints $q_0,q_1$, one defines the associated action value function
\begin{equation}\label{eq:WdW_action_value}
    c_T(q_0,q_1):=-\inf_{\gamma}\left\{
        \int_0^1 L_\ast(T,\gamma,\dot\gamma)\,\mathrm d T:\gamma(0)=q_0,\,\gamma(1)=q_1,\,\dot\gamma(T)\in  \Omega_{T,\gamma(T)}\right\}\,,
\end{equation}
with the constraint
\begin{equation}\label{eq:admissible_velocities_WdW}
    \Omega_{T,q}=\left\{\dot q^\alpha  : G_{\alpha\beta}(T,q)\dot q^\alpha \dot q^\beta \le 1\right\}\,,
\end{equation}
so that the Lagrangian remains real. This defines the corresponding static OT problem: given two probability distributions $\mu_0,\mu_1$, one minimises
\begin{equation}\label{eq:MK_WdW_lorentzian}
    \inf_{\pi\in \Gamma(\mu_0,\mu_1)} \int c_T(q_0,q_1)\,\mathrm d\pi(q_0,q_1)\,.
\end{equation}

For concreteness, we translate the previous section into the setting of a four-dimensional spatially flat FLRW spacetime with lapse function $N(t)$ and scale factor $a(t)=e^{\alpha(t)}$,
\begin{equation}
ds^2=-N(t)^2\,dt^2+a(t)^2\,d\vec x^{\,2}\,,
\end{equation}
together with a homogeneous scalar field $\phi(t)$ subject to a potential $V(\phi)$. The canonical WDW formulation in terms of the Hamiltonian constraint \eqref{eq:WdW_quantum_constr} and the complementary scalar-field equations of motion, can  ---  at least in the case of 1D minisuperspace reductions of such highly symmetric backgrounds --- be obtained from a variational principle, with an action functional on minisuperspace that reads 
\begin{equation}\label{eq:mini_action_L}
    S[N;q] = \int \mathrm{d}\tau \left( \frac{1}{2N}G_{ij}(q)\dot q^i \dot q^j - N V(q)\right)\,.
\end{equation}
In the concrete setting, the reduced action takes the form
\begin{equation}\label{eq:scalar_gravity_mini}
S[N;a,\phi]=
\int dt
\left[
-\frac{3a\dot a^2}{N}
+\frac{a^3\dot\phi^2}{2N}
-N\,a^3V(\phi)
\right]\,= \int dt\left(p_a\dot a+p_\phi\dot\phi-N\mathcal H\right)\,,
\end{equation}
with associated canonical momenta and Hamiltonian constraints $p_a = -\frac{6a\dot a}{N}$ and $p_\phi=\frac{a^3\dot\phi}{N}$. 
In terms of $\alpha(t)$, and in the gauge $N=1$, which we will adopt throughout the rest of this section, the Hamiltonian constraint becomes
\begin{equation}
\mathcal H
=
-\frac{p_\alpha^2}{12a^3}
+\frac{p_\phi^2}{2a^3}
+a^3V(\phi)
=
e^{-3\alpha}
\left(
-\frac{p_\alpha^2}{12}
+\frac{p_\phi^2}{2}
+e^{6\alpha}V(\phi)
\right)\,,
\end{equation}
where $p_\alpha$ is obtained from the canonical transformation.

Let us now consider the reduced Lagrangian described in \eqref{eq:reduced_Lagrangian}. Assuming that $\alpha$ is monotonic and choosing, by convention, the expanding branch $p_\alpha<0$, one may solve the Hamiltonian constraint and explicitly evaluate~\eqref{eq:reduced_Lagrangian} to
\begin{equation}\label{eq:off-shell_dist1}
L_\alpha^{\mathrm{off}}(\alpha,\phi,v_\phi^{(\alpha)}) = -\sqrt{12}\,e^{3\alpha}\sqrt{V(\phi)}\,
\sqrt{1-\frac{1}{6}\left(v_\phi^{(\alpha)}\right)^2}\,,
\end{equation}
with reality condition $\left|v_\phi^{(\alpha)}\right|\le \sqrt{6}$, or equivalently the requirement that $\dot{\alpha}^2>0$. The reduced Lagrangian \eqref{eq:off-shell_dist1} is proportional to the (square root of the) potential. It therefore degenerates in the zero-potential limit, analogously to the massless limit of a relativistic particle action. In particular, the DBI-type cost is adapted to the deparametrised gravitational dynamics, not to the intrinsic moduli-space distance. Recovering the latter in the
$V=0$ limit requires choosing a different cost\footnote{For example, by working directly with the minisuperspace action functional~\eqref{eq:scalar_gravity_mini}, which admits a non-trivial $V \to 0$ limit.}.

Paths are encoded in two essential ingredients. First, in the presence of gravity the Hamiltonian is a first-class constraint and therefore vanishes on shell. Second, in Lorentzian signature, the kinetic term of the scale factor enters with the opposite sign relative to that of the scalar field.\footnote{It is easy to show that in Euclidean signature, and --- in the absence of a scalar potential --- the equations of motion cannot be solved in terms of real momenta.}
Working with the on-shell action, i.e., $S^{on}$ viewed as a function of the internal clock or time parameter $\alpha$ (and not as a functional) the one-form along a classical trajectory determined by the reduced action~\eqref{eq:scalar_gravity_mini} reads 
\begin{equation}\label{two_separated_distances}
\mathrm d S^{\mathrm{on}}=
\left(
-h_\alpha+p_\phi v_\phi^{(\alpha)}
\right)\mathrm d \alpha\,.
\end{equation}
Thus, after choosing the branch $p_\alpha=-h_\alpha$, the constrained minisuperspace action and the reduced Legendre transform \eqref{eq:off-shell_dist1} agree --- as expected --- on shell. This means that this cost function is relative to the full solution space, which comprises also the metric space, and not to the single scalar field space.

\paragraph{A projected cost in scalar field space.} Since the cost function defined in \eqref{eq:off-shell_dist1} is not smoothly connected to the standard moduli-space distance, we propose\footnote{This is motivated by the observation that in the absence of a potential, eq.~\eqref{two_separated_distances} suggests that individual contributions to the reduced Lagrangian, i.e., the metric and scalar field parts, remain non-vanishing.} to define a different cost in scalar field space using \eqref{eq:off-shell_dist1}, and evaluating on solutions of the full variational problem. This cost will no longer be connected to an OT-formulation, but will admit an interpretation in terms of a Maupertuis principle in minisuperspace.

More concretely, starting from \eqref{two_separated_distances}, we consider the adapted cost 
\begin{equation}
\mathrm d S_{\phi}^{\mathrm{on}}
=
p_\phi\,\mathrm d\phi
=
p_\phi v_\phi^{(\alpha)}\,\mathrm d \alpha\,,
\end{equation}
such that the total cost reads
\begin{equation}
\mathcal C_{M,\phi}^{\mathrm{on}}
=
\int \mathrm d\alpha\,\mathrm d\phi\;
\rho_\alpha\,p_\phi v_\phi^{(\alpha)}.
\end{equation}
We want to stress again that this is an on-shell quantity with respect to the full (unreduced) dynamics, and not an OT cost. It hence must be interpreted as a projected cost onto the scalar-field component. In particular, neither is the total cost obtained in this way optimal nor invariant under the choice of projection and hence clock.

According to this definition, the cost can be written as
\begin{equation}
p_\phi v_\phi^{(\alpha)}
=
\frac{6p_\phi^2}{h_\alpha}
=
h_\alpha-\frac{12e^{6\alpha}V(\phi)}{h_\alpha}
=
h_\alpha+L_\alpha^{\mathrm{on}},
\end{equation}
where the last term is the on-shell evaluation of the cost in eq. \eqref{eq:off-shell_dist1}. Finally, we may then write
\begin{equation}
p_\phi v_\phi^{(\alpha)}
=
v_\phi^{(\alpha)} \frac{h_\alpha}{\sqrt{6}}
\sqrt{\left( 1 - \frac{\mathcal{V}}{(h_\alpha/\sqrt{6})^2}\right)}
=
v_\phi^{(\alpha)} \sqrt{\left( \frac{h_\alpha^2}{6} - \mathcal{V}\right)}\,,
\end{equation}
and, defining $\Lambda(\phi) = (h_\alpha/\sqrt{6})^2$ and $\mathcal{V} = 2 e^{6\alpha} V(\phi)$, the proposed cost becomes
\begin{equation}\label{MMVV_solution}
    \mathcal C_{M,\mathrm{ext}}^{\mathrm{on}}
    =
    \int_{\gamma_{on}} \mathrm d\alpha \,\rho_\alpha \, v_\phi^{(\alpha)} \sqrt{\Lambda(\phi) - \mathcal{V}}
    =
    \int_{\gamma_{on}} \mathrm d\phi \,\rho_\alpha \sqrt{\Lambda(\phi) - \mathcal{V}}\,.
\end{equation}
In the last expression, we slightly abuse notation by denoting the on-shell variation of the scalar field by $\mathrm d\phi$, whereas in the previous section the same symbol denoted the volume element on scalar field space. The reason for doing so is that, in this notation --- which is valid only in the full deterministic limit --- eq.~\eqref{MMVV_solution} matches the proposed expression for the tension of a thick domain wall in scalar field space given in~\cite{Mohseni:2024njl}. Hence, depending on the choice of cost, this functional seems to measure the energy required to move the WDW configuration from one point to another in scalar field space. 

\medskip

In~\cite{Mohseni:2024njl}, they argued that equation~\eqref{MMVV_solution} cannot be seen as a proper definition of distance in scalar field space. The present discussion indeed relates the above formula to a projected part of the cost \eqref{eq:off-shell_dist1} in $\mathcal{P}(\mathcal{M)}$. 
In the same work, due to an analogy to the QFT counterpart, it has been proposed that a possible definition of distance in scalar field space would be given by a normalisation of \eqref{MMVV_solution} with respect to the dynamical (kinetic) energy of the scale factor $h_\alpha^2$. In this part of the section, we will give an interpretation of such a definition according to our perspective. In the perspective of~\cite{Mohseni:2024njl}, the normalisation by $\dot{\alpha}$ of the proposed thick domain-wall tension density can be argued by noticing that in the non-gravitational limit the would-be $T_{DW}$ is proportional by the square root of energy of the system $E$. In the presence of dynamical gravity, there is no such concept of fixed energy due to the WDW equation. However, it is interesting to note that this definition is exactly the kinetic terms of the reduced Lagrangian $|v_\phi^{(\alpha)}|^2$ due to the on-shell relations
\begin{equation}
 L_\alpha^{\text{on}} \mathrm d\alpha= |v_\phi^{(\alpha)}|^2 \,\mathrm d\alpha = \frac{6}{h_\alpha} v_\phi^{(\alpha)} p_\phi \, \mathrm d\alpha = 6\frac{L_{\phi}^{on}}{h_\alpha} \, \mathrm d\alpha= \mathrm d\phi  \, \sqrt{1- \frac{V(\phi)}{\Lambda(\phi)}}\,.
\end{equation}
Hence we see that in order to obtain the cost proposed in~\cite{Mohseni:2024njl} one has to impose the on-shell relations such that it cannot be directly captured by a generic off-shell functional.

\paragraph{Cost from minisuperspace dynamics.} As mentioned above, the canonical WDW formulation in terms of the Hamiltonian constraint \eqref{eq:WdW_quantum_constr} can be equivalently captured by the action functional on minisuperspace~\eqref{eq:mini_action_L}. Then the standard formulation given in terms of the Hamiltonian constraint $\mathcal{H}(q,p)=0$, which in the case of a FLRW spacetime encodes the (first) Friedmann equation, follows by variation with respect to the lapse function $N$,  while the remaining (Euler-Lagrange) equations of the scalar fields follow from variation with respect to $q^i$.

Note that the formulation of the dynamics in terms of the Hamiltonian $h_T$, discussed in the preceding section, is on-shell equivalent and hence generates the same set of equations of motion as both the Hamiltonian constraint and the above minisuperspace action. Therefore, it can be easily shown that the reduced Lagrangian \eqref{eq:reduced_Lagrangian} and \eqref{eq:mini_action_L} give the same value when evaluated on on-shell solutions. In particular, first eliminating the lapse function gives
\begin{equation}
    S[q] = \int \mathrm{d}\tau \sqrt{2V}\sqrt{G_{ij}\dot q^i \dot q^j}\,.
\end{equation}
Using the constraint or equation of motion, this can be written as (taking $G_{TT}=1$)
\begin{equation}
    S[q] = \int \mathrm{d}\tau \dot T\sqrt{2V}\sqrt{1+G_{ab}\partial_t q^i \partial_T q^j}=\int  L_\ast(q) \mathrm{d}T\,.
\end{equation}
Hence, they provide the same (on-shell) cost. However, $S[N,q]$ is much easier to handle as a variational problem since it is of quadratic --- or at least Tonelli --- form. Also, the $V \to 0$ limit is much more transparent form this perspective. However, the clear disadvantage is that there is no preferred notion of time, and hence it is much less clear how to define the associated Hamilton-Jacobi and continuity equation for the probability density needed to define the distance.
Just like the DBI-like cost discussed earlier, the functional $S[N;q]$ generically suffers from a indefinite metric $G_{ij}$ such that the cost is not directly monotonic. This is particularly obvious in the free case with $V=0$, where on-shell cosmological solutions give rise to a vanishing action when evaluating on-shell\footnote{As mentioned before, there are situation/higher dimensional backgrounds where $G_{ij}$ can turn out to be positive definite after the reduction, however in general this is not the case. One could then alternatively Wick rotate and study the Euclidean system. We leave such questions for further investigations.}.

Furthermore, note that one could also apply the strategy of restricting or projecting onto the scalar-field part of the system directly at the level of the minisuperspace action~\eqref{eq:mini_action_L}. The resulting OT transport inspired cost function --- or more precisely its square root form --- can then be directly related to the notion of distance proposed in~\cite{Debusschere:2024rmi}. On-shell this will give rise to a 1-Wasserstein-like distance in terms of the (fake) superpotential $W$; for details on the motivation for this distance, we refer to the original publication~\cite{Debusschere:2024rmi}.

\paragraph{Quadratic Wasserstein distance.} In contrast to the dynamically induced and inspired cost discussed above, one may instead choose a cost which is not derived from the reduced Hamiltonian $h_T$ or the more general minisuperspace dynamics, but instead impose directly impose a transport problem. A simple and mathematically well-defined choice is the usual quadratic cost 
\begin{equation}
    c_2(q_0,q_1)=\frac{1}{2}d(q_0,q_1)^2\,,
\end{equation}
where $d(q_0,q_1)$ is a distance on configuration space to be chosen ad hoc (e.g., the euclidean distance in one dimension or the minisuperspace metric $G_{ab}$ or its restriction $G_{\alpha \beta}$ to the reduced dynamics). The associated OT problem \eqref{eq:optimal_transport_cost_problem} then defines the quadratic Wasserstein distance \eqref{eq:BB_formula}. Equivalently, one may also use the dynamical Benamou-Brenier formulation \eqref{eq:W22}. 

The quadratic cost has the advantage that it places the problem in the well-known terrain of OT where in some cases one can use a closed form formula. The trade-off is that the cost is no longer derived from the underlying gravitational dynamics but is imposed as an external choice, and hence OT is no longer along geodesics of the induced geometry by the Wasserstein metric.

\newpage
\section{Applications and examples}\label{sec:appl}

In this section, we apply the Optimal Transport framework developed above to two settings where semiclassical densities naturally arise. 

First, in Section \ref{sec:fake_SUSY_HJ}, we recast the fake supergravity flow equations for black holes/domain walls as finite-dimensional Hamilton-Jacobi systems on the associated minisuperspace and use the associated characteristic function $\mathcal W$ to define WKB densities on the configuration space \cite{Trigiante:2012eb, Andrianopoli:2007gt, Andrianopoli:2009je}. In this Section, we will not explicitly compute all the various notions of distance introduced in the previous section, as some of them partially overlap with those obtained in \cite{Mohseni:2024njl, Debusschere:2024rmi, Li:2023gtt, Basile:2023rvm}, to which we refer the reader for the relevant formulae. Indeed, our emphasis here is instead on the role and consequences of considering a genuinely semiclassical distribution, rather than limiting the discussion to the classical $\hbar^0$ approximation.  
In Section \ref{sec:fakeSUGRA_examples}, we then evaluate these densities in explicit attractor examples, extracting their Gaussian widths from the Hessian of $\mathcal W$ and tracking controlled limits in which the probability density becomes sharply localised. We include in Appendix~\ref{app:gauged_fake_sugra} the analogous construction for gauged D0-D4 systems with a cubic prepotential, where the FI gauging introduces an additional scalar potential and modifies the attractor equations. Second, in Section \ref{sec:QC}, we implement the same logic in examples in cosmology using the minisuperspace approximation. Starting from the Wheeler-DeWitt constraints, we identify the induced minisuperspace geometry and construct semiclassical wave packets on the WKB branches. We then develop the formalism through some examples, focusing, in particular, on the Gaussian approximation. In contrast with the usual pointwise distance on moduli space, the (generalised) Wasserstein distance between the corresponding Gaussian approximations depends both on the separation of their centres and on the difference of their widths.

\subsection{Fake supergravity as a Hamilton-Jacobi system}\label{sec:fake_SUSY_HJ}

In this subsection, we review the minimal structure of (fake) supergravity needed for our purposes, namely its formulation as a finite-dimensional Hamilton-Jacobi system governing first-order flows on field space~\cite{Freedman:2003ax,Celi:2004st,Skenderis:2006jq,Ceresole:2007wx,Andrianopoli:2009je}. 
For simplicity, consider a four-dimensional (bosonic) action with scalars $\phi^a$ and Abelian gauge fields $A_\mu^A$ with curvature $F_{\mu\nu}^A$. Following the conventions of~\cite{Andrianopoli:2009je}, it takes the form 
\begin{equation}\label{eq:Lagrangian_BH}
S=\!\! \int \!\! d^4x\sqrt{-g}\left(
\frac12 R \!-\! \frac12 G_{ab}(\phi)\partial_\mu\phi^a\partial^\mu\phi^b \!+\! \frac{1}{4}\mathcal I_{AB}(\phi)F^A_{\mu\nu}F^{B\,\mu\nu} \!\!+\! \frac{1}{4}\mathcal R_{AB}(\phi)F^A_{\mu\nu}\tilde F^{B\,\mu\nu}
\!\right),\!\!
\end{equation}
where $\mathcal I_{AB}$ and $\mathcal R_{AB}$ are the imaginary and real parts of the gauge kinetic matrix $\mathcal{N}_{AB}$.

For static and spherically symmetric configurations, one may adopt a metric ansatz of the form~\cite{Andrianopoli:2009je}
\begin{equation}
	\mathrm ds^2=e^{2U(\tau)}\mathrm dt^2-e^{-2U(\tau)}\left( \frac{c^4}{\sinh^4{(c\tau)}
    }\mathrm d\tau^2+ \frac{c^2}{\sinh^2{(c \tau)}}\mathrm d\Omega_{2}^2\right)\,,
\end{equation}
together with electric and magnetic charges, collected into a symplectic vector $Q=(p^\Lambda,q_\Lambda)$. In this notation, $c$ is the extremality parameter which is usually defined as $c = 2ST$, where $S$ and $T$ are the black hole entropy and temperature, respectively\footnote{As per usual convention, the coordinate $\tau$ is connected to the radial coordinate $r$ via
\begin{equation}\dotag{fneq:aa_4}
\frac{c^2}{\sinh^2{(c \tau)}} = (r-r_0)^2 - c^2\,,    
\end{equation}
where $r_0 = 2M$ is the associated black hole mass.}.
As is well-known, the equations of motion are given by a set of second-order differential equations for the $q^i(\tau) = (U(\tau), \phi^a(\tau))$. However, as argued in~\cite{Andrianopoli:2009je}, it is possible to define a set of first-order equations that return the same equations of motion using the Hamilton-Jacobi theory introduced in previous sections
\begin{equation}
    p_i = \frac{\partial S_0}{\partial q_i}\,,\qquad -H = \frac{\partial S_0}{\partial \tau}\,,\qquad S_0(q,\tau) = \mathcal{W}(q) - c^2 \tau\,.
\end{equation}
Here, $S_0$ is again the Hamilton principal function of Section \ref{sec:basics_HJ} (we omit the momentum dependence). We slightly adapt our notation for the characteristic function, which is now denoted by $\mathcal{W}(q)$ instead of $W(q)$, the reason for which will become apparent in a moment.
The equation of motion of \eqref{eq:Lagrangian_BH}, can then equivalently be obtained from the Hamilton-Jacobi system by setting
\begin{equation}\label{eq:HPFandBHs}
 H = \frac{1}{2} G^{ij} \partial_i \mathcal{W}  \partial_j \mathcal{W} - \mathcal{V}(q) = c^2\,,
\end{equation}
where we have
\begin{equation}
    G_{ij}(q) = \begin{pmatrix}
        4 && 0 \\
        0 && G_{ab}(\phi)
    \end{pmatrix}\,,
\end{equation}
and  $\mathcal{V}(q)$ is the potential
\begin{align}
    \frac{1}{2}\mathcal{W}(q) &=  e^{U(\tau)} W(\phi, \tau) + c^2\tau\,,\\  
    \mathcal{V}(q) = e^{2U}\,V_{\rm BH}(\phi,Q) 
    &\,,\quad V_{\rm BH}(\phi, Q) = -\tfrac12\, Q^T \mathcal M(\phi)\, Q\,.
\end{align}
Here $\mathcal{M}=\mathcal{M}(\mathcal{N})$ is the symmetric symplectic matrix, which is negative definite so that $V_{BH}>0$~\cite{Andrianopoli:2007gt, Andrianopoli:2009je, Andrianopoli:2010bj}. 
With respect to the characteristic function, the equations of motion are 
\begin{equation}\label{eq:first_order_eq}
\dot{U} = e^U W\,,\quad \dot{\phi}^a = 2 e^U G^{ab} \partial_b W\,, \quad \partial_\tau \mathcal{W}= - c^2 e^{-U}\,.
\end{equation}
In view of the equivalence proposed in the previous section, the characteristic function $\mathcal W$ plays a role directly analogous to the (leading) WKB phase in the WDW equation.  In particular, it defines a congruence of classical trajectories on field space along which semiclassical localisation may be analysed.  In particular, using \eqref{eq:HPFandBHs}, we can always write the characteristic function as 
\begin{equation}
    \mathcal{W}(q) = \mathcal{W}_0 + 2\int_{\tau_0}^\tau (c^2 + \mathcal{V}(q))d\tau^\prime\,,
\end{equation}
which, as noted in~\cite{Andrianopoli:2009je}, is reminiscent of the eikonal equation of the wave front $\mathcal{W} = \text{const.}$ propagating in a medium with refractive index 
\begin{equation}
    n = \sqrt{2(c^2 + \mathcal{V})} = \sqrt{G^{ij} \partial_i \mathcal{W} \partial_j \mathcal{W}_j}\,,
\end{equation}
This is analogous to the on-shell version of eq.~\eqref{eq:off-shell_dist1}.
In the examples below we will focus on extremal black hole backgrounds ($c=0$), keeping in mind that it can be associated with a flat domain wall background as explained in~\cite{Freedman:2003ax,Trigiante:2012eb,Cribiori:2023swd}.
Following the general WKB logic reviewed in Section \ref{sec:associatedWKB}, we may associate to the background a semiclassical density on field space using the WKB construction reviewed in Section \ref{sec:associatedWKB}.

In particular, for Euclidean branches one formally obtains a normalisable density of the schematic form
\begin{equation}\label{eq:rhoE_fake_bridge}
\rho_E(q)\propto\frac{1}{\sqrt{\det G(q)}\,|\nabla \mathcal W(q)|}\exp\!\left[-\frac{2}{\hbar}\,\mathcal W(q)\right]\,.
\end{equation}
For simplicity, let us now focus on the exponential part. Up to slowly varying prefactors, expanding $\mathcal W$ around a critical point $\phi_\ast$ as in Section~\ref{sec:gaussian_limit}
\begin{equation}
\mathcal W(q)=\mathcal W(q_\ast)+\frac12\,\delta q^i\,\partial_i\partial_j\mathcal W\,\delta q^j\Big|_{\phi_\ast}+\cdots\,,
\end{equation}
one finds that the Hessian of $\mathcal W$ controls the Gaussian width. We will use this approximation to perform explicit calculations in controlled examples.
This construction provides a natural probability density on scalar field space, intrinsic to the Hamilton-Jacobi structure of the system as in~\cite{Ooguri:2005vr}. Distances between families of semiclassical solutions can therefore be defined using OT (Wasserstein) distances built from $\rho(q)$ and the metric $G_{ij}$ (for the details of the formulae we refer to Section \ref{sec:gaussian_limit}). For instance, in this particular case, we may ask what the optimal transport distance is  between two black hole backgrounds with charge $Q$ and $\tilde{Q}$.

\paragraph{Comments}

\begin{itemize}
	\item The Hamilton-Jacobi formulation applies most directly to extremal black-hole and flat domain-wall backgrounds, for which the equations of motion admit a genuine first-order gradient flow generated by a characteristic function $\mathcal W(q)$. In this case, the dynamics are fully captured by a flow equation together with the Hamilton-Jacobi constraint. In particular, for extremal solutions ($c=0$) the function  $\mathcal W$ behaves as a Lyapunov function along the radial flow, implying monotonic  behaviour toward the attractor; see, e.g.,~\cite{Andrianopoli:2009je}. Away from extremality ($c\neq0$), such monotonicity need not persist.   
     While this does affect the detailed supergravity equations, it does not obstruct the construction relevant here. Whenever a real Hamilton-Jacobi function $\mathcal W(q)$ exists, it defines a distinguished congruence of classical trajectories on configuration space and hence a natural semiclassical density. In particular, on Euclidean or tunnelling branches, one obtains a normalisable WKB density of the form \eqref{eq:rhoE_fake_bridge}, up to the standard transport prefactor discussed in Section \ref{sec:associatedWKB}. This density provides the distribution on field space entering our definition of Wasserstein distances between semiclassical backgrounds.

	\item In the examples discussed in Sections \ref{sec:fakeSUGRA_examples} and \ref{sec:QC}, we either work on the extended space $q=(U,\phi)$ or reduce to the scalar manifold $ M$ by (i) restricting to a transverse slice $\Sigma_\tau$ (e.g. $U=U_0$) and using the induced cost, or (ii) pushing forward the semiclassical measure along $\pi(U,\phi)=\phi$ and reducing the ground cost by taking the fibre-infimum $c_{ M}(\phi_0,\phi_1)=\inf_{U_0,U_1}c((U_0,\phi_0),(U_1,\phi_1))$. In Wheeler-DeWitt examples, the slice $\Sigma_\tau$ is chosen to be transverse to the WKB current\footnote{The interpretation of semiclassical probabilities in minisuperspace, their conservation along Hamilton-Jacobi flows, and the projection onto hypersurfaces transverse to these flows are standard in quantum cosmology; see, e.g.,~\cite{Halliwell:1989myn,kiefer2006quantum}.}. \label{comment:Ufixed}
\end{itemize}

\subsection{Applications in supergravity and black holes}\label{sec:fakeSUGRA_examples}

In this subsection, we illustrate the framework developed in the previous sections on a set of controlled fake supergravity examples.  In particular, we first consider a one-modulus attractor model in which all steps can be carried out analytically.  We then embed the same logic into a standard special-geometry realisation of the D0-D4 and D2-D6 sectors with a cubic prepotential, first in the ungauged theory as a baseline. We then consider FI-gauged models where a genuine scalar potential modifies the effective dynamics and can lead to multiple extrema in Appendix~\ref{app:gauged_fake_sugra}.

\subsubsection{One-modulus attractor flows}

We first consider the Ceresole-Dall'Agata one-modulus model~\cite{Ceresole:1995ca}, which will serve as a minimal model for the constructions outlined in previous sections.  In addition, we will consider representative charge sectors (D0-D4 and D2-D6) in which the semiclassical widths admit a direct interpretation in terms of the Kaluza-Klein and species scale (cf.~\cite{Dvali:2007hz,Dvali:2007wp,Dvali:2009ks,Dvali:2010vm,vandeHeisteeg:2022btw,Cribiori:2022nke,Cribiori:2023ffn,Castellano:2023aum,vandeHeisteeg:2023dlw,Calderon-Infante:2025ldq, Herraez:2025clp}). 

\medskip

The special K\"ahler manifold is $SU(1,1)/U(1)$, parametrised by $t=x+i\tau$ with $\tau>0$. 
Restricting to the axion-less slice $x=0$ and introducing the logarithmic coordinate $y=\log\tau$, the metric reads
\begin{equation}
\mathrm ds^{2}=g_{t\bar t}\,\mathrm dt\,\mathrm d\bar t=\frac{\mathrm d\tau^{2}}{4\tau^{2}}
=\frac14\,\mathrm dy^{2}\,,
\end{equation}
where we used $K=-\log(2\,\mathrm{Im}\,t)=-\log(2\tau)$ and $g_{t\bar t}=\partial_t\partial_{\bar t}K=(4\tau^2)^{-1}$. For generic dyonic charges $Q=(p^0,p^1;q_0,q_1)$, the superpotential is~\cite{Ceresole:2007wx}
\begin{equation}\label{eq:superpotential_CR}
W(y)= \frac{e^{-y}}{2}\left((-q_0 + i p^1) + (q_1 - i p^0)e^y \right)\,,
\end{equation}
such as the black hole potential is $V_{\mathrm{BH}}(y)= |W(y)|^2+4\,g^{yy} |\partial_y W|^2$ with $g^{yy}=4$.
Defining $A_\ast=(p^0)^2+(q_1)^2$, $B_\ast=(p^1)^2+(q_0)^2$ and $M=p^0p^1+q_0q_1$, the attractor $\partial_y V_{BH}=0$ occurs at
\begin{equation}\label{eq:attractor_pt_DC}
e^{2y_\ast}=\frac{B_\ast}{A_\ast}\,,
\end{equation}
and the superpotential and its curvature at the attractor $y_\ast$ are
\begin{equation}
W_\ast= |W(y_\ast)|=\sqrt{\frac{M+\sqrt{A_\ast B_\ast}}{2}}\,,\qquad \kappa_y=\partial_y^2W\big|_{y_\ast}=\frac{\sqrt{A_\ast B_\ast}}{4W_\ast}\,.
\end{equation}
The Bekenstein-Hawking entropy is then
\begin{equation}\label{eq:entrop_A_CD}
S_{\mathrm{BH}}=\pi\,V_{\mathrm{BH}}(y_\ast)=\pi\,W_\ast^2=\frac{\pi}{2}\left(M+\sqrt{A_\ast B_\ast}\right)\,.
\end{equation}
The Hamilton-Jacobi principal function factorises as $S(U,y)=\mathcal{W}(U,y)=e^{U}W(y)$. On a fixed slice $U=U_0$, the associated Euclidean WKB density \eqref{eq:gaussapprox_var_clean} then reads
\begin{equation}
\rho_{\mathrm E}(y)=Z^{-1}\exp\left(-\frac{2e^{U_0}}{\hbar}\,W(y)\right)\,,
\end{equation}
where $Z$ normalises $\rho_E$ with respect to the measure $\frac12\,\mathrm dy$. 
Expanding $W(y)=W_\ast+\tfrac12\kappa_y(y-y_\ast)^2+\cdots$ gives the local Gaussian approximation
\begin{equation}
\rho_E(y)\approx Z^{-1}\exp\left(-\frac{2e^{U_0}}{\hbar}\,W_\ast\right)\, \exp\left(-\frac{e^{U_0}\kappa_y}{\hbar}\,(y-y_\ast)^2\right)\,,
\end{equation}
with variance
\begin{equation}\label{eq:CD_var}
\sigma_y^{2}=\frac{\hbar}{2e^{U_0}\kappa_y}\,.
\end{equation} 
In this special case, it reads $\sigma_y^{2} =\frac{2\hbar\,W_\ast}{e^{U_0}\sqrt{A_\ast B_\ast}}$. Before proceeding, let us first clarify the meaning of this approximation.
Our aim is to isolate the purely scalar-field-space contribution to the wavefunction, evaluated on the attractor geometry, in order to extract information about black holes from the Wheeler-DeWitt wavefunction. However, as argued above, the phase of the Wheeler-DeWitt wavefunction is determined by $\mathcal{W}(U,\phi)$ rather than by $W(\phi)$ alone. It is well known, for instance, that at the attractor the black hole entropy is given by $S_{\mathrm{BH}} = \pi |W_\star|^2$, while the requirement of a finite horizon area implies $\lim_{\tau \rightarrow \infty} e^{-2U(\tau)} \sim \frac{|Z_{BH}|^2}{\rho^2} \rightarrow +\infty\,,$
where $\rho$ is the near-horizon coordinate defined by $r = r_h + \rho$ \cite{Ferrara:1995ih, Ferrara:2008hwa}. Therefore, in order to extract meaningful information from the wavefunction in the sense relevant for this work, one should either integrate out the variable $U(\tau)$, thus obtaining a probability density depending only on $\phi$, or instead fix a slice $U=U_0$ and expand the scalar fields near the horizon. The latter procedure amounts to fixing a position $\rho_0(|Z_{BH}|)$, which in general depends on the charges, and then sending the charges to infinity while keeping a fixed relative distance from the horizon, so as to maintain $U_0$ fixed.

\medskip

Having set up the form of the Gaussian approximation and in view of taking particular limits of the charges to access infinite distances, let us distinguish three cases. Notably, in these examples, the same mechanism repeats: along infinite-distance charge limits\footnote{For a systematic classification of charge limits and their interpretation in terms of towers and EFT cut-offs, see~\cite{Calderon-Infante:2025pls}.} the Euclidean WKB density becomes sharply localised, and the resulting width of the distribution is parametrically comparable to the species scale or the associated light Kaluza-Klein mass in this regime.

\paragraph{Entropy-driven collapse at infinite distance.} This first example isolates an  infinite distance limit in which the attractor runs to the boundary while the Euclidean WKB density collapses, so that the Wasserstein distance is dominated by the separation of attractor points (and width effects parametrically suppressed). More concretely, we first consider the large-$|q_1|$ regime at fixed $(p^0,p^1,q_0)$.
The attractor position \eqref{eq:attractor_pt_DC} obeys
\begin{equation}
e^{2y_\ast}=\frac{B_\ast}{A_\ast}=\frac{(p^1)^2+(q_0)^2}{(p^0)^2+(q_1)^2}\,,
\end{equation}
so for $|q_1|\to\infty$, one has $e^{y_\ast}\sim |q_1|^{-1}$, and hence $y_\ast\to-\infty$. 
The moduli-space distance diverges since $\mathrm ds=\tfrac12|\mathrm dy|$:
\begin{equation}
\Delta(y_0,y_\ast)=\int_{y_0}^{y_\ast}\frac12\,|\mathrm dy|\ \longrightarrow\ \infty\,.
\end{equation}
Moreover, $W_\ast^2=\tfrac12(M+\sqrt{A_\ast B_\ast})$ grows linearly in $|q_1|$ for fixed $(p^0,p^1,q_0)$, and therefore $S_{\mathrm{BH}}=\pi W_\ast^2\ \propto\ |q_1|$ as $|q_1|\to\infty$. Using \eqref{eq:CD_var} and $\kappa_y=\sqrt{A_\ast B_\ast}/(4W_\ast)$, the variance scales as
\begin{equation}\label{eq:variance_exampleACD}
\sigma_y^2\sim \frac{\hbar}{e^{U_0}}\frac{1}{\sqrt{|q_1|}}\ \propto\ S_{\mathrm{BH}}^{-1/2}\,,
\end{equation}
so the density becomes sharply localised in this infinite-distance limit. Finally, using the  coordinate $z=y/2$, in the Gaussian approximation, the (quadratic) Wasserstein distance \eqref{eq:wasserstein_1dgaussian} gives
\begin{equation}
W_2^2(\rho_{q_1},\rho_{q_1'})
\approx \frac14\Big((\Delta_g y_\ast)^2+(\Delta_g\sigma_y)^2\Big)\,,
\end{equation}
with $\sigma_z=\sigma_y/2$ and $\Delta_g$ denoting the distance measured with respect to the underlying metric on scalar field space, here $\mathrm d s^2=
\mathrm d y^2/4$. In the large-$|q_1|$ regime, the mean-separation term dominates, while the width correction is subleading due to \eqref{eq:variance_exampleACD}.

\paragraph{D0-D4 attractors and the species scale.}\label{sec:D0D4examples} In this case let us work in a simpler set-up, i.e. four-dimensional $\mathcal N=2$ supergravity. In the large-volume limit, the vector-multiplet moduli space is special K\"ahler with cubic prepotential
\begin{equation}
F(X) = -\frac{1}{6}\,C_{ijk}\,\frac{X^i X^j X^k}{X^0}\,,
\end{equation}
where $C_{ijk}=\int_Y \omega_i\wedge\omega_j\wedge\omega_k$ are the triple intersection numbers and $X^I=(X^0,X^i)$ are homogeneous special-geometry coordinates. We focus on the D0-D4 charge sector, where only the D0 charge $q_0$ and D4 charges $p^i$ are non-vanishing, while $p^0=q_i=0$. This example will illustrate how, in the infinite-distance limit associated with the D0-D4 attractor flow, the semiclassical localisation of the WKB density on moduli space becomes controlled by the species scale.

\medskip

For this example we will follow~\cite{Kallosh:2006bt,Andrianopoli:2010bj}. For simplicity, we restrict ourselves to a one-modulus truncation with K\"ahler modulus $t=B+i\phi$, $\phi>0$, and a single non-vanishing intersection number $C_{111}=-6$.  The prepotential reduces to
\begin{equation}
F(X)=\frac{(X^1)^3}{X^0}\,,\qquad t=\frac{X^1}{X^0}=i\phi\,.
\end{equation}
On the axion-free slice $B=0$, the K\"ahler potential and metric are
\begin{equation}
K=-\log(4\phi^3)\,,\qquad \mathrm ds^2=\frac{3}{4}\,\frac{\mathrm d\phi^2}{\phi^2} = \frac{3}{4}\,\mathrm du^2\,.
\end{equation}
while we have introduced the variable $u=\log\phi$.
In the D0-D4 charge sector $(q_0,p^1)\neq0$, a real fake superpotential may be chosen as~\cite{Kallosh:2006bt,Andrianopoli:2010bj}
\begin{equation}
W(\phi)=\frac{1}{2\,\phi^{3/2}}\left(q_0+3\,p^1\,\phi^2\right)\,.
\end{equation}
The attractor point is determined by $\partial_\phi W=0$, giving
\begin{equation}
\phi_\ast^2=\frac{q_0}{p^1}\,,\qquad W_\ast=W(\phi_\ast)=2\,q_0^{1/4}(p^1)^{3/4}\,.
\end{equation}
Expanding $W$ to quadratic order around the attractor and introducing the canonically normalised coordinate $\xi=\frac{\sqrt3}{2}u$, one finds
\begin{equation}
\partial_\xi^2 W\big|_{\xi_*} = W_\ast\,.
\end{equation}
On a fixed-$U$ slice, the associated Euclidean WKB density, therefore, admits a Gaussian approximation with variance (see \eqref{eq:CD_var})
\begin{equation}
\sigma_u^2=\frac{\hbar}{2\,e^{U_0}\,W_\ast}\,.
\end{equation}
Using the standard relation $S_{\rm BH}=\pi W_\ast^2$, this can be written as
\begin{equation}
\sigma_\xi \propto S_{\rm BH}^{-1/4}\,.
\end{equation}
Thus, in large-charge limits where the attractor approaches the boundary of moduli space and the entropy diverges, the semiclassical density becomes sharply localised.  In this regime, Wasserstein distances \eqref{eq:wasserstein_1dgaussian} between associated distributions are dominated by the separation of attractor points, with the width effects parametrically suppressed.

\medskip

We can now interpret this result by studying the behaviour of the K\"ahler moduli at the horizon, which for D0-D4 systems take the form (see, e.g.,~\cite{Behrndt:1996jn,Bonnefoy:2019nzv, Cribiori:2022nke, Calderon-Infante:2025pls, Castellano:2025yur, Castellano:2025ljk} for a partial overview)
\begin{equation}
t_h^a = \mathrm{Im}\!\left(\frac{C X^a}{C X^0}\right)\Big|_h = p^a \sqrt{\frac{|q_0|}{D_{abc}p^a p^b p^c}}\,.
\end{equation}
In a one-modulus truncation, this implies $t_h \sim \sqrt{q_0/p^1}$. In the semi-classical regime, due to the fact that underlying Calabi-Yau volume behaves as $\mathcal V_h \sim t_h^3 \sim \left(q_0/p^1\right)^{3/2}$, the large $q_0$ charge limit is therefore realised by taking
\begin{equation}
q_0 \gg p^1,\qquad q_0\to\infty\,,
\end{equation}
which corresponds to an M-theory limit in which the M-theory circle decompactifies faster than any normalised four-cycle volume. Along this trajectory, we have $W_\ast^2 \sim \sqrt{(p^1)^3 q_0}$ and the Bekenstein-Hawking entropy scales as
\begin{equation}
 S_{\rm BH}\sim \sqrt{q_0 (p^1)^3} \sim t_h\,,
\end{equation}
and thus diverges as $q_0\rightarrow 
\infty$. In this case, it is simple to argue that the Kaluza-Klein scale evaluated at the horizon scales as
\begin{equation}
m_{\mathrm{KK}} \propto \frac{1}{\sqrt{\mathcal{V}_h}}\sim \left(\frac{q_0}{p^1}\right)^{-3/4}\hspace{-20pt}\,.
\end{equation}
At fixed $p^1$, this implies that $S_{\rm BH} \propto m_{\mathrm{KK}}^{-2/3}$.

\noindent
In this set-up, the species scale at the horizon is parametrically controlled by the five-dimensional Planck scale and scales as
\begin{equation}
\Lambda_{\rm{sp}} = m_{\mathrm{KK}}^{\frac{1}{3}} \sim  \left(
\frac{q_0}{p^1}\right)^{-1/4}\hspace{-20pt}\,.
\end{equation}
Using the tree-level entropy formula $S_{\rm BH}=\pi r_h^2$, the horizon radius satisfies
\begin{equation}
 r_h\sim S_{\rm BH}^{1/2} \sim  \Lambda_{\mathrm{sp}}^{-1}\,.
\end{equation}
Finally, the Gaussian width of the Euclidean WKB density on moduli space given in \eqref{eq:CD_var} behaves as
\begin{equation}
\sigma_{\mathrm{BH}}^2 \sim \frac{1}{\sqrt{S_{\rm BH}}} \sim \Lambda_{\mathrm sp}\,.
\end{equation}
In this infinite-distance limit, the Euclidean WKB density collapses on moduli space with a variance that is parametrically controlled by $\Lambda_{\mathrm{sp}}$. Roughly speaking, this also defines the precision of the solution around the attractor point in moduli space. This result suggests a nice interplay between the width associated with the black hole wavefunction in moduli space, and the species scale in the infinite distance limit; see also~\cite{Anchordoqui:2025izb,Anchordoqui:2026nit} for a similar interplay between the species scale and wavefunctions.

\paragraph{D2-D6 attractors and the Kaluza-Klein scale.} The purpose of this example is to show that, for D2-D6 attractor flows approaching infinite-distance limits, the localisation scale of the semiclassical wave function is now controlled by the Kaluza-Klein scale. This is just a new implication of known limits studied in~\cite{Calderon-Infante:2025pls}.

For a D2-D6 black hole, the non-vanishing charges are $Q=(p^0,0;q_a,0)$. Restricting for simplicity to a one-modulus truncation in~\cite{Calderon-Infante:2025pls}, the K\"ahler modulus at the horizon scales as
\begin{equation}
t_h \sim \left(\frac{q_1}{p^0}\right)^{1/2}\hspace{-12pt}\,.
\end{equation}
Imposing the large-volume regime at the attractor therefore requires $q_1\gg p^0$, which corresponds to an infinite-distance limit in moduli space $t_h \rightarrow \infty$. In this regime, the Kaluza-Klein scale behaves as
\begin{equation}
    m_{\mathrm{KK}} \sim \left(\frac{q_1}{p^0}\right)^{-3/4}\hspace{-20pt}\,.
\end{equation} 
As a result, the tree-level entropy $S_{\rm BH} \sim \sqrt{p^0\,q_1^{3}}$ scales as
\begin{equation}
S_{\rm BH}\sim m_{\mathrm{KK}}^{-2}\,.
\end{equation}
Using the relation $\sigma^2 \sim S_{\rm BH}^{-1/2}$ for the Gaussian width of the semiclassical distribution \eqref{eq:CD_var}, this leads to
\begin{equation}
\sigma^2_{\mathrm{BH}} \sim m_\mathrm{KK}\,.
\end{equation}
In the large-volume limit, the WKB density collapses as $m_\mathrm{KK}\rightarrow 0$. This indeed suggests that the resolution of the black hole horizon cannot be less than $R_{\mathrm{KK}}$, differently from the previous case.

This distinction, however, does not affect the leading behaviour of the Wasserstein distance, which is then dominated by the separation of the corresponding attractor points.

\paragraph{Comments on the general case.} More generally, in extremal black-hole attractor flows, the inverse covariance of the semiclassical distribution, $\mathbf{A}=\mathbf{\Sigma}^{-1}/\hbar$, is controlled by the Hessian of the superpotential $W(\phi)$ evaluated at the attractor point $\phi_*$. Using
\begin{equation}\label{eq:VBH}
V_{\rm BH}=W^2+4\,g^{i\bar j}\partial_i W\partial_{\bar j}W\,,
\end{equation}
together with the attractor condition $\partial_i W|_{\phi_*}=0$, one finds $V_{\rm BH}(\phi_*)=W_*^2$ and hence $S_{\rm BH}=\pi W_*^2$. In one-dimensional reductions, this implies that the confining curvature along the attractor direction grows with the black-hole entropy, driving $\mathbf{\Sigma}\to 0$ in the large-entropy limit.

\subsection{Other examples in minisuperspace} \label{sec:QC}

We illustrate our general construction on simple models where the Wheeler-DeWitt equation can be solved (or controlled semiclassically) in closed form. 
In the solvable models we consider below, we will see how a choice of WKB branch fixes Hamilton-Jacobi trajectories and induces a positive, conserved semiclassical density on hypersurfaces transverse to the associated current. As an illustrative example, we consider the free scalar field in a flat background ($K=0,V(\phi)=0$), in a non-flat background ($K\neq 0$, $V(\phi)=0$); cf. also~\cite{Kan:2021yoh}.

To start, let us consider a homogeneous and isotropic universe and a single scalar field $\phi$. We assume the FLRW metric ansatz in four dimensions
\begin{equation}\label{metric_ans}
    \mathrm ds^2 = - N^2(t)\, \mathrm dt^2 + a^2(t)\, \mathrm d\Omega^2\,.
\end{equation}
With respect to this ansatz, and up to overall convention-dependent normalisations, the minisuperspace Lagrangian takes the schematic form
\begin{equation}\label{Lagrangian_scalar_gravity_free}
    L \supset -\frac{1}{2N}\, a \dot{a}^2 + \frac{1}{2N}\, a^3 \dot{\phi}^2 - N\,U(a,\phi)\,,    \quad U(a,\phi)=a^3 V(\phi) - \frac{1}{2}K a\,,
\end{equation}
where $K$ parametrizes the intrinsic curvature of the spatial slices (we keep $K\neq 0$ for now). The lapse function $N$ plays the role of a Lagrange multiplier that enforces the Hamiltonian constraint $H=0$. The canonical momenta are
\begin{equation}
    \pi_a = -\frac{a \dot{a}}{N}\,,\qquad 
    \pi_\phi = \frac{a^3 \dot{\phi}}{N}\,,
\end{equation}
and the Hamiltonian constraint may be written (again, up to conventions) as
\begin{equation}\label{eq:constraint_FLRW}
    H = -\frac{1}{2}\frac{\pi_a^2}{a}+\frac{1}{2}\frac{\pi_\phi^2}{a^3}+U(a,\phi)\approx 0\,.
\end{equation}
In a quantum theory of gravity, the Hamiltonian constraint becomes the Wheeler-DeWitt (WDW) equation by promoting $\pi_\omega \rightarrow - i \frac{\partial}{\partial \omega}$ for a generic dynamical variable $\omega$. The factor-ordering ambiguity for quadratic operators will be addressed using the Laplace-Beltrami operator on minisuperspace $\pi_\omega^2\rightarrow \Delta$, which is diffeomorphism invariant.

\subsubsection{Free particle with $K=0$} 

We first consider the flat case $K=0$ with a vanishing scalar potential $V(\phi)=0$. 
The Wheeler-DeWitt equation obtained from \eqref{eq:constraint_FLRW} becomes 
\begin{equation}\label{eq:WdW_free_flat}
\left( \Delta_a-\partial_\phi^2 \right)\Psi(a,\phi)=0\,,
\end{equation}
which is a massless Klein-Gordon equation on the two-dimensional minisuperspace spanned by $\{\log a,\phi\}$. It admits exact separated solutions\footnote{Here and in the following examples, for ease of notation, we ignore constant shifts of the form $e^{i\nu\phi_0}$ which can be reabsorbed into the classical solutions $\phi_\ast$.} labelled by the conserved momentum $\nu$ conjugate to $\phi$
\begin{equation}\label{eq:free_part_Knz_wave}
\Psi(a,\phi)=\int_{\mathbb R} \mathrm d\nu\,A(\nu)\,\psi_\nu(a)\,e^{i\nu \phi}\,,
\end{equation}
with $\psi_\nu(a)=C(a)a^{\pm i\nu}$ and $A(\nu)$ real. We can extract the corresponding semiclassical WKB phase $S_0$, such that
\begin{equation}
    \Psi(a,\phi) \sim \int_{\mathbb{R}}\mathrm  d\nu\, A(\nu)C(a) 
    e^{iS_0}\,,
\end{equation}
by identifying 
\begin{equation}\label{eq:S_0_flat}
S_0(a,\phi)=\nu \phi\pm\nu\log a\,.
\end{equation}
The classical trajectories $\phi_\ast$ leading to the saddle point of the integral are then given by
\begin{equation}\label{eq:free_part_Knz_classtraj}
\phi-\phi_\ast=\phi \pm \log a+\text{const.} =0\,,
\end{equation}
which are straight null lines in minisuperspace, reflecting the absence of a potential.

It is customary to build a semiclassical state by choosing a Gaussian packet in $\nu$
\begin{equation}\label{eq:nu_packet_K0}
    A(\nu)=A_0 \exp\!\left[-\frac{(\nu-\nu_0)^2}{4\sigma^2_\nu}\right]\,.
\end{equation}
At fixed\footnote{Fixing the scale factor $a$ provides a natural choice of slice and corresponds to working on hypersurfaces transverse to the Hamilton-Jacobi flow; cf. also the example of $K \neq 0$ below.} $a$, the wave function in \eqref{eq:free_part_Knz_wave} along \eqref{eq:free_part_Knz_classtraj} reduces to the Fourier transform of $A(\nu)$,
\begin{equation}
\Psi(a,\phi)\propto \int \mathrm d\nu\,A(\nu)C(a)\,e^{i\nu(\phi\pm\log a)}\,,
\end{equation}
yielding an exactly Gaussian profile,  
\begin{equation}\label{eq:distr_freepart_K0}
\rho=|\Psi(a,\phi)|^2 \sim \exp\!\left[-\frac{(\phi-\phi_\ast)^2}{2\sigma_\phi^2}\right]\,, \quad \sigma_\phi=\frac{1}{2\sigma_\nu}\,.
\end{equation}
We see that the width $\sigma_\phi$ is independent of the position along the classical trajectory.  Thus, although the wave packet follows the classical solution, it does not become increasingly localised as the scalar field runs to infinite distance.  This absence of localisation enhancements or collapse is not surprising considering the free propagation in flat minisuperspace. We will see shortly that the situation is very different from the behaviour of non flat backgrounds $K\neq 0$ discussed below.

\subsubsection{Free particle with $K\neq 0$}

We again consider a free particle, i.e., $V(\phi)=0$, but now allow for intrinsic curvature $K\neq 0$. After an Eisenhart-Duval lift\footnote{The Eisenhart-Duval lift\label{footnote_ED_lift} provides a systematic way to rewrite a WDW equation with potential as a potential-free equation on an extended minisuperspace with one additional cyclic coordinate. Starting from an operator of the schematic form
\begin{equation}\dotag{fneq:aa_1}
	\Big(\Delta_G-2\,U(q)\Big)\Psi(q)=0\,,
\end{equation} 
one introduces an auxiliary coordinate $\chi$ and considers instead the lifted equation
\begin{equation}\dotag{fneq:aa_2}
	\Big(\Delta_G+2\,U(q)\,\partial_\chi^2\Big)\Psi(q,\chi)=0\,.
\end{equation}
Upon separation of variables $\Psi(q,\chi)=e^{ip\chi}\Psi(q)$, so that $\partial_\chi^2\Psi=-p^2\Psi$, the lifted equation reduces back to the original Wheeler-DeWitt equation, with the potential $U(q)$ encoded through the conserved momentum $p$ along the $\chi$ direction. See~\cite{Kan:2021yoh} and the reference therein for additional details on the Eisenhart-Duval lift.}, introducing an auxiliary cyclic variable $\chi$, the WDW equation can be written in the convenient form
\begin{equation}\label{eq:WdW_ED_freeK}
    \left( a\frac{\partial}{\partial a}\, a\frac{\partial}{\partial a} -\frac{\partial^2}{\partial \phi^2} + K a^4 \frac{\partial^2}{\partial \chi^2}\right)\Psi(a,\phi,\chi)=0\,,
\end{equation}
where $\chi$ is the Eisenhart-Duval coordinate. We separate variables by taking\footnote{As detailed in~\cite{Kan:2021yoh}, one recovers conventional WDW by setting $p^2=1$.}
\begin{equation}
    \Psi(a,\phi,\chi)=e^{ip\chi}\,\Psi(a,\phi)\,,  \quad  \frac{\partial^2}{\partial\chi^2}\Psi=-p^2\Psi\,,
\end{equation}
so that the $(a,\phi)$ wave function solves
\begin{equation}\label{eq:WdW_freeK_reduced}
    \left( \Delta_a -\partial_\phi^2  -K p^2 a^4  \right)\Psi(a,\phi)=0\,,
\end{equation}
with $\Delta_a =  a\partial_a\,a\partial_a $. As argued in~\cite{Hawking:1990in, Kiefer:1988tr, Kan:2021yoh, Andrianov:2018wdx}, a useful basis of separated solutions is labelled by a real parameter $\nu$ (conjugate to $\phi$). Writing a superposition
\begin{equation}\label{eq:gensolsFP}
    \Psi(a,\phi,\chi)=\int_{\mathbb{R}}\mathrm  d\nu\, A(\nu)\,\psi_{\nu,p}(a,\phi)\,e^{ip\chi}\,,
\end{equation}
$\psi_{\nu,p}$ satisfies \eqref{eq:WdW_freeK_reduced} as
\begin{equation}
    \psi_{\nu,p}(a,\phi)\sim
    \begin{cases}
        K_{i \nu/2}\!\left( \frac{\sqrt{K}|p|}{2}\,a^2\right)e^{i\nu \phi}\,,& K>0\,,\\
        J_{\pm i \nu/2}\!\left( \frac{\sqrt{|K|}|p|}{2}\,a^2\right)e^{i\nu\phi}\,,& K<0\,,
    \end{cases}
\end{equation}
where $\phi_\ast$ is a constant that will parametrize the classical family selected by the wave packet, $J_\alpha$ are Bessel functions of the first kind and $K_\alpha$ are modified Bessel functions of the second kind. In what follows, we will restrict the discussion to the $K>0$ branch. We again choose $A(\nu)$ to be a Gaussian wave packet peaked at $\nu_\ast$ as in \eqref{eq:nu_packet_K0} and evaluate \eqref{eq:gensolsFP} by the stationary phase in the WKB regime. The relevant WKB phase is controlled by the large-order asymptotics of the Bessel functions. Introducing $y = \sqrt{K}\,|p|\,a^2$ for $K>0$, we extract a ($\chi$ independent) semiclassical phase $S_0$ such that
\begin{equation}
    \Psi(a,\phi,\chi) \sim \int_{\mathbb{R}}\mathrm  d\nu\, 
    C(a)e^{iS_0}\,e^{ip\chi}\,,
\end{equation}
with 
\begin{equation}\label{eq:Snu_corrected}
    S_0(a,\phi)=    \frac{\nu}{2}\cosh^{-1}\!\left(\frac{\nu}{y}\right)    -\frac12\sqrt{\nu^2-y^2}    -\frac{\pi}{4}    +\nu\phi\,,    \qquad (\nu>y)\,.
\end{equation}
The peak of the wave packet is determined by the stationary condition
\begin{equation}\label{eq:stationary_nu}
    \left.\frac{\partial S_0(a,\phi)}{\partial \nu}\right|_{\nu=\nu_\ast}=0\,,
\end{equation}
which selects a classical trajectory in minisuperspace
\begin{equation}
   \phi_\ast \equiv \phi(a,\nu_\ast) = -\frac{1}{2}\cosh^{-1}\left(\frac{\nu_\ast}{y}\right)\,.
\end{equation}
Thus, the Gaussian packet in $\nu$ selects a family of classical histories along the trajectories defined by $\nu_\ast$ and $\phi_\ast$.

To make contact with localisation properties of the wave packet, as in the previous section, we move on to extract the width of the semiclassical state in the scalar direction at fixed $a$. In minisuperspace, fixing the scale factor $a$ provides a canonical slice, physically corresponding to comparing scalar configurations at fixed spatial volume.  This is thus a natural choice, as in many approaches to defining probability densities from the WDW equation; we already remarked in Section \ref{sec:ADM_WdW} that these are meaningfully defined only on hypersurfaces transverse to the Hamilton-Jacobi flow.  However, one has to be careful that this choice is only consistent, provided the hypersurface $a=\mathrm{const.}$ is transverse to the Hamilton-Jacobi flow generated by $S_0$. Equivalently, the WKB current must have non-vanishing flux through the slice $J^a\sim |A|^2G^{aa}\partial_a S_0 \neq0$, which reduces to the condition $\partial_a S_0\neq 0$. For the phase \eqref{eq:Snu_corrected}, one finds
\begin{equation}
\partial_a S_0= -\frac{\sqrt{\nu^2-y^2}}{a}\,,
\end{equation}
where $y=\sqrt{K}|p|a^2$. Hence, throughout the WKB regime $\nu>y$ in \eqref{eq:Snu_corrected}, the fixed-$a$ slices are indeed transverse, and the induced probability distribution in the scalar direction is well defined. The construction only degenerates at the boundary $\nu=y$, where the Bessel-function asymptotics also break down.

To obtain the width on this chosen hypersurface in minisuperspace; $a=\mathrm{const.}$ or equivalently $y=\sqrt{K}|p|a^2$ constant, the saddle condition \eqref{eq:stationary_nu} defines the on-shell relation
\begin{equation}\label{eq:saddle_cond_freepart}
\nu_\ast(a,\phi_\ast)=y\,\cosh\!\big(2\phi_\ast\big)\,.
\end{equation}

We now expand $S_0$ around $\nu_\ast$
\begin{equation}\label{eq:saddle_expansion_Knot0}
    S_0(a,\phi,\nu_\ast +\delta\nu) = S_\ast + \partial_\nu S_\ast \delta \nu + \frac{1}{2} \partial_\nu^2 S_\ast (\delta \nu)^2+\dots\,,
\end{equation}
which, upon inserting into $\Psi(a,\phi)$, gives a Gaussian integral that can be easily evaluated and leads to a distribution of the form 
\begin{equation}
    \rho=|\Psi(a,\phi)|^{2} \sim \exp\!\left[- \frac{(\partial_\nu S_\ast)^2}{2 \mathfrak{Re}(\sigma_\phi^2)} \right]\,, \qquad \sigma_\phi^{-2} = \frac{1}{2\sigma_\nu^2} - i \partial_\nu^2 S_\ast\,.
\end{equation}
Explicitly, we obtain
\begin{align}
\begin{aligned}
    S_\ast&=S_0(a,\phi_,\nu_\ast)\\
    \partial_\nu S_\ast &=\partial_\nu S_0(a,\phi,\nu)|_{\nu_\ast}= \frac{1}{2}\cosh^{-1}\Bigr(\frac{\nu_\ast}{y}\Bigl)+\phi=\phi-\phi_\ast\\
     \partial_\nu^2 S_\ast &=\partial_\nu^2 S_0(a,\phi,\nu)|_{\nu_\ast}= \frac{1}{2\sqrt{\nu_\ast^2-y^2}}\,,
\end{aligned}
\end{align}
and hence
\begin{equation}\label{eq:distr_free_Kl0}
    \rho=|\Psi(a,\phi)|^2 \sim \exp\!\left[- \frac{(\phi-\phi_\ast)^2}{2 \bar\sigma_\phi^2} \right]\,, \qquad \bar\sigma_\phi^2 =\frac{|\sigma_\phi^2|^2}{\mathfrak{Re} (\sigma_\phi^2)}= \frac{1}{2\sigma_\nu^2} +  \frac{\sigma_\nu^2 \,\mathrm{csch}^2(2\phi_\ast)}{4y^2}\,.
\end{equation}
Note that since $y=\sqrt{K}|p|a^2$, one could naively expect that in the limit $y \to 0$, one should obtain the flat $K=0$ expression. However, this is not the case since the expansion of the Bessel functions and the associated solution is only well defined for $K\neq 0$. For $K=0$ one has to first reduce to the simpler WDW equation \eqref{eq:WdW_free_flat} and then work with the associated $S_0$. However, qualitatively, one sees that this corresponds to reducing $S_0$ to the linear form of \eqref{eq:S_0_flat} and therefore one can view the limit $K\to 0$ as sending $\partial_\nu^2 S \to 0$, in which case the associated variance indeed reduces to the free case.

\section{Relation to other notions of distance on field space}\label{sec:rels}

In this section, we relate the WKB-derived distributions and their Wasserstein distances introduced in this work to earlier approaches in the literature that attempt to quantify the separation between effective theories. These approaches fall broadly into two classes. The first concerns modifications of field-space distance due to scalar potentials or dynamics, particularly black-hole physics in the context of the Swampland Distance Conjecture. The second concerns constructions that associate density distributions or weights to vacua or moduli-space configurations.
Our framework intersects with both directions, but differs from both in its basic object --- density measures on scalar field space --- and in the way the distance between such objects is defined.
We focus on approaches that are structurally closest to our framework. Many other proposals address related questions from different perspectives and are not reviewed in detail here. See, for example~\cite{Raml:2025yrb} for a detailed discussion of the SDC on scalar field space in the presence of a non-trivial potential, as well as a summary of recent developments on the topic.

\subsection{Related approaches in the presence of a potential}

Several recent proposals refine the geodesic distance on scalar field space in order to incorporate scalar potentials or dynamical constraints. A common feature of these constructions is that the distance is generated by a cost accumulated along trajectories rather than by a fixed background metric. In this subsection, we explain how several of such approaches relate to the transport framework developed in this work.  To keep the discussion concise, we defer technical elaborations and more detailed comparisons with some of the complementary notions of distance to Appendix \ref{app:details_connections}.

\paragraph{Relation to hyper-distinguishability.} Our construction should be contrasted with the hyper-distinguishability proposal of~\cite{Stout:2021ubb}, where infinite-distance limits are diagnosed through the local sensitivity of EFT observables to variations of moduli. In that framework, moduli $\phi$ parameterise families of probability distributions over observables $x$, 
\begin{equation}
\mathcal M_{\mathrm{obs}} = \{\, p(x|\phi)\mid \phi \in \mathcal M \,\}\,,
\end{equation}
and the Fisher information metric
\begin{equation}\label{eq:Fisher_info}
g_{{\rm F},ab}(\phi)= \int p(x|\phi)\,\partial_a \log p(x|\phi)\,\partial_b \log p(x|\phi)\,\mathrm dx\,
\end{equation}
measures the infinitesimal distinguishability of nearby EFTs. Divergences of $g_{\rm F}$ signal infinite-distance limits and the onset of light towers. This notion of distance is intrinsically local, probing the response of observables to small displacements in moduli space.

By contrast, we associate to each EFT background a semiclassical density $\rho_E(\phi|\theta)$ directly on moduli space and define distances as Wasserstein distances between these measures,
\begin{equation}
W_c(\rho_E^{(1)},\rho_E^{(2)}) = \inf_{\pi\in\Pi(\rho_E^{(1)},\rho_E^{(2)})} \int_{\mathcal M\times\mathcal M} c(\phi,\phi')\,\mathrm d\pi(\phi,\phi')\,,
\end{equation}
with cost $c(\phi, \phi^\prime)$ determined by the moduli-space geometry (or its deformation in the presence of a potential). In the limit where $\rho_E$ collapses to a Dirac delta, both approaches reduce to measuring the displacement of a representative point in moduli space. Away from this limit, however, the Fisher information distance and the Wasserstein distance do not coincide in general and probe different structures: hyper-distinguishability captures local EFT sensitivity, whereas the transport-based distance introduced here measures global separation of semiclassical backgrounds.

For a more explicit comparison with hyper-distinguishability, see Appendix \ref{app:hyperdist}.

\paragraph{Jacobi-Maupertuis distance.}\label{sec:JM_dist} We now turn to the proposal of~\cite{Mohseni:2024njl} to define distances in the presence of scalar potentials using the Euclidean Jacobi-Maupertuis principle in the absence of dynamical gravity. We now explain how it embeds naturally within our OT framework in a rigorous way. Consider a scalar field configuration $\phi(\tau)$ in Euclidean signature, with 
\begin{equation}\label{eq:scalarEucl}
S_E[\phi] = \int \mathrm d\tau \left( \tfrac12 g_{ij}(\phi)\dot\phi^i \dot\phi^j + V(\phi) \right)\,.
\end{equation}
Instead of fixing the Euclidean time interval, one may formulate the problem at fixed energy density $\rho_E$. Subsequently, the on-shell functional is obtained by introducing the conserved energy 
\begin{equation}
    \rho_E = \tfrac12 g_{ij}(\phi)\dot\phi^i \dot\phi^j - V(\phi) 
\end{equation}
as Legendre constraint in \eqref{eq:scalarEucl}. This leads to
\begin{equation}\label{eq:appJM_onshellact}
S_{\mathrm{on\text{-}shell}} = \int \mathrm d\tau\, g_{ij}\dot\phi^i \dot\phi^j = \int_{\phi_i}^{\phi_f} \sqrt{2(\rho_E+V(\phi))\, g_{ij}(\phi)\, \mathrm d\phi^i \mathrm d\phi^j}\,,
\end{equation}
which, up to normalisation, is the so-called Jacobi-Maupertuis functional. 

\noindent
A proposal to incorporate scalar potentials using the Jacobi metric into scalar-field distances was put forward in~\cite{Mohseni:2024njl}, where distances are defined precisely via the Euclidean Jacobi-Maupertuis principle. For a moduli space $(\mathcal M,g)$ endowed with a potential $V(\phi)$ and a fixed Euclidean energy scale $\rho_E$, as defined through \eqref{eq:scalarEucl}, the distance between two configurations $\phi_i,\phi_f$ is defined by the minimised Jacobi functional
\begin{equation}
\Delta(\rho_E)=\frac{1}{\sqrt{\rho_E}}\int_{\gamma:\,\phi_i\to\phi_f}\sqrt{g_{J,ij}(\phi)\, \mathrm d\phi^i \mathrm d\phi^j}\,.
\end{equation}
where one identifies from \eqref{eq:appJM_onshellact} the conformally rescaled Jacobi metric $g_{J,ij}$, defined in \eqref{eq:Jacobi_metric}. Hence, the extremal trajectories satisfy the geodesic equation on $(\mathcal M, \mathbf g_J)$. The solutions of the equations of motion at fixed energy $\rho_E$ thus determine the geodesics of the Jacobi metric.

From the viewpoint of OT, this construction corresponds to a particular choice of cost function on field space. Instead of the quadratic cost induced by the metric $g$, one considers the action or Ma\~n\'e cost associated with the Tonelli Lagrangian
\begin{equation}
L(\phi,\dot\phi)=\tfrac12\|\dot\phi\|_g^2 + V(\phi)\,,
\end{equation}
leading to the point-to-point cost
\begin{equation}
c_L(\phi_i,\phi_f)=\inf_{T>0}\inf_{\gamma} \int_0^T L(\gamma,\dot\gamma)\,\mathrm dt\,.
\end{equation}
As reviewed in Section~\ref{sec:OTpotential}, minimisers of this cost coincide with the geodesics of the Jacobi metric, so that the transport problem with cost $c_L$ is equivalent to a Wasserstein problem on $(\mathcal M, \mathbf g_J)$.

In the limit where the measures collapse to Dirac distributions, $\mu_0=\delta_{\phi_i}$ and $\mu_1=\delta_{\phi_f}$ --- or, for instance, for two Gaussians with the same variance --- the OT distance reduces to the underlying point-to-point cost:
\begin{equation}
W_{2,L}(\mu_0,\mu_1)=c_L(\phi_i,\phi_f)=d_J(\phi_i,\phi_f)\,.
\end{equation}
Thus, the Jacobi-Maupertuis distance appears as the deterministic (zero-variance) limit of the transport geometry used in the main text. In this sense, the proposal of~\cite{Mohseni:2024njl} is recovered as a special case of the more general framework in which configurations are described by semiclassical distributions rather than points.

Section~\ref{sec:cost_and_gravity} discusses how such scalar-sector expressions can be related to the reduced gravitational dynamics. In particular, it distinguishes costs defined on the enlarged configuration space from projected on-shell quantities in scalar field space, which are useful for comparison with several proposals put forward in~\cite{Mohseni:2024njl}.

\paragraph{Ricci-flow-based distances.} The Ricci Flow Distance Conjecture (RFC)~\cite{Kehagias:2019akr} proposes that flowing along the gradient of geometric data toward a fixed point at infinite distance is accompanied by the emergence of an infinite tower of states in quantum gravity, in close analogy with the Swampland Distance Conjecture, see also~\cite{DeBiasio:2022zuh,DeBiasio:2022nsd,DeBiasio:2022omq,Velazquez:2022eco}. In its original formulation, the relevant gradient flow is realised by Ricci flow together with Perelman's entropy functional, and the associated notion of distance is inferred from the scaling behaviour of curvature and entropy along the flow. More recent work~\cite{Demulder:2024glx} extends this framework to flux-supported internal spaces and generalised geometric flows, where infinite-distance limits are characterised through entropy functionals and canonical measures arising along the flow. In these settings, distance is not a geodesic notion on a fixed configuration space, but is instead encoded in an effective cost accumulated along the flow, thereby connecting to the cost-based viewpoint adopted in this work.

The relation between (generalised) Ricci flow and OT is therefore conceptually close but should be interpreted in a flow-time-dependent sense. In the RFC, distances between backgrounds are defined through costs generated dynamically by the geometric evolution of the compactification manifold. Concretely, the evolution is governed by
\begin{equation}
\partial_t g_{ij} = -2\,\mathrm{Ric}_{ij}\,,
\end{equation}
which can be formulated as the gradient flow generated by Perelman's entropy functional
\begin{equation}\label{eq:entropy_functional}
\mathcal F(g,f)
\equiv
\int_M \mathrm d^n x\,\sqrt{g}\,
e^{-f}\bigl(R+|\nabla f|^2\bigr)\,.
\end{equation}
In the context of the RFC, $\mathcal F$ does not define a distance in the metric sense, nor does it generate geodesics; rather, it provides a functional whose monotonicity along the flow allows one to define an effective, direction-dependent cost associated with evolving background configurations. It is this accumulated cost, rather than a geodesic length, that underlies the notion of distance relevant for the RFC.

This interpretation aligns naturally with the philosophy of OT, where the primary object is a cost functional rather than a metric. Ricci flow equips the evolving manifold $(M,\mathbf g(\tau))$ with a normalised density $\rho(x,\tau)$ that satisfies the (conjugated) heat equation
\begin{equation}
\partial_\tau \rho=\Delta_{g(\tau)}\rho-\tfrac12\mathrm{tr}\!\left(\partial_\tau g\right)\rho\,.
\end{equation}
As shown in~\cite{mccann2010ricci,topping2009ricci}, contraction does not hold for the density alone with respect to a fixed Wasserstein distance; instead, the evolution of the coupled metric-measure leads to a transport problem with a time-dependent, flow-adapted cost.

This picture extends to generalised Ricci flows~\cite{Demulder:2024glx}, where additional fields are incorporated into the flow, and recent work~\cite{kopfer2024optimal} formulates these structures within OT in terms of evolving costs.

\subsection{Related measures on moduli spaces of vacua}

A complementary strand of the literature associates weights to moduli-space configurations or to entire flux sectors, typically motivated by semiclassical wavefunctions, entropy functionals, or statistical distinguishability of theories. Unlike the approaches discussed in the previous subsection, the primary goal of these approaches is not to define a distance between backgrounds, but rather to identify preferred loci, extrema, or locally distinguishable directions in moduli space. Here, we briefly clarify the relation to our construction.

\paragraph{Semiclassical wavefunctions and entropic selection.}
The Hartle-Hawking-type wavefunction for flux compactifications proposed in~\cite{Ooguri:2005vr} assigns to each flux sector $(p,q)$ a semiclassical weight
\begin{equation}
\Psi_{p,q}(X,\bar X)\sim e^{-S_E(X,\bar X;p,q)}\,,
\end{equation}
written in homogeneous special-geometry variables $X$. Because the Euclidean action $S_E$ is homogeneous under an overall rescaling of periods, $|\Psi_{p,q}|^2$ is naturally viewed as a formal weight on an extended configuration space rather than as a normalisable probability density on projective moduli space. The follow-up work~\cite{Gukov:2005bg} (see also~\cite{Gukov:2005iy}) made this point precise and advocated the entropic principle: after fixing the homogeneous scaling direction, one studies extrema of an entropy functional on the constrained space, thereby selecting special loci such as attractor points.

Our framework is conceptually orthogonal. We do not use a semiclassical wavefunction to define a landscape measure or an extremisation principle. Instead, we fix the discrete data (e.g. fluxes or charges) and construct a normalised density on the relevant field space (or an appropriate slice thereof), which can then be compared across different choices of discrete data.

\paragraph{Information-theoretic proximity of QFTs.} A different notion of proximity between effective theories is provided by information geometry, where one quantifies distinguishability via relative entropy between Euclidean path-integral measures~\cite{Balasubramanian:2014bfa,Heckman:2013kza}. In the infinitesimal limit this reduces to a Fisher information metric on parameter space and reproduces standard structures such as the Zamolodchikov metric for conformal manifolds~\cite{Balasubramanian:2014bfa}. More recent work develops functional versions of these constructions as Riemannian geometries on spaces of probability functionals~\cite{Floerchinger:2023ekw}. These approaches are fundamentally local: they diagnose sensitivity and distinguishability under small deformations.

By contrast, our distances are global: we compare semiclassical densities on field space using quadratic OT (and, when relevant, potential-deformed Tonelli-costs). This yields a bona fide metric on the space of measures and captures finite separations between backgrounds, including the effect of support, tails, and covariance, rather than only infinitesimal distinguishability.

In summary, this work constitutes a step toward assigning a precise geometric meaning to the notion of distance between effective field theories in quantum gravity. By using the well-established framework of OT, we aim to provide a more geometric and analytic characterisation of distance from first principles.
In particular, OT provides a useful language for separating three pieces of data: the semiclassical distribution associated with a background, the configuration space on which this distribution is defined, and the cost used to compare two such distributions.
More broadly, our work suggests that the wavefunction itself may serve as a useful tool for probing new aspects of the Swampland program and may offer fresh insight into existing Swampland conjectures.

\section{Conclusions}\label{sec:conclusions}

In this work, we have explored how the framework of Optimal Transport can be used to formulate notions of distance between effective field theories in the presence of scalar potentials and, more generally, dynamical backgrounds. The central idea is to replace points in field or configuration space by semiclassical probability distributions, and to compare such distributions by means of suitable transport costs. In the strictly localised limit of delta distribution, these distances reduce to the corresponding point-to-point costs between the centres of the distributions. Away from this limit, they retain information about the spread and shape of the underlying distribution, which in turn captures additional information encoded by the associated physical configuration.

The key bridge to OT is provided by the semiclassical expansion of the relevant wave equation. In the WKB expansion of Schr\"odinger-type problems, the leading phase satisfies a Hamilton-Jacobi equation, whereas the leading amplitude obeys a continuity equation. These are precisely the two structures that appear in the dynamical Benamou-Brenier formulation of quadratic OT, and they underlie the dictionary developed in Section~\ref{sec:WKB_HJprobs}. On this basis, the associate distances can be associated with the resulting semiclassical densities. For the quadratic cost this gives the usual Wasserstein distance, while in the presence of scalar potentials the Tonelli or Jacobi-Maupertuis costs~\cite{bernard2006monge} provide a controlled way of incorporating the potential into the transport problem. In this setting, the transport distance is not simply the Riemannian distance on field space, but a cost-dependent distance on the space of probability measures.

\medskip

In the presence of dynamical gravity, the dictionary becomes less straightforward. The Wheeler-DeWitt equation does not define evolution with respect to a preferred external time and, therefore, does not single out a canonical dynamical transport problem. We showed how a semiclassical transport interpretation can nevertheless be obtained after choosing a WKB branch and an internal clock, which induces a continuity equation for the remaining variables. This construction is necessarily more choice-dependent than in ordinary Schr\"odinger problem. Therefore, we distinguished between costs motivated by the reduced dynamics, such as those obtained from reduced Hamiltonians, and other costs, such as a quadratic cost, where we need to impose the on-shell conditions externally.

For dynamically motivated WDW costs, solving the gravitational constraint after choosing an internal clock generically leads to reduced square-root-type cost functions. These define legitimate action-type costs, but they are naturally costs on the reduced configuration space, rather than intrinsic scalar-field distances. In particular, in the zero-potential limit, such costs can degenerate, reflecting the null structure of the constrained minisuperspace dynamics. We also discussed projected scalar-sector costs, motivated by the question of how to extract a scalar-field distance relevant for the SDC from the full construction. These quantities isolate the scalar contribution of the on-shell action and provide diagnostics that match proposals such as~\cite{Mohseni:2024njl,Debusschere:2024rmi}. In particular, after an appropriate normalisation, they take the same structural form as the proposed the domain-wall tension functional and scalar field distance, suggesting a geometric interpretation from the this perspective.

\noindent
Finally, we applied the framework to several examples. First, we showed that (fake) supergravity provides a natural setting for setting up the optimal transport problem, through its Hamilton-Jacobi formulation. In the extremal black-hole examples, this structure allows one to define transport costs naturally adapted to the effective one-dimensional radial flow. Semiclassical densities near attractor points can then be approximated by Gaussians, and the associated transport distances are controlled by the behaviour of the covariance matrix. In large-entropy regimes, the localisation of the distribution sharpens, so that the associated distance approaches the corresponding deterministic field-space distance. In particular, the variance in scalar field space of the associated distributions is controlled by the entropy and scales as the species scale or the KK scale, depending on the class of black-hole solutions under consideration. We also studied free WDW solutions with and without extrinsic curvature.

\medskip

In summary, this work constitutes a step toward assigning a precise geometrical meaning to the notion of distance between effective field theories in quantum gravity. By using the well-established framework of OT we aim to provide a more geometric and analytic characterisation of distances from first principles. Additionally, the perspective presented in this paper also bears analogies to earlier uses of wave functions and probability distributions in String Theory and QFT~\cite{Ooguri:2005vr,Gukov:2005bg,Gukov:2005iy,Balasubramanian:2014bfa,Heckman:2013kza}. More broadly, our work suggests that considering semiclassical probability distributions on field space may provide a useful tool for probing new aspects of the Swampland program and may offer fresh insight into existing Swampland conjectures.

\bigskip

The results and observations highlighted above suggest several interesting directions for future investigation:

\paragraph{Quantum Wasserstein.} Our framework compares semiclassical scalar field space densities $\rho(\phi;\,\alpha)$ arising from the WKB approximation, rather than fully quantum states. Nevertheless, one may ask whether, if an underlying quantum state $\hat{\Psi}$ on an appropriate Hilbert space is defined, a corresponding notion of distance between states — a ``quantum Wasserstein" distance — can encode physically relevant information beyond that captured by classical transport costs. Such a concept would have to generalise the one developed in this work in order to naturally include transitions between different vacua that are not connected by semiclassical motion. Several inequivalent proposals for quantum Optimal Transport and quantum Wasserstein metrics exist~\cite{Cole:2021jwk,DePalma:2021zvy,Li:2022exc,Beatty:2024nkq}, typically requiring additional structure beyond the classical setting. Understanding whether any of these quantum distances admits a meaningful interpretation in the context of the SDC, and more generally in quantum gravity, and how they reduce to the semiclassical distances used here, is an interesting direction for future work.

\paragraph{Relation to Fisher-Rao metric and information geometry.} The Wasserstein distances used in this work and the Fisher-Rao information metric underlying the hyper-distinguishability approaches such as~\cite{Heckman:2013kza,Balasubramanian:2014bfa,Stout:2021ubb} arise from distinct geometric frameworks on spaces of probability measures. Wasserstein geometry encodes a global transport cost tied to the underlying field-space metric, whereas Fisher-Rao captures local statistical distinguishability. While these metrics do not coincide in general, there exist precise mathematical constructions that relate them, such as entropy-regularised transport and interpolating distances like the Wasserstein-Fisher-Rao metric, which reduce to either structure in suitable limits~\cite{khan2022optimal}. In this work, we focus on the pure Wasserstein framework as it naturally reflects transport costs relevant for comparing semiclassical densities across different semiclassical backgrounds.

\paragraph{Higher dimensional scalar field spaces and their geometry.} The framework developed in this work should be viewed as a first step towards a more general analysis of semiclassical wave functionals on scalar field space and their role in characterising distances relevant for quantum gravity. In the present work, we focus primarily on examples with a single scalar field. A natural extension is to consider higher-dimensional moduli spaces with multiple coupled scalar fields, where the geometry of field space and the structure of the associated transport problem would become considerably richer. In particular, it would  be interesting to connect the concept developed in this work to the spectral and Laplacian properties of moduli space discussed in~\cite{Aoufia:2025ppe}.

\paragraph{Wasserstein distance and the volume of moduli space} In full generality, the analysis developed here suggests that the Wasserstein distance should be computed between probability measures induced by semiclassical wavefunctions on the appropriate moduli or configuration space. From this perspective, the volume of moduli space enters through the normalisability and integrability properties of these induced densities. In particular, it would be interesting to understand whether bounds on the volume of moduli space in infinite-distance limits are related to the integrability properties of WDW wavefunctions defined on the same space~\cite{Delgado:2024skw, Grimm:2025lip}.

\bigskip
\subsection*{Acknowledgments}
We thank Ivano Basile, Leonardo Bersigotti, Nicol\`o de Ponti, Muldrow Etheredge, Bernardo Fraiman, Alvaro Herr\'aez, Joaquin Masias, and Matteo Zatti for interesting discussions and comments on the draft.
S.D. and C.M. would like to thank DESY and Universit\"at Hamburg, where part of this work was carried out, for the kind hospitality. T.R. thanks CUNEF Universidad for the kind hospitality during the final stages of this work.  The work of D.L. is supported by the German-Israel-Project (DIP) on Holography and the Swampland.

\newpage
\appendix
\addtocontents{toc}{\protect\setcounter{tocdepth}{1}}

\section{Comments on Otto calculus}\label{app:Otto}

It was already briefly mentioned in Section \ref{sec:revOT_OC} that the Wasserstein distance can be given a very explicit geometric interpretation by viewing the space of probability measures $\mathcal{P}(M)$ on a smooth manifold $M$ formally as a (infinite-dimensional) Riemannian manifold with metric $g_\rho^W(\cdot,\cdot)$, which was developed by Otto~\cite{otto2001geometry,otto2005eulerian} and is known as Otto calculus. 

Starting from the space of probability densities $\rho(x)\, \mathrm{d}V_M \in \mathcal{P}(M)$, Otto endowed it with a Riemannian (Wasserstein) metric	
\begin{equation}\label{eq:inn_Otto}
	g^W_\rho(\delta \rho_0, \delta \rho_1) = \int_M \nabla \phi_0  \nabla \phi_1 \rho \, \mathrm{d} V_M\,,
\end{equation}
where the variations $\delta \rho_i$ are defined\footnote{Due to this definition, we are guaranteed ``mass-conservation'' Indeed, at linear order, the conservation condition demands that $\int_M \delta \rho \, dx = 0$, which --- due to the total divergence --- is automatically satisfied.} by $\delta \rho_i = \nabla \cdot  (\rho \nabla \phi_i)$, where $\phi$ is the potential featured in the continuity equation \eqref{eq:continuity} as the potential for $v$, i.e. $v=\nabla \phi$. This defines a positive-definite bilinear form on the tangent bundle $T\mathcal{P}$. While the geodesics of the Wasserstein metric are precisely the length-minimising curves of the 2-Wasserstein distance $W_2$, one can also define a gradient associated with $g^W$, called the Wasserstein gradient $\mathrm{grad}_W$, which, acting on a functional $\mathcal{F}$, can be defined as\footnote{See also~\cite{villani2008optimal} eq. (15.2).}
\begin{equation}\label{eq:W_grad}
    \mathrm{grad}_W \mathcal{F}(\rho) \equiv - \mathrm{div}\left(\rho \nabla \frac{\delta}{\delta \rho}\mathcal{F}(\rho)\right)\,.
\end{equation}
Using this gradient, it is possible to interpret certain classes of diffusion equations as gradient flows of a given generating (entropy) functional $\mathcal{E}(\rho)$
\begin{equation}
	\partial_t \rho = -\mathrm{grad}_W \mathcal{E}(\rho)\,,
\end{equation}
in the space of probability measures equipped with the Wasserstein distance $W_2$.

The prime example illustrating the gradient flow formulation associated with the Wasserstein metric is given by 
\begin{equation}
    \mathcal{E}(\rho) = \int \rho \ln(\rho)\mathrm{d}x\,.
\end{equation}
In this case, we find 
\begin{equation}
    \mathrm{grad}_W \mathcal{E}= -\mathrm{div}\left((\rho \vec{\nabla} (1+\ln(\rho))\right) = -\nabla^2 \rho\,,
\end{equation}
such that the associated flow equation becomes
\begin{equation}
    \partial_t \rho =-\mathrm{grad}_W \mathcal{E} = \Delta \rho\,,
\end{equation}
which is nothing else than the well-known heat equation.

\section{Hyper-distinguishability and Optimal Transport}\label{app:details_connections}\label{app:hyperdist}

In this appendix, we compare --- in more detail --- our framework and the notion of hyper-distinguishability introduced in~\cite{Stout:2021ubb}. First, discrete parameters $\theta$ (such as fluxes, charges, or gaugings) label different compactification sectors, and hence different underlying UV realisations. Second, the moduli fields $\phi$ parameterise the scalar field space $\mathcal M$ associated with a given sector. Finally, effective observables $x$ correspond to quantities defined in the low-energy EFT (such as couplings or spectra), which arise as functions of the moduli. 

In the framework of~\cite{Stout:2021ubb}, one considers a probability distributions $p(x|\phi)$ over observables, with $\phi$ playing the role of parameters. This defines a statistical manifold
\begin{equation}\label{eq:probs_obs}
\mathcal M_{\mathrm{obs}} = \{\, p(x|\phi)\mid \phi \in \mathcal M \,\}\,,
\end{equation}
equipped with the Fisher information metric \eqref{eq:Fisher_info}. This metric measures the distinguishability of nearby distributions in $\mathcal M_{\mathrm{obs}}$, and induces a Riemannian geometry on moduli space encoding the local sensitivity of EFT observables to variations in $\phi$. In this setting, infinite-distance limits are detected through divergences of the Fisher metric along trajectories in moduli space, corresponding to a breakdown of distinguishability of nearby theories. This probe is intrinsically local as it depends on derivatives of the observables with respect to the moduli.

By contrast, in our approach we associate, for each fixed choice of discrete data $\theta$, a probability density directly on moduli space,
\begin{equation}
\rho_E(\phi|\theta)\,,
\end{equation}
obtained from a Euclidean WKB or Hamilton-Jacobi analysis. This density is assumed to be normalised, i.e., $\int_{\mathcal M} \rho_E(\phi|\theta)\,\mathrm d\mu(\phi)=1$, and encodes the intrinsic semiclassical spread of a given EFT background in field space.  Distances between EFT backgrounds are then defined as Wasserstein distances between these measures, with cost determined by the geometry of the moduli space (or by its Jacobi deformation in the presence of a scalar potential). The basic object is therefore not a family of probability distributions on observable space, but a family of semiclassical distributions on moduli space itself.

Although distinct, one can transport one class to the other and ask how they compare directly. To relate the two settings, we introduce a map from moduli space to observable space.  In the simplest deterministic setting, when the measures are Dirac distributions, this is a map
\begin{equation}
f:\mathcal M \to \mathcal X\,,
\end{equation}
sending a moduli configuration $\phi$ to a set of EFT observables $x=f(\phi)$.  More generally, this relation need not be deterministic: a given $\phi$ may correspond  to a distribution of observables. This is described by a (Markov) kernel $K(x|\phi)$,  i.e. a conditional probability density on $\mathcal X$ for each $\phi\in\mathcal M$,  which reduces to $K(x|\phi)=\delta(x-f(\phi))$ in the deterministic limit. Given a semiclassical density $\rho_E(\phi|\theta)$, such a kernel induces a distribution over observables
\begin{equation}\label{eq:distr_from_kernel}
P(x|\theta)=\int_{\mathcal M} K(x|\phi)\,\rho_E(\phi|\theta)\,\mathrm d\mu(\phi)\,.
\end{equation}
However, this conceptually differs from the distributions $p(x|\phi)$ that appear within the information-geometric framework~\eqref{eq:probs_obs}. There, the moduli $\phi$ label a family of probability distributions over observables, and one studies how these distributions vary as $\phi$ is varied.  In contrast, in our construction $\phi$ is itself distributed according to $\rho_E(\phi|\theta)$, and $P(x|\theta)$ is obtained as a mixture over moduli configurations. 

Note that the Fisher metric is local and measures infinitesimal distinguishability of nearby EFTs in observable space, while the Wasserstein distance measures global separation between semiclassical distributions on moduli space. Finally, even if the family $\rho_E(\phi|\theta)$ evolves on moduli space by a transport equation, one should not expect this structure to descend to observable space in any simple way after pushforward by $f$ or $K$. In particular, the pushforward of an OT problem on moduli space does not, in general, define another OT problem on observable space with the same couplings or geodesics. 

In the deterministic limit in which the WKB density collapses to a Dirac measure, $\rho_E(\phi|\theta) \approx \delta(\phi-\phi(\theta))$,  one obtains from \eqref{eq:distr_from_kernel}, $P(x|\theta) \approx K(x|\phi(\theta))$, and both frameworks reduce to comparing the location of the corresponding background in moduli space.

\section{Details on the WKB expansion \& Wheeler-DeWitt equation}\label{sec:HJ_conteq_WdW}

In this appendix, we show how, when a time is fixed in a given WDW problem, a continuity equation naturally emerges.

Using coordinates $q^i$ (e.g. $q^i = \left[h_{kl},\phi^a\right]$) on configuration space $Q_q$ with metric $G_{ij}(q)$ and potential $V(q)$, the WDW equation can simply be recast as
\begin{equation}\label{eq:WdWeq}
\mathcal H \Psi(q)=0\,, \quad   \mathcal H = -\frac{1}{2} \nabla^2 + V(q)\,.
\end{equation}
Here $\nabla^2$ denotes the Laplace-Beltrami operator associated with the full minisuperspace metric, given  by
\begin{equation}\label{laplace-beltrami}
\nabla^2 \Psi = \frac{1}{\sqrt{|G|}}\partial_i \left( \sqrt{|G|} G^{ij}\partial_j \Psi \right)\,.
\end{equation}
Applying the WKB expansion (reviewed in Section \ref{sec:associatedWKB}) to the Wheeler-DeWitt equation yields, at leading order, a Hamilton-Jacobi equation on configuration space together with a conserved current for the associated semiclassical density.

A crucial difference from the systems discussed previously is that the Wheeler-DeWitt equation does not involve an external time parameter, reflecting the reparametrisation invariance of the gravitational theory. The resulting Hamilton-Jacobi system is therefore intrinsically timeless. As a result, the construction of a Wasserstein distance based on time-like evolution does not directly apply in a full-quantum gravitational setting. However, in the WKB regime this obstruction can be partially circumvented by introducing an intrinsic parameter along classical trajectories in configuration space.

Concretely, starting from the WKB approximation and its associated Hamilton-Jacobi description of the gravitational sector~\cite{DeWitt:1967ub,Banks:1984cw,Vilenkin:1988yd} we can write the minisuperspace Hamiltonian as\footnote{This corresponds to a minisuperspace action of the form $S[N,q]=\int \mathrm{d}t \left(\frac{1}{2N}G_{ij}\dot q^i \dot q^j - N V(q) \right)$.}
\begin{equation}\label{eq:Ham_standard}
    \mathcal{H}(q,p)=\frac{1}{2}G^{ij}p_i p_j + V(q)\,,
\end{equation}
together with the Hamiltonian constraint $\mathcal{H}(q,p)=0$. The momentum is given by
\begin{equation}
    p_i = \frac{1}{N}G_{ij}\dot q^j\,,
\end{equation}
where the dot denotes differentiation with respect to the cosmological time $t$. 

The idea is to choose one of the minisuperspace coordinates as a reference clock $T$ for the evolution. In most cosmological applications, a convenient choice is to use the scale factor $T=U=\ln(a)$ or a scalar field $T=\phi$. This motivates a split of configuration space variables $q$ into 
\begin{equation}
    q^i = (T; q^\alpha)\,.
\end{equation}
The identification of $T$ as a time variable should be understood as a relational construction valid on a given semiclassical branch. It requires that $T$ be monotonic along classical trajectories, i.e. $\dot T\neq 0$, so that it can serve as a good local clock.

For convenience, we assume in the following that the split can be done in a way such that the diagonal elements $G_{Ti}=0$. In this case, the constraint reads
\begin{equation}\label{eq:constraint_time}
    \mathcal{H}(T,p_T; q^\alpha,p_\alpha)= \frac{1}{2}G^{TT}p_T p_T+ \frac{1}{2}G^{\alpha \beta}p_\alpha p_\beta+V(q)=0\,,
\end{equation}
with 
\begin{equation}\label{eq:Hamilton_equation}
    \begin{aligned}
        \dot T &= N \frac{\partial \mathcal{H}}{\partial p_T}\,, \qquad  & \dot q^\alpha &= N \frac{\partial \mathcal{H}}{\partial p_\alpha}\,,\\
        \dot p_T &= -N \frac{\partial \mathcal{H}}{\partial T}\,, & \dot p_\alpha &= -N \frac{\partial \mathcal{H}}{\partial q^\alpha}\,.
    \end{aligned}
\end{equation}

The constraint \eqref{eq:constraint_time} can be solved for $p_T$ in order to define a new Hamiltonian $h_T$ with respect to the effective time $T$
\begin{equation}\label{eq:H_ast}
    p_T + h_T(T,q^\alpha,p_\alpha)=0\,,
\end{equation}
where $h_T$ may depend on $T$ through $V$ or non-trivial $G^{\alpha \beta}$.  Since $T$ and $p_T$ are conjugated variables, $\{T,p_T\}=1$, $p_T$ generates translations in $T$. This then implies, using $\dot T = \{T, N\mathcal{H}\}=N$, that $h_T$ generates evolution in $T$ via
\begin{equation}
    \frac{\mathrm{d}}{\mathrm{d}T}f = \frac{\partial}{\partial T}f + \{f,h_T\}\,.
\end{equation}
Hence, $T$ can be interpreted as an emergent time parameter with respect to the reduced Hamiltonian $h_T$, which governs the flow on the reduced phase space. The associated Hamilton equations read
\begin{equation}
    \frac{\mathrm{d}q^\alpha}{\mathrm{d}T}  = \frac{\partial h_T}{\partial p_\alpha}\,, \qquad 
    \frac{\mathrm{d}p_\alpha}{\mathrm{d}T}  = -\frac{\partial h_T}{\partial q^\alpha}\,.
\end{equation}
Now defining a Hamiltonian principal function $S_0(T,q^\alpha)$ such that 
\begin{equation}
    p_\beta = \frac{\partial}{\partial q^\beta} S_0(T,q^\alpha)\,,
\end{equation}
we obtain the reduced Hamilton-Jacobi equation
\begin{equation}
    \partial_T S_0 + h_T(T,q^\alpha, \nabla_\alpha S_0)=0\,,
\end{equation}
which can be interpreted as the leading-order WKB limit of an effective Schr\"odinger equation
\begin{equation}\label{eq:eff_Schroedinger_equation}
    i \hbar \frac{\partial}{\partial T}\Psi = h_T \Psi\,,\qquad \text{with}\quad \Psi_0= A e^{\frac{i}{\hbar} S_0}\,.
\end{equation}
To obtain the associated continuity equation, we start from the conserved Wheeler-DeWitt current. The latter is of Klein-Gordon type (see also the remark on p.\,\pageref{remark:KG_WdW})
\begin{equation}
J^i=\frac{i}{2}\bigl(\Psi^\ast \nabla^i \Psi-\Psi \nabla^i \Psi^\ast\bigr)\,, \qquad \nabla_i J^i=0\,.
\end{equation}
In the WKB regime $\Psi = A_\circ e^{\frac{i}{\hbar}S_0}$, and at leading order in $\hbar$, this reduces to
\begin{equation}
J^i = A_\circ^2\, G^{ij}\nabla_j S_0\,,
\end{equation}
where $A_\circ$ denotes the WKB amplitude in the Wheeler-DeWitt description (to distinguish it from the amplitude $A$ appearing in the reduced Schr\"odinger equation \eqref{eq:eff_Schroedinger_equation}; the phase $S_0$ given by the principal function is common to both by definition.
Assuming that $T$ defines a good clock ($J^T\neq 0$), we define a density and velocity field by
\begin{equation}
    \rho := J^T\,, \qquad 
    v^\alpha := \frac{J^\alpha}{J^T} = \frac{G^{\alpha\beta}\partial_\beta S_0}{G^{TT}\partial_T S_0}\,.
\end{equation}
The conservation equation $\nabla_i J^i=0$ then induces the continuity equation
\begin{equation}
    \partial_T \rho + \nabla_\alpha(\rho\,v^\alpha)=0\,.
\end{equation}
Note that using $\dot T = G^{TT}\partial_T S_0$, the density can be written as $\rho = A_\circ^2\, \dot T$. In the reduced Schr\"odinger description, one instead has $\rho = |A|^2$. Comparing the two expressions yields the relation between amplitudes
\begin{equation}\label{norm_wf}
    A_\circ = A/\sqrt{\dot T}\,.
\end{equation}
Similarly, the velocity fields satisfy
\begin{equation}
    v^\alpha = \frac{1}{\dot T} G^{\alpha\beta}\partial_\beta S_0\,, \qquad 
    v^\alpha_\circ = G^{\alpha\beta}\partial_\beta S_0 = \dot T\, v^\alpha\,.
\end{equation}

We still need to determine the role of the reduced Hamiltonian $h_T$. Solving the constraint for $p_T=-h_T$ and differentiating with respect to $p_\alpha$ at fixed $(T,q^\alpha)$ gives
\begin{equation}
    \frac{\partial h_T}{\partial p_\alpha} =  \frac{\partial_{p_\alpha}\mathcal H}{\partial_{p_T}\mathcal H}
    = \frac{G^{\alpha\beta}p_\beta}{G^{TT}p_T}\,.
\end{equation}
Evaluating on $p_i=\partial_i S_0$, we obtain
\begin{equation}
    v^\alpha =\frac{\partial h_T}{\partial p_\alpha}\Big|_{p=\nabla S_0}\,.
\end{equation}
Hence, the continuity equation can equivalently be written as
\begin{equation}\label{eq:cont_eq_Hstar}
    \partial_T \rho  +\nabla_\alpha\!\left(\rho\,\frac{\partial h_T}{\partial p_\alpha}\right)=0\,,
\end{equation}
demonstrating that, although $h_T$ is not the fundamental Hamiltonian of the timeless Wheeler-DeWitt system, it governs the induced semiclassical flow once the internal clock $T$ has been chosen.

\section{Examples in gauged supergravity}\label{app:gauged_fake_sugra} 

In this appendix, we extend the black-hole examples of Section~\ref{sec:fakeSUGRA_examples} to FI-gauged supergravity. The aim is to illustrate how the Gaussian WKB construction is modified when the radial fake-supergravity system contains a genuine scalar potential in addition to the black-hole potential. We keep the one-modulus cubic model and the D0-D4 charge sector used in the main text. The gauging changes the fake superpotential and the attractor equations, but the local semiclassical density is obtained in the same way expanding the corresponding characteristic function around a stable critical point.

\subsubsection{Cubic prepotential attractors for gauged D0-D4 systems}\label{sec:D0D4examplesgauged}

We now turn to the gauged theory, introducing a genuine scalar potential in addition to the effective black hole potential.  We keep the same one-modulus cubic model and the axion-free slice $t=i\phi$, $\phi>0$, and consider Fayet-Iliopoulos (FI) gaugings of a $U(1)$ in the $SU(2)_R\times U(1)_R$ R-symmetry of $N=2$ supergravity.  In the vector-multiplet sector, this induces a scalar potential
\begin{equation}
V_{\rm gauge} =g^2\left(G^{\phi\bar\phi}\,D_\phi \mathcal{L}\,\overline{D_\phi \mathcal{L}} -3\,|\mathcal{L}|^2\right)\,,
\end{equation}
where
$\mathcal{L}=\langle\mathcal{V},\xi\rangle$ is constructed from the covariantly holomorphic symplectic section $\mathcal{V}$ and the constant FI vector $\xi^\Lambda$ representing purely electric gauging.  This is the standard FI scalar potential in gauged $N=2$ vector multiplet supergravity~\cite{DallAgata:2010ejj,Andrianopoli:2009je}. 

On the axion-free slice, choosing the special coordinate gauge $X^0=1$ and $X^1=i\phi$, one has
\begin{equation}
F_0=-t^3=i\phi^3,\qquad F_1=-3t^2=-3\phi^2\,,
\end{equation}
so that
\begin{equation}
\mathcal V=e^{K/2} \begin{pmatrix}
1\\ i\phi\\ i\phi^3\\ -3\phi^2
\end{pmatrix}\,,\qquad \mathcal L(\phi)=\langle\mathcal V,\xi\rangle =e^{K/2}\big(\xi_0+i\,\xi_1\,\phi\big)\,,
\end{equation}
where $K$ is the K\"ahler potential of the cubic model. The full one-dimensional effective potential governing static radial flows is then~\cite{DallAgata:2010ejj}
\begin{equation}
V_{\rm eff}(\phi)=V_{\rm BH}(\phi)+V_{\rm gauge}(\phi),
\end{equation}
with $V_\mathrm{BH}$, the usual black-hole effective potential in special geometry and critical points are determined by $\partial_\phi V_\mathrm{eff}=0$. 

To organise the first-order flows in the gauged theory, adopt a static, spherically symmetric metric ansatz with independent warp functions
\begin{equation}
\mathrm ds^2 = - e^{2U(r)} \mathrm dt^2
   + e^{-2U(r)} \left( \mathrm dr^2 + e^{2\psi(r)} \mathrm d\Omega^2 \right) \, .
\end{equation}
so that the radial Hamiltonian for the coupled system of $U(r)$, $\psi(r)$, and $\phi(r)$ can be written in Hamilton-Jacobi form.  In this setup, a convenient choice of a fake superpotential (Hamilton's characteristic function) driving the flow is
\begin{equation}\label{eq:fakeW_gauged}
\mathcal W(\phi,U,\psi) =e^{U}\,\Big|\,Z(\phi)-i\,e^{2(\psi-U)}\,\mathcal L(\phi)\,\Big|\,,
\end{equation}
where $Z(\phi)=\langle\mathcal V,P\rangle$ is the central charge built from the black hole charge vector $P=(p^\Lambda,q_\Lambda)$~\cite{DallAgata:2010ejj,Andrianopoli:2009je}.  On the axion-free slice, $\mathcal W$ reduces to a real positive function of $\phi$, $U$, and $\psi$, and in the setup of~\cite{DallAgata:2010ejj}, the system admits a first-order form
\begin{equation}\label{eq:DG_fakeW_general}
\dot U=-G^{UU}\,\frac{\partial\mathcal W}{\partial U}\,,\qquad \dot\psi=-G^{\psi\psi}\,\frac{\partial\mathcal W}{\partial\psi}\,,\qquad \dot\phi=-G^{\phi\phi}\,\frac{\partial\mathcal W}{\partial\phi}\,,
\end{equation}
ensuring that solutions of these first-order equations solve the full second-order Einstein- scalar system subject to $V_\mathrm{eff}$.  The function $\mathcal W$ here plays the role of Hamilton's principal function in the radial Hamilton-Jacobi formalism for gauged supergravity flows.

\paragraph{Gauged D0-D4 system: purely electric FI parameter ($\xi_0=0$ case).} We first set $\xi_0=0$ and keep $\xi_1\neq0$.  In this case, $\mathcal W$ admits a globally real representation and thus allows a consistent Euclidean WKB construction. We work on the axion-free slice $t=i\phi$ with $\phi>0$, and consider a charge vector
\begin{equation}
Q=(0,p^1;q_0,0)\,, \qquad q_0>0\,,\; p^1>0\,.
\end{equation}
Following~\cite{DallAgata:2010ejj}, on a fixed $(U,\psi)$ slice, one may use a real fake superpotential of the form
\begin{equation}
W(\phi)=\frac{e^{U_0}}{2\sqrt{2}}\left(q_0\,\phi^{-3/2}+3p^1\,\phi^{1/2}+\alpha\,\xi_1\,\phi^{-1/2}\right)\,,
\end{equation}
where $\alpha=e^{2(\psi_0-U_0)}$ is a positive constant fixed by the $(U,\psi)$ slice. The corresponding effective potential satisfies the first-order relation \eqref{eq:VBH}.

We first determine the critical point and associated stability. Introducing the logarithmic variable $u=\log\phi$, the scalar metric becomes $\mathrm ds^2=\frac{3}{4}\,\mathrm du^2$, and the superpotential reads
\begin{equation}
W(u)=\frac{e^{U_0}}{2\sqrt{2}}\left(q_0 e^{-3u/2}+3p^1 e^{u/2}+\alpha\xi_1 e^{-u/2}\right)\,.
\end{equation}
The attractor equation $\partial_u W=0$ reduces to $3p^1\,\phi^2-\alpha\xi_1\,\phi-3q_0=0$ with the physical solution
\begin{equation}
\phi_\ast=\frac{\alpha\xi_1+\sqrt{\alpha^2\xi_1^2+36p^1q_0}}{6p^1}\,.
\end{equation}
For $q_0>0$, $p^1>0$, and $\xi_1\ge0$, this critical point is a strict minimum. The second derivative at the attractor is
\begin{equation}
\partial_u^2 W\big|_\ast =\frac{e^{U_0}}{2\sqrt{2}} \left( 3q_0\,\phi_\ast^{-3/2} +\frac12\,\alpha\xi_1\,\phi_\ast^{-1/2} \right)>0\,,
\end{equation}
confirming local stability.

Moving on to the Gaussian approximation \eqref{eq:approx_gaussian}, the Euclidean WKB density at fixed $U=U_0$ is $\rho_E(u)\propto \exp\!\left(-\frac{2}{\hbar}W(u)\right)$. Expanding $W$ to quadratic order around $u_\ast$, the resulting Gaussian has a variance determined by the metric-raised Hessian
\begin{equation}\label{eq:HessiangD0D4}
\kappa=g^{uu}\,\partial_u^2 W\big|_\ast=\frac{4}{3}\,\partial_u^2 W\big|_\ast\,.
\end{equation}
The variance in the canonically normalised coordinate is therefore
\begin{equation}
\sigma^2=\frac{\hbar}{2\kappa}=\frac{3\hbar}{8}\,\frac{1}{\partial_u^2 W|_\ast}=\frac{3\sqrt{2}\,\hbar}{4e^{U_0}\left( 3q_0\,\phi_\ast^{-3/2} +\tfrac12\alpha\xi_1\,\phi_\ast^{-1/2} \right)}\,.
\end{equation}

\paragraph{Scaling regimes.} In the weak-gauging limit $\alpha\xi_1\ll \sqrt{p^1q_0}$, one finds
\begin{equation}
\phi_\ast=\sqrt{\frac{q_0}{p^1}}+O(\alpha\xi_1)\,,\qquad \sigma_u^2\sim \frac{\hbar}{e^{U_0}}\frac{1}{q_0^{1/4}(p^1)^{3/4}}\,,
\end{equation}
Therefore, the width shrinks with increasing charge, as in the ungauged case. More generally, for large $q_0$ at fixed $\xi_1$, the standard deviation scales as
\begin{equation}
\sigma_u\sim q_0^{-1/8}\,,
\end{equation}
while the attractor moves to an infinite distance in moduli space. For two such backgrounds, the 2-Wasserstein distance between the associated semiclassical densities admits, in the Gaussian approximation with potential given \eqref{eq:Wasserstein_Gaussian_Jacobi}, the decomposition
\begin{equation}
W_{2,J}^2(\rho_1,\rho_2) = \left(\int\!\sqrt{g_J(s)}\,\mathrm ds\right)^{\!2} + \left( \sqrt{g_J}\,\sigma_{y,1} - \sqrt{g_J}\,\sigma_{y,2} \right)^{\!2}\,,
\end{equation}
where the integration bounds correspond to the moduli space centres of masses $\phi_1$ and  $\phi_2$ of the distributions $\rho_1$ and $\rho_2$.
In this example of a gauged theory, the appropriate transport cost is determined by the Jacobi-Maupertuis metric $g_J$ associated with the full effective potential $V_{\rm eff}=V_{\rm BH}+V_{\rm gauge}$. For sharply localised semiclassical densities near a stable attractor, this locally reduces to the Jacobi-Maupertuis distance.

\paragraph{FI parameter: $\xi_0\neq 0$ case.} We keep the one-modulus cubic model with axion fixed to zero, $t=i\phi$, $\phi>0$, and the same K\"ahler data as above.  We consider the D0-D4 charge sector $Q=(0,p^1;q_0,0)$ and purely electric FI gauging $\vec{\xi}=(\xi_0,\xi_1;0,0)^T$ with $\xi_0\neq 0\,,\ \xi_1\neq 0$. On the axion-free slice, the central charge is real
\begin{equation}
Z(\phi)=e^{K/2}\Big(q_0 - p^1 F_1\Big)=e^{K/2}\Big(q_0+3p^1\phi^2\Big)\,,
\end{equation}
while the gauging superpotential is
\begin{equation}
\mathcal L(\phi)=\langle \mathcal V,\xi\rangle=e^{K/2}\big(\xi_0+\xi_1 t\big)=e^{K/2}\big(\xi_0+i\xi_1\phi\big)\,.
\end{equation}
Following~\cite{DallAgata:2010ejj}, the (real) fake superpotential governing the first-order flow can be taken as in \eqref{eq:DG_fakeW_general}.
On the axion-free slice, this becomes explicit.
\begin{equation}
\mathcal W(\phi;U,\psi)= e^{U}e^{K/2}\sqrt{\Big(q_0+3p^1\phi^2 + e^{2(\psi-U)}\xi_1\phi\Big)^2 +\Big(e^{2(\psi-U)}\xi_0\Big)^2}\,,
\end{equation}
which is manifestly real and strictly positive for $\phi>0$. Fix a $(U,\psi)$ slice, $U=U_0$, $\psi=\psi_0$, and introduce the positive constant $\alpha=e^{2(\psi_0-U_0)}$. With $K=-\log(4\phi^3)$, one has $e^{K/2}=(4\phi^3)^{-1/2}= (2\phi^{3/2})^{-1}$.  It is convenient to define
\begin{equation}
A(\phi)=q_0+3p^1\phi^2+\alpha\xi_1\phi\,,\qquad B=\alpha\xi_0\,,
\end{equation}
so that the real fake superpotential reads
\begin{equation}\label{eq:W_xi0neq0}
\mathcal W(\phi)=\frac{e^{U_0}}{2\phi^{3/2}}\sqrt{A(\phi)^2+B^2}\,.
\end{equation}
We first determine the critical points, i.e., solutions to the attractor equation are $\partial_\phi \mathcal W=0$.  Using
\begin{equation}
\partial_\phi \log \mathcal W = -\frac{3}{2}\frac{1}{\phi}
+\frac{A(\phi)A'(\phi)}{A(\phi)^2+B^2}\,,\qquad A'(\phi)=6p^1\phi+\alpha\xi_1\,,
\end{equation}
one finds the equivalent algebraic condition
\begin{equation}\label{eq:attractor_eq_general}
2\phi\,A(\phi)A'(\phi)=3\big(A(\phi)^2+B^2\big)\,.
\end{equation}
Expanding this gives a quartic equation for $\phi_\ast$,
\begin{equation}\label{eq:quartic_phi}
9(p^1)^2\phi^4-\Big(6p^1q_0+(\alpha\xi_1)^2\Big)\phi^2-4q_0(\alpha\xi_1)\phi-3\Big(q_0^2+(\alpha\xi_0)^2\Big)=0\,.
\end{equation}
For $\xi_0\to 0$ (i.e. $B\to 0$) this factorises as
\begin{equation}
\Big(3p^1\phi^2-\alpha\xi_1\phi-3q_0\Big)\Big(3p^1\phi^2+\alpha\xi_1\phi+q_0\Big)=0\,,
\end{equation}
and the physical root reduces to the $\xi_0=0$ attractor discussed above.  For $\xi_0\neq 0$, the quartic \eqref{eq:quartic_phi} can admit one or more positive roots

As before, the normalisability of the Euclidean WKB density on a fixed $(U,\psi)$ slice requires that the critical point of $\mathcal W$ be a local minimum in the physical domain $\phi>0$ (equivalently $\partial_u^2\mathcal W|_{u_\ast}>0$ with $u=\log\phi$), so that the quadratic expansion of $\mathcal W$ defines a confining Gaussian, as discussed around \eqref{eq:HessiangD0D4} and \eqref{eq:gaussapprox_var_clean}. In the present $\xi_0\neq 0$ case, $\mathcal W(\phi)$ is strictly positive for $\phi>0$ and diverges as $\phi\to0^+$ and $\phi\to+\infty$, so any critical point in $\phi>0$ is necessarily a local minimum. We will not attempt to extract analytic large-charge scaling in this branch; the purpose of this subcase is simply to illustrate that the normalisability and Gaussian construction persist when $\xi_0\neq 0$. One can verify numerically that for a range of parameters, \eqref{eq:quartic_phi} admits a positive root $\phi_\ast$, and the construction then proceeds exactly as in the $\xi_0=0$ case: expanding $\mathcal W(u)$ about $u_\ast$ gives a normalisable Gaussian with variance fixed by the metric-raised Hessian \eqref{eq:HessiangD0D4}.

\bibliography{references}

@article{otto2005eulerian,
  title={Eulerian calculus for the contraction in the Wasserstein distance},
  author={Otto, Felix and Westdickenberg, Michael},
  journal={SIAM journal on mathematical analysis},
  volume={37},
  number={4},
  pages={1227--1255},
  year={2005},
  publisher={SIAM}
}

@article{Grimm:2025lip,
    author = "Grimm, Thomas W. and Prieto, David and van Vliet, Mick",
    title = "{Tame embeddings, volume growth, and complexity of moduli spaces}",
    eprint = "2503.15601",
    archivePrefix = "arXiv",
    primaryClass = "hep-th",
    doi = "10.1103/d51c-j1s9",
    journal = "Phys. Rev. D",
    volume = "112",
    number = "10",
    pages = "106015",
    year = "2025"
}

@article{Trigiante:2012eb,
    author = "Trigiante, Mario and Van Riet, Thomas and Vercnocke, Bert",
    title = "{Fake supersymmetry versus Hamilton-Jacobi}",
    eprint = "1203.3194",
    archivePrefix = "arXiv",
    primaryClass = "hep-th",
    reportNumber = "IPHT-T12-019",
    doi = "10.1007/JHEP05(2012)078",
    journal = "JHEP",
    volume = "05",
    pages = "078",
    year = "2012"
}

@article{Luscher:1985iu,
    author = "Luscher, M.",
    editor = "Haag, R.",
    title = "{Schrödinger representation in Quantum Field Theory}",
    doi = "10.1016/0550-3213(85)90210-X",
    journal = "Nucl. Phys. B",
    volume = "254",
    pages = "52--57",
    year = "1985"
}

@article{Symanzik:1981wd,
    author = "Symanzik, K.",
    title = "{Schrodinger Representation and Casimir Effect in Renormalizable Quantum Field Theory}",
    reportNumber = "DESY-81-004",
    doi = "10.1016/0550-3213(81)90482-X",
    journal = "Nucl. Phys. B",
    volume = "190",
    pages = "1--44",
    year = "1981"
}

@article{Ferrara:2008hwa,
    author = "Ferrara, Sergio and Hayakawa, Kuniko and Marrani, Alessio",
    editor = "Zichichi, Antonino",
    title = "{Lectures on Attractors and Black Holes}",
    eprint = "0805.2498",
    archivePrefix = "arXiv",
    primaryClass = "hep-th",
    reportNumber = "CERN-PH-TH-2008-002, UCLA-08-TEP-11",
    doi = "10.1002/prop.200810569",
    journal = "Fortsch. Phys.",
    volume = "56",
    number = "10",
    pages = "993--1046",
    year = "2008"
}

@article{Andrianopoli:2007gt,
    author = "Andrianopoli, Laura and D'Auria, Riccardo and Orazi, Emanuele and Trigiante, Mario",
    title = "{First order description of black holes in moduli space}",
    eprint = "0706.0712",
    archivePrefix = "arXiv",
    primaryClass = "hep-th",
    doi = "10.1088/1126-6708/2007/11/032",
    journal = "JHEP",
    volume = "11",
    pages = "032",
    year = "2007"
}

@article{Andrianopoli:2009je,
    author = "Andrianopoli, L. and D'Auria, R. and Orazi, E. and Trigiante, M.",
    title = "{First Order Description of D=4 static Black Holes and the Hamilton-Jacobi equation}",
    eprint = "0905.3938",
    archivePrefix = "arXiv",
    primaryClass = "hep-th",
    doi = "10.1016/j.nuclphysb.2010.02.020",
    journal = "Nucl. Phys. B",
    volume = "833",
    pages = "1--16",
    year = "2010"
}

@article{Ceresole:2007wx,
    author = "Ceresole, Anna and Dall'Agata, Gianguido",
    title = "{Flow Equations for Non-BPS Extremal Black Holes}",
    eprint = "hep-th/0702088",
    archivePrefix = "arXiv",
    doi = "10.1088/1126-6708/2007/03/110",
    journal = "JHEP",
    volume = "03",
    pages = "110",
    year = "2007"
}

@inproceedings{Halliwell:1989myn,
    author = "Halliwell, Jonathan J.",
    title = "{Introductory lectures on quantum cosmology}",
    booktitle = "{7th Jerusalem Winter School for Theoretical Physics: Quantum Cosmology and Baby Universes}",
    eprint = "0909.2566",
    archivePrefix = "arXiv",
    primaryClass = "gr-qc",
    reportNumber = "MIT-CTP-1845",
    year = "1989"
}

@article{Ooguri:2005vr,
    author = "Ooguri, Hirosi and Vafa, Cumrun and Verlinde, Erik P.",
    title = "{Hartle-Hawking wave-function for flux compactifications}",
    eprint = "hep-th/0502211",
    archivePrefix = "arXiv",
    reportNumber = "CALT-68-2543, HUTP-05-A005, ITFA-2005-05",
    doi = "10.1007/s11005-005-0022-x",
    journal = "Lett. Math. Phys.",
    volume = "74",
    pages = "311--342",
    year = "2005"
}

@book{villani2008optimal,
  title={Optimal transport: old and new},
  author={Villani, C{\'e}dric and others},
  volume={338},
  year={2008},
  publisher={Springer}
}

@article{Mohseni:2024njl,
    author = "Mohseni, Amineh and Montero, Miguel and Vafa, Cumrun and Valenzuela, Irene",
    title = "{On measuring distances in the quantum gravity landscape}",
    eprint = "2407.02705",
    archivePrefix = "arXiv",
    primaryClass = "hep-th",
    reportNumber = "IFT-24-097, CERN-TH-2024-101",
    doi = "10.1007/JHEP12(2024)168",
    journal = "JHEP",
    volume = "12",
    pages = "168",
    year = "2024"
}

@article{otto2001geometry,
  title={THE GEOMETRY OF DISSIPATIVE EVOLUTION EQUATIONS: THE POROUS MEDIUM EQUATION.},
  author={Otto, Felix},
  journal={Communications in Partial Differential Equations},
  volume={26},
  year={2001}
}

@article{khan2022optimal,
  title={When optimal transport meets information geometry},
  author={Khan, Gabriel and Zhang, Jun},
  journal={Information Geometry},
  volume={5},
  number={1},
  pages={47--78},
  year={2022},
  publisher={Springer}
}

@article{benamou2000computational,
  title={A computational fluid mechanics solution to the Monge-Kantorovich mass transfer problem},
  author={Benamou, Jean-David and Brenier, Yann},
  journal={Numerische Mathematik},
  volume={84},
  number={3},
  pages={375--393},
  year={2000},
  publisher={Springer-Verlag Berlin/Heidelberg}
}

@article{bernard2006monge,
  title={The Monge problem for supercritical Man{\'e} potentials on compact manifolds},
  author={Bernard, Patrick and Buffoni, Boris},
  journal={Advances in Mathematics},
  volume={207},
  number={2},
  pages={691--706},
  year={2006},
  publisher={Elsevier}
}

@article{fathi2010optimal,
  title={Optimal transportation on non-compact manifolds},
  author={Fathi, Albert and Figalli, Alessio},
  journal={Israel Journal of Mathematics},
  volume={175},
  number={1},
  pages={1--59},
  year={2010},
  publisher={Springer}
}

@article{elamvazhuthi2024benamou,
  title={Benamou-Brenier formulation of optimal transport for nonlinear control systems on Rd},
  author={Elamvazhuthi, Karthik},
  journal={arXiv preprint arXiv:2407.16088},
  year={2024}
}

@article{agrachev2009optimal,
  title={Optimal transportation under nonholonomic constraints},
  author={Agrachev, Andrei and Lee, Paul},
  journal={Transactions of the American Mathematical Society},
  volume={361},
  number={11},
  pages={6019--6047},
  year={2009}
}

@article{Cribiori:2023swd,
    author = {Cribiori, Niccol{\`o} and Gnecchi, Alessandra and L{\"u}st, Dieter and Scalisi, Marco},
    title = "{On the correspondence between black holes, domain walls and fluxes}",
    eprint = "2302.03054",
    archivePrefix = "arXiv",
    primaryClass = "hep-th",
    reportNumber = "LMU-ASC 09/23, MPP-2023-26",
    doi = "10.1007/JHEP05(2023)033",
    journal = "JHEP",
    volume = "05",
    pages = "033",
    year = "2023"
}

@article{Debusschere:2024rmi,
    author = "Debusschere, C{\'e}dric and Tonioni, Flavio and Van Riet, Thomas",
    title = "{A distance conjecture beyond moduli?}",
    eprint = "2407.03715",
    archivePrefix = "arXiv",
    primaryClass = "hep-th",
    doi = "10.1007/JHEP03(2025)140",
    journal = "JHEP",
    volume = "03",
    pages = "140",
    year = "2025"
}

@article{Delgado:2022dkz,
    author = "Delgado, Matilda and Montero, Miguel and Vafa, Cumrun",
    title = "{Black holes as probes of moduli space geometry}",
    eprint = "2212.08676",
    archivePrefix = "arXiv",
    primaryClass = "hep-th",
    reportNumber = "IFT-UAM/CSIC-22-151",
    doi = "10.1007/JHEP04(2023)045",
    journal = "JHEP",
    volume = "04",
    pages = "045",
    year = "2023"
}

@article{Kallosh:2006bt,
    author = "Kallosh, Renata and Sivanandam, Navin and Soroush, Masoud",
    title = "{The Non-BPS black hole attractor equation}",
    eprint = "hep-th/0602005",
    archivePrefix = "arXiv",
    reportNumber = "SLAC-PUB-11660, SU-ITP-06-03",
    doi = "10.1088/1126-6708/2006/03/060",
    journal = "JHEP",
    volume = "03",
    pages = "060",
    year = "2006"
}

@article{Gukov:2005bg,
    author = "Gukov, Sergei and Saraikin, Kirill and Vafa, Cumrun",
    title = "{The Entropic principle and asymptotic freedom}",
    eprint = "hep-th/0509109",
    archivePrefix = "arXiv",
    reportNumber = "HUTP-05-A041, ITEP-TH-55-05",
    doi = "10.1103/PhysRevD.73.066010",
    journal = "Phys. Rev. D",
    volume = "73",
    pages = "066010",
    year = "2006"
}

@article{Grimm:2019ixq,
	Archiveprefix = {arXiv},
	Author = {Grimm, Thomas W. and Li, Chongchuo and Valenzuela, Irene},
	Doi = {10.1007/JHEP06(2020)009},
	Eprint = {1910.09549},
	Journal = {JHEP},
	Note = {[Erratum: JHEP 01, 007 (2021)]},
	Pages = {009},
	Primaryclass = {hep-th},
	Title = {{Asymptotic Flux Compactifications and the Swampland}},
	Volume = {06},
	Year = {2020}}

@article{Lee:2019wij,
    author = "Lee, Seung-Joo and Lerche, Wolfgang and Weigand, Timo",
    title = "{Emergent strings from infinite distance limits}",
    eprint = "1910.01135",
    archivePrefix = "arXiv",
    primaryClass = "hep-th",
    reportNumber = "CERN-TH-2019-159",
    doi = "10.1007/JHEP02(2022)190",
    journal = "JHEP",
    volume = "02",
    pages = "190",
    year = "2022"
}

@article{Stout:2020uaf,
	Archiveprefix = {arXiv},
	Author = {Stout, John},
	Eprint = {2012.11605},
	Month = {12},
	Primaryclass = {hep-th},
	Title = {{Instanton Expansions and Phase Transitions}},
	Year = {2020}}

@article{Stout:2021ubb,
	Archiveprefix = {arXiv},
	Author = {Stout, John},
	Eprint = {2106.11313},
	Month = {6},
	Primaryclass = {hep-th},
	Title = {{Infinite Distance Limits and Information Theory}},
	Year = {2021}}

@article{Vafa:2005ui,
	Archiveprefix = {arXiv},
	Author = {Vafa, Cumrun},
	Eprint = {hep-th/0509212},
	Primaryclass = {hep-th},
	Reportnumber = {HUTP-05-A043},
	Slaccitation = {%%CITATION = HEP-TH/0509212;%%},
	Title = {{The String landscape and the swampland}},
	Year = {2005}}

@article{Ooguri:2006in,
	Archiveprefix = {arXiv},
	Author = {Ooguri, Hirosi and Vafa, Cumrun},
	Doi = {10.1016/j.nuclphysb.2006.10.033},
	Eprint = {hep-th/0605264},
	Journal = {Nucl.Phys.},
	Pages = {21-33},
	Primaryclass = {hep-th},
	Reportnumber = {CALT-68-2600, HUTP-06-A017},
	Slaccitation = {%%CITATION = HEP-TH/0605264;%%},
	Title = {{On the Geometry of the String Landscape and the Swampland}},
	Volume = {B766},
	Year = {2007}}

@article{Vilenkin:1988yd,
    author = "Vilenkin, Alexander",
    title = "{The Interpretation of the Wave Function of the Universe}",
    reportNumber = "TUTP-88-3",
    doi = "10.1103/PhysRevD.39.1116",
    journal = "Phys. Rev. D",
    volume = "39",
    pages = "1116",
    year = "1989"
}

@article{Banks:1984cw,
    author = "Banks, Tom",
    title = "{TCP, Quantum Gravity, the Cosmological Constant and All That...}",
    reportNumber = "SLAC-PUB-3376",
    doi = "10.1016/0550-3213(85)90020-3",
    journal = "Nucl. Phys. B",
    volume = "249",
    pages = "332--360",
    year = "1985"
}

@article{Valenzuela:2016yny,
	Archiveprefix = {arXiv},
	Author = {Valenzuela, Irene},
	Doi = {10.1007/JHEP06(2017)098},
	Eprint = {1611.00394},
	Journal = {JHEP},
	Pages = {098},
	Primaryclass = {hep-th},
	Reportnumber = {MPP-2016-319},
	Slaccitation = {%%CITATION = ARXIV:1611.00394;%%},
	Title = {{Backreaction Issues in Axion Monodromy and Minkowski 4-forms}},
	Volume = {06},
	Year = {2017}}

@article{Baume:2016psm,
	Archiveprefix = {arXiv},
	Author = {Baume, Florent and Palti, Eran},
	Doi = {10.1007/JHEP08(2016)043},
	Eprint = {1602.06517},
	Journal = {JHEP},
	Pages = {043},
	Primaryclass = {hep-th},
	Slaccitation = {%%CITATION = ARXIV:1602.06517;%%},
	Title = {{Backreacted Axion Field Ranges in String Theory}},
	Volume = {08},
	Year = {2016}}

@article{Blumenhagen:2017cxt,
	Archiveprefix = {arXiv},
	Author = {Blumenhagen, Ralph and Valenzuela, Irene and Wolf, Florian},
	Doi = {10.1007/JHEP07(2017)145},
	Eprint = {1703.05776},
	Journal = {JHEP},
	Pages = {145},
	Primaryclass = {hep-th},
	Reportnumber = {MPP-2017-34},
	Slaccitation = {%%CITATION = ARXIV:1703.05776;%%},
	Title = {{The Swampland Conjecture and F-term Axion Monodromy Inflation}},
	Volume = {07},
	Year = {2017}}

@article{Dvali:2007hz,
	Archiveprefix = {arXiv},
	Author = {Dvali, Gia},
	Doi = {10.1002/prop.201000009},
	Eprint = {0706.2050},
	Journal = {Fortsch. Phys.},
	Pages = {528-536},
	Primaryclass = {hep-th},
	Slaccitation = {%%CITATION = ARXIV:0706.2050;%%},
	Title = {{Black Holes and Large N Species Solution to the Hierarchy Problem}},
	Volume = {58},
	Year = {2010}}

@article{Dvali:2007wp,
	Archiveprefix = {arXiv},
	Author = {Dvali, Gia and Redi, Michele},
	Doi = {10.1103/PhysRevD.77.045027},
	Eprint = {0710.4344},
	Journal = {Phys. Rev.},
	Pages = {045027},
	Primaryclass = {hep-th},
	Slaccitation = {%%CITATION = ARXIV:0710.4344;%%},
	Title = {{Black Hole Bound on the Number of Species and Quantum Gravity at LHC}},
	Volume = {D77},
	Year = {2008}}

@article{Dvali:2009ks,
	Archiveprefix = {arXiv},
	Author = {Dvali, Gia and L\"ust, Dieter},
	Doi = {10.1002/prop.201000008},
	Eprint = {0912.3167},
	Journal = {Fortsch. Phys.},
	Pages = {505-527},
	Primaryclass = {hep-th},
	Reportnumber = {CERN-PH-TH-2009-243, MPP-2009-205, LMU-ASC-56-09},
	Slaccitation = {%%CITATION = ARXIV:0912.3167;%%},
	Title = {{Evaporation of Microscopic Black Holes in String Theory and the Bound on Species}},
	Volume = {58},
	Year = {2010}}

@article{Dvali:2010vm,
	Archiveprefix = {arXiv},
	Author = {Dvali, Gia and Gomez, Cesar},
	Eprint = {1004.3744},
	Primaryclass = {hep-th},
	Reportnumber = {CERN-PH-TH-2010-069, IFT-UAM-CSIC-10-25},
	Slaccitation = {%%CITATION = ARXIV:1004.3744;%%},
	Title = {{Species and Strings}},
	Year = {2010}}

@article{Ceresole:1995ca,
    author = "Ceresole, Anna and D'Auria, R. and Ferrara, S.",
    editor = "Gava, E. and Narain, K. S. and Vafa, C.",
    title = "{The Symplectic structure of N=2 supergravity and its central extension}",
    eprint = "hep-th/9509160",
    archivePrefix = "arXiv",
    reportNumber = "POLFIS-TH-10-95, CERN-TH-95-244",
    doi = "10.1016/0920-5632(96)00008-4",
    journal = "Nucl. Phys. B Proc. Suppl.",
    volume = "46",
    pages = "67--74",
    year = "1996"
}

@article{Palti:2019pca,
    author = "Palti, Eran",
    title = "{The Swampland: Introduction and Review}",
    eprint = "1903.06239",
    archivePrefix = "arXiv",
    primaryClass = "hep-th",
    reportNumber = "MPP-2019-53",
    doi = "10.1002/prop.201900037",
    journal = "Fortsch. Phys.",
    volume = "67",
    number = "6",
    pages = "1900037",
    year = "2019"
}

@article{vandeHeisteeg:2022btw,
    author = "van de Heisteeg, Damian and Vafa, Cumrun and Wiesner, Max and Wu, David H.",
    title = "{Moduli-dependent species scale}",
    eprint = "2212.06841",
    archivePrefix = "arXiv",
    primaryClass = "hep-th",
    doi = "10.4310/bpam.2024.v1.n1.a1",
    journal = "Beijing J. Pure Appl. Math.",
    volume = "1",
    number = "1",
    pages = "1--41",
    year = "2024"
}

@article{vandeHeisteeg:2023dlw,
    author = "van de Heisteeg, Damian and Vafa, Cumrun and Wiesner, Max and Wu, David H.",
    title = "{Species scale in diverse dimensions}",
    eprint = "2310.07213",
    archivePrefix = "arXiv",
    primaryClass = "hep-th",
    doi = "10.1007/JHEP05(2024)112",
    journal = "JHEP",
    volume = "05",
    pages = "112",
    year = "2024"
}

@article{Castellano:2023aum,
    author = "Castellano, Alberto and Herr\'aez, Alvaro and Ib\'a\~nez, Luis E.",
    title = "{On the species scale, modular invariance and the gravitational EFT expansion}",
    eprint = "2310.07708",
    archivePrefix = "arXiv",
    primaryClass = "hep-th",
    doi = "10.1007/JHEP12(2024)019",
    journal = "JHEP",
    volume = "12",
    pages = "019",
    year = "2024"
}

@article{vanBeest:2021lhn,
    author = "van Beest, Marieke and Calder\'on-Infante, Jos\'e and Mirfendereski, Delaram and Valenzuela, Irene",
    title = "{Lectures on the Swampland Program in String Compactifications}",
    eprint = "2102.01111",
    archivePrefix = "arXiv",
    primaryClass = "hep-th",
    doi = "10.1016/j.physrep.2022.09.002",
    journal = "Phys. Rept.",
    volume = "989",
    pages = "1--50",
    year = "2022"
}

@article{Ferrara:1995ih,
    author = "Ferrara, Sergio and Kallosh, Renata and Strominger, Andrew",
    title = "{N=2 extremal black holes}",
    eprint = "hep-th/9508072",
    archivePrefix = "arXiv",
    reportNumber = "CERN-TH-95-211, SU-ITP-95-16",
    doi = "10.1103/PhysRevD.52.R5412",
    journal = "Phys. Rev. D",
    volume = "52",
    pages = "R5412--R5416",
    year = "1995"
}

@article{Cribiori:2022nke,
    author = {Cribiori, Niccol\`o and L\"ust, Dieter and Staudt, Georgina},
    title = "{Black hole entropy and moduli-dependent species scale}",
    eprint = "2212.10286",
    archivePrefix = "arXiv",
    primaryClass = "hep-th",
    reportNumber = "LMU-ASC 56/22, MPP-2022-289",
    doi = "10.1016/j.physletb.2023.138113",
    journal = "Phys. Lett. B",
    volume = "844",
    pages = "138113",
    year = "2023"
}

@article{Calderon-Infante:2025ldq,
    author = "Calder\'on-Infante, Jos\'e and Castellano, Alberto and Herr\'aez, Alvaro",
    title = "{The Double EFT Expansion in Quantum Gravity}",
    eprint = "2501.14880",
    archivePrefix = "arXiv",
    primaryClass = "hep-th",
    month = "1",
    year = "2025"
}

@article{Castellano:2025yur,
    author = {Castellano, Alberto and L\"ust, Dieter and Montella, Carmine and Zatti, Matteo},
    title = "{Quantum Calabi-Yau Black Holes and Non-Perturbative D0-brane Effects}",
    eprint = "2505.15920",
    archivePrefix = "arXiv",
    primaryClass = "hep-th",
    reportNumber = "EFI-25-06,MPP-2025-107,LMU-ASC 13/25",
    month = "5",
    year = "2025"
}

@article{Hartle:1972ya,
    author = "Hartle, J. B. and Hawking, S. W.",
    title = "{Solutions of the Einstein-Maxwell equations with many black holes}",
    doi = "10.1007/BF01645696",
    journal = "Commun. Math. Phys.",
    volume = "26",
    pages = "87--101",
    year = "1972"
}

@article{Delgado:2024skw,
    author = "Delgado, Matilda and van de Heisteeg, Damian and Raman, Sanjay and Torres, Ethan and Vafa, Cumrun and Xu, Kai",
    title = "{Finiteness and the emergence of dualities}",
    eprint = "2412.03640",
    archivePrefix = "arXiv",
    primaryClass = "hep-th",
    reportNumber = "MPP-2024-224, CERN-TH-2024-204",
    doi = "10.21468/SciPostPhys.19.2.047",
    journal = "SciPost Phys.",
    volume = "19",
    number = "2",
    pages = "047",
    year = "2025"
}

@article{Calderon-Infante:2025pls,
    author = {Calder\'on-Infante, Jos\'e and Delgado, Matilda and Li, Yixuan and L\"ust, Dieter and Uranga, Angel M.},
    title = "{Classical Black Hole Probes of UV Scales}",
    eprint = "2502.03514",
    archivePrefix = "arXiv",
    primaryClass = "hep-th",
    reportNumber = "CERN-TH-2025-033, MPP-2025-10, LMU-ASC 04/25, IFT-UAM/CSIC-25-009",
    month = "2",
    year = "2025"
}

@article{Cribiori:2023ffn,
    author = {Cribiori, Niccol\`o and L\"ust, Dieter and Montella, Carmine},
    title = "{Species entropy and thermodynamics}",
    eprint = "2305.10489",
    archivePrefix = "arXiv",
    primaryClass = "hep-th",
    reportNumber = "LMU-ASC 18/23, MPP-2023-97",
    doi = "10.1007/JHEP10(2023)059",
    journal = "JHEP",
    volume = "10",
    pages = "059",
    year = "2023"
}

@article{Castellano:2025ljk,
    author = "Castellano, Alberto and Zatti, Matteo",
    title = "{Black hole entropy, quantum corrections and EFT transitions}",
    eprint = "2502.02655",
    archivePrefix = "arXiv",
    primaryClass = "hep-th",
    doi = "10.1007/JHEP08(2025)112",
    journal = "JHEP",
    volume = "08",
    pages = "112",
    year = "2025"
}

@article{DeWitt:1967ub,
    author = "DeWitt, Bryce S.",
    editor = "Hsu, Jong-Ping and Fine, D.",
    title = "{Quantum Theory of Gravity. 2. The Manifestly Covariant Theory}",
    doi = "10.1103/PhysRev.162.1195",
    journal = "Phys. Rev.",
    volume = "162",
    pages = "1195--1239",
    year = "1967"
}

@article{Weinberg:1982id,
    author = "Weinberg, Steven",
    title = "{Does Gravitation Resolve the Ambiguity Among Supersymmetry Vacua?}",
    doi = "10.1103/PhysRevLett.48.1776",
    journal = "Phys. Rev. Lett.",
    volume = "48",
    pages = "1776--1779",
    year = "1982"
}

@article{Agrachev2018,
   title={Curvature: A Variational Approach},
   volume={256},
   ISSN={1947-6221},
   url={http://dx.doi.org/10.1090/memo/1225},
   DOI={10.1090/memo/1225},
   number={1225},
   journal={Memoirs of the American Mathematical
                Society},
   publisher={American Mathematical Society (AMS)},
   author={Agrachev, A. and Barilari, D. and Rizzi, L.},
   year={2018},
   month=nov, pages={0–0} }

@article{Kantorovich1960,
  author    = {Kantorovich, L. V.},
  title     = {Mathematical Methods of Organizing and Planning Production},
  journal   = {Management Science},
  volume    = {6},
  issue     = {4},
  pages     = {366--422},
  year      = {1960},
  doi       = {10.1287/mnsc.6.4.366},
  note      = {English translation of the 1939 Russian paper.},
}

@article{Kan:2021yoh,
    author = "Kan, Nahomi and Aoyama, Takuma and Hasegawa, Taiga and Shiraishi, Kiyoshi",
    title = "{Eisenhart-Duval lift for minisuperspace quantum cosmology}",
    eprint = "2105.09514",
    archivePrefix = "arXiv",
    primaryClass = "gr-qc",
    doi = "10.1103/PhysRevD.104.086001",
    journal = "Phys. Rev. D",
    volume = "104",
    number = "8",
    pages = "086001",
    year = "2021"
}

@article{Hawking:1990in,
    author = "Hawking, S. W. and Page, Don N.",
    title = "{The spectrum of wormholes}",
    reportNumber = "NSF-ITP-90-76",
    doi = "10.1103/PhysRevD.42.2655",
    journal = "Phys. Rev. D",
    volume = "42",
    pages = "2655--2663",
    year = "1990"
}

@article{Kiefer:1988tr,
    author = "Kiefer, Claus",
    title = "{Wave Packets in Minisuperspace}",
    reportNumber = "HD-TVP-88-1",
    doi = "10.1103/PhysRevD.38.1761",
    journal = "Phys. Rev. D",
    volume = "38",
    pages = "1761--1772",
    year = "1988"
}

@article{Andrianov:2018wdx,
    author = "Andrianov, Alexander A. and Lan, Chen and Novikov, Oleg O. and Wang, Yi-Fan",
    title = "{Integrable Minisuperspace Models with Liouville Field: Energy Density Self-Adjointness and Semiclassical Wave Packets}",
    eprint = "1802.06720",
    archivePrefix = "arXiv",
    primaryClass = "hep-th",
    doi = "10.1140/epjc/s10052-018-6255-5",
    journal = "Eur. Phys. J. C",
    volume = "78",
    number = "9",
    pages = "786",
    year = "2018"
}

@article{Gukov:2005iy,
    author = "Gukov, Sergei and Saraikin, Kirill and Vafa, Cumrun",
    title = "{A Stringy wave function for an S**3 cosmology}",
    eprint = "hep-th/0505204",
    archivePrefix = "arXiv",
    reportNumber = "HUTP-05-A024, ITEP-TH-37-05",
    doi = "10.1103/PhysRevD.73.066009",
    journal = "Phys. Rev. D",
    volume = "73",
    pages = "066009",
    year = "2006"
}

@article{brenier1987decomposition,
  title={D{\'e}composition polaire et r{\'e}arrangement monotone des champs de vecteurs},
  author={Brenier, Yann},
  journal={CR Acad. Sci. Paris S{\'e}r. I Math.},
  volume={305},
  pages={805--808},
  year={1987}
}

@article{knott1984optimal,
  title={On the optimal mapping of distributions},
  author={Knott, Martin and Smith, Cyril S},
  journal={Journal of Optimization Theory and Applications},
  volume={43},
  number={1},
  pages={39--49},
  year={1984},
  publisher={Springer}
}

@book{ambrosio2021lectures,
  title={Lectures on optimal transport},
  author={Ambrosio, Luigi and Bru{\'e}, Elia and Semola, Daniele and others},
  year={2021},
  publisher={Springer}
}

@article{Heckman:2013kza,
    author = "Heckman, Jonathan J.",
    title = "{Statistical Inference and String Theory}",
    eprint = "1305.3621",
    archivePrefix = "arXiv",
    primaryClass = "hep-th",
    doi = "10.1142/S0217751X15501602",
    journal = "Int. J. Mod. Phys. A",
    volume = "30",
    number = "26",
    pages = "1550160",
    year = "2015"
}

@article{Balasubramanian:2014bfa,
    author = "Balasubramanian, Vijay and Heckman, Jonathan J. and Maloney, Alexander",
    title = "{Relative Entropy and Proximity of Quantum Field Theories}",
    eprint = "1410.6809",
    archivePrefix = "arXiv",
    primaryClass = "hep-th",
    doi = "10.1007/JHEP05(2015)104",
    journal = "JHEP",
    volume = "05",
    pages = "104",
    year = "2015"
}

@article{Beatty:2024nkq,
    author = "Beatty, Emily and Fran{\c{c}}a, Daniel Stilck",
    title = "{Order $p$ quantum Wasserstein distances from couplings}",
    eprint = "2402.16477",
    archivePrefix = "arXiv",
    primaryClass = "quant-ph",
    doi = "10.1007/s00023-025-01557-z",
    month = "2",
    year = "2024"
}

@article{Li:2022exc,
    author = "Li, Lu and Bu, Kaifeng and Koh, Dax Enshan and Jaffe, Arthur and Lloyd, Seth",
    title = "{Wasserstein Complexity of Quantum Circuits}",
    eprint = "2208.06306",
    archivePrefix = "arXiv",
    primaryClass = "quant-ph",
    doi = "10.1088/1751-8121/ade381",
    journal = "J. Phys. A",
    volume = "58",
    pages = "265302",
    year = "2025"
}

@article{Cole:2021jwk,
    author = "Cole, Sam and Eckstein, Micha{\l} and Friedland, Shmuel and {\.Z}yczkowski, Karol",
    title = "{On Quantum Optimal Transport}",
    eprint = "2105.06922",
    archivePrefix = "arXiv",
    primaryClass = "quant-ph",
    doi = "10.1007/s11040-023-09456-7",
    journal = "Math. Phys. Anal. Geom.",
    volume = "26",
    number = "2",
    pages = "14",
    year = "2023"
}

@article{DePalma:2021zvy,
    author = "De Palma, Giacomo and Trevisan, Dario",
    title = "{Quantum optimal transport with quantum channels}",
    eprint = "1911.00803",
    archivePrefix = "arXiv",
    primaryClass = "math-ph",
    doi = "10.1007/s00023-021-01042-3",
    month = "2",
    year = "2021"
}

@article{Kehagias:2019akr,
    author = {Kehagias, Alex and L{\"u}st, Dieter and L{\"u}st, Severin},
    title = "{Swampland, Gradient Flow and Infinite Distance}",
    eprint = "1910.00453",
    archivePrefix = "arXiv",
    primaryClass = "hep-th",
    reportNumber = "MPP-2019-198, LMU-ASC 32/19, IPhT-T19/133, CPHT-RR055.092019",
    doi = "10.1007/JHEP04(2020)170",
    journal = "JHEP",
    volume = "04",
    pages = "170",
    year = "2020"
}

@article{Demulder:2024glx,
    author = {Demulder, Saskia and L\"ust, Dieter and Raml, Thomas},
    title = "{Navigating string theory field space with geometric flows}",
    eprint = "2412.10364",
    archivePrefix = "arXiv",
    primaryClass = "hep-th",
    doi = "10.1007/JHEP05(2025)030",
    journal = "JHEP",
    volume = "05",
    pages = "030",
    year = "2025"
}

@article{dowson1982frechet,
  title={The Fr{\'e}chet distance between multivariate normal distributions},
  author={Dowson, DC and Landau, BV666017},
  journal={Journal of multivariate analysis},
  volume={12},
  number={3},
  pages={450--455},
  year={1982},
  publisher={Elsevier}
}

@article{takatsu2011wasserstein,
author = {Asuka Takatsu},
title = {{Wasserstein geometry of Gaussian measures}},
volume = {48},
journal = {Osaka Journal of Mathematics},
number = {4},
publisher = {The University of Osaka and Osaka Metropolitan University, Departments of Mathematics},
pages = {1005 -- 1026},
year = {2011},
}

@article{DallAgata:2010ejj,
    author = "Dall'Agata, Gianguido and Gnecchi, Alessandra",
    title = "{Flow equations and attractors for black holes in N = 2 U(1) gauged supergravity}",
    eprint = "1012.3756",
    archivePrefix = "arXiv",
    primaryClass = "hep-th",
    reportNumber = "DFPD-10-TH-21",
    doi = "10.1007/JHEP03(2011)037",
    journal = "JHEP",
    volume = "03",
    pages = "037",
    year = "2011"
}

@article{kiefer2006quantum,
  title={Quantum gravity: general introduction and recent developments},
  author={Kiefer, Claus},
  journal={Annalen der Physik},
  volume={518},
  number={1-2},
  pages={129--148},
  year={2006},
  publisher={Wiley Online Library}
}

@article{Vilenkin:1987kf,
    author = "Vilenkin, Alexander",
    title = "{Quantum Cosmology and the Initial State of the Universe}",
    reportNumber = "TUTP-87-13a",
    doi = "10.1103/PhysRevD.37.888",
    journal = "Phys. Rev. D",
    volume = "37",
    pages = "888",
    year = "1988"
}

@article{Demulder:2023vlo,
    author = {Demulder, Saskia and L\"ust, Dieter and Raml, Thomas},
    title = "{Topology change and non-geometry at infinite distance}",
    eprint = "2312.07674",
    archivePrefix = "arXiv",
    primaryClass = "hep-th",
    doi = "10.1007/JHEP06(2024)079",
    journal = "JHEP",
    volume = "06",
    pages = "079",
    year = "2024"
}

@article{Shiu:2022oti,
    author = "Shiu, Gary and Tonioni, Flavio and Van Hemelryck, Vincent and Van Riet, Thomas",
    title = "{AdS scale separation and the distance conjecture}",
    eprint = "2212.06169",
    archivePrefix = "arXiv",
    primaryClass = "hep-th",
    doi = "10.1007/JHEP05(2023)077",
    journal = "JHEP",
    volume = "05",
    pages = "077",
    year = "2023"
}

@article{Andrianopoli:2010bj,
    author = "Andrianopoli, L. and D'Auria, R. and Ferrara, S. and Trigiante, M.",
    title = "{Fake Superpotential for Large and Small Extremal Black Holes}",
    eprint = "1002.4340",
    archivePrefix = "arXiv",
    primaryClass = "hep-th",
    doi = "10.1007/JHEP08(2010)126",
    journal = "JHEP",
    volume = "08",
    pages = "126",
    year = "2010"
}

@article{Bonnefoy:2019nzv,
    author = {Bonnefoy, Quentin and Ciambelli, Luca and L{\"u}st, Dieter and L{\"u}st, Severin},
    title = "{Infinite Black Hole Entropies at Infinite Distances and Tower of States}",
    eprint = "1912.07453",
    archivePrefix = "arXiv",
    primaryClass = "hep-th",
    reportNumber = "LMU-ASC 54/19, IPhT-T19/163, CPHT096.122019, MPP-2019-257, DESY 19-226, DESY-19-226, LMU{\textendash}ASC 54/19",
    doi = "10.1016/j.nuclphysb.2020.115112",
    journal = "Nucl. Phys. B",
    volume = "958",
    pages = "115112",
    year = "2020"
}

@article{Floerchinger:2023ekw,
    author = "Floerchinger, Stefan",
    title = "{Functional information geometry of Euclidean quantum fields}",
    eprint = "2303.04081",
    archivePrefix = "arXiv",
    primaryClass = "hep-th",
    doi = "10.1103/PhysRevD.110.125027",
    journal = "Phys. Rev. D",
    volume = "110",
    number = "12",
    pages = "125027",
    year = "2024"
}

@article{gelbrich1990formula,
  title={On a formula for the L2 Wasserstein metric between measures on Euclidean and Hilbert spaces},
  author={Gelbrich, Matthias},
  journal={Mathematische Nachrichten},
  volume={147},
  number={1},
  pages={185--203},
  year={1990},
  publisher={Wiley Online Library}
}

@article{Visser:1992pz,
    author = "Visser, Matt",
    title = "{van Vleck determinants: Geodesic focusing and defocusing in Lorentzian space-times}",
    eprint = "hep-th/9303020",
    archivePrefix = "arXiv",
    reportNumber = "WASH-U-HEP-92-76",
    doi = "10.1103/PhysRevD.47.2395",
    journal = "Phys. Rev. D",
    volume = "47",
    pages = "2395--2402",
    year = "1993"
}

@article{monge1781memoire,
  title={M{\'e}moire sur la th{\'e}orie des d{\'e}blais et des remblais},
  author={Monge, Gaspard},
  journal={Mem. Math. Phys. Acad. Royale Sci.},
  pages={666--704},
  year={1781}
}

@article{kantorovich2006translocation,
  title={On the Translocation of Masses.},
  author={Kantorovich, Leonid V},
  journal={Journal of mathematical sciences},
  volume={133},
  number={4},
  year={2006}
}

@book{bao2012introduction,
  title={An introduction to Riemann-Finsler geometry},
  author={Bao, David and Chern, S-S and Shen, Zhongmin},
  volume={200},
  year={2012},
  publisher={Springer Science \& Business Media}
}

@book{landau1960mechanics,
  title     = {Mechanics},
  author    = {Landau, Lev D. and Lifshitz, Evgenii M.},
  volume    = {1},
  year      = {1960},
  publisher = {Pergamon Press}
}

@book{arnol2013mathematical,
  title={Mathematical methods of classical mechanics},
  author={Arnol'd, Vladimir Igorevich},
  volume={60},
  year={2013},
  publisher={Springer Science \& Business Media}
}

@book{goldstein1950classical,
  title={Classical mechanics},
  author={Goldstein, Herbert and Poole, Charles P and Safko, John},
  volume={2},
  year={1950},
  publisher={Addison-wesley Reading, MA}
}

@phdthesis{Raml:2025yrb,
    author = "Raml, Thomas",
    title = "{Potentials on scalar field space and the Swampland} (\href{https://edoc.ub.uni-muenchen.de/36026/}{10.5282/edoc.36026})",
    doi = "10.5282/edoc.36026",
    school = "Munich U., LMU Munich",
    month = "11",
    year = "2025",
}

@article{Behrndt:1996jn,
    author = {Behrndt, Klaus and Lopes Cardoso, Gabriel and de Wit, Bernard and Kallosh, Renata and L\"ust, Dieter and Mohaupt, Thomas},
    title = "{Classical and quantum N=2 supersymmetric black holes}",
    eprint = "hep-th/9610105",
    archivePrefix = "arXiv",
    reportNumber = "CERN-TH-96-276, HUB-EP-96-53, SU-ITP-96-41, THU-96-34",
    doi = "10.1016/S0550-3213(97)00028-X",
    journal = "Nucl. Phys. B",
    volume = "488",
    pages = "236--260",
    year = "1997"
}

@article{Klawer:2021ltm,
    author = {Kl{\"a}wer, Daniel},
    title = "{Modular curves and the refined distance conjecture}",
    eprint = "2108.00021",
    archivePrefix = "arXiv",
    primaryClass = "hep-th",
    reportNumber = "MITP/21-034",
    doi = "10.1007/JHEP12(2021)088",
    journal = "JHEP",
    volume = "12",
    pages = "088",
    year = "2021"
}

@article{Grimm:2020ouv,
    author = "Grimm, Thomas W. and Li, Chongchuo",
    title = "{Universal axion backreaction in flux compactifications}",
    eprint = "2012.08272",
    archivePrefix = "arXiv",
    primaryClass = "hep-th",
    doi = "10.1007/JHEP06(2021)067",
    journal = "JHEP",
    volume = "06",
    pages = "067",
    year = "2021"
}

@article{Grimm:2022sbl,
    author = "Grimm, Thomas W. and Lanza, Stefano and Li, Chongchuo",
    title = "{Tameness, Strings, and the Distance Conjecture}",
    eprint = "2206.00697",
    archivePrefix = "arXiv",
    primaryClass = "hep-th",
    doi = "10.1007/JHEP09(2022)149",
    journal = "JHEP",
    volume = "09",
    pages = "149",
    year = "2022"
}

@article{Grimm:2025cpq,
    author = "Grimm, Thomas W. and van de Heisteeg, Damian and Revello, Filippo",
    title = "{Axion-Scalar Systems and Dynamical Distances}",
    eprint = "2510.12879",
    archivePrefix = "arXiv",
    primaryClass = "hep-th",
    month = "10",
    year = "2025"
}

@article{Grana:2021zvf,
	title        = {{The Swampland Conjectures: A Bridge from Quantum Gravity to Particle Physics}},
	author       = {Gra\~na, Mariana and Herr\'aez, Alvaro},
	year         = {2021},
	journal      = {Universe},
	volume       = {7},
	number       = {8},
	pages        = {273},
	doi          = {10.3390/universe7080273},
	eprint       = {2107.00087},
	archiveprefix = {arXiv},
	primaryclass = {hep-th}
}

@article{Agmon:2022thq,
	title        = {{Lectures on the string landscape and the Swampland}},
	author       = {Agmon, Nathan Benjamin and Bedroya, Alek and Kang, Monica Jinwoo and Vafa, Cumrun},
	year         = {2022},
	month        = {12},
	eprint       = {2212.06187},
	archiveprefix = {arXiv},
	primaryclass = {hep-th}
}

@article{Palti:2024voy,
	title        = {{A positive metric over DGKT vacua}},
	author       = {Palti, Eran and Petri, Nicol\`o},
	year         = {2024},
	journal      = {JHEP},
	volume       = {06},
	pages        = {019},
	doi          = {10.1007/JHEP06(2024)019},
	eprint       = {2405.01084},
	archiveprefix = {arXiv},
	primaryclass = {hep-th}
}

@article{Li:2023gtt,
	title        = {{Towards AdS distances in string theory}},
	author       = {Li, Yixuan and Palti, Eran and Petri, Nicol\`o},
	year         = {2023},
	journal      = {JHEP},
	volume       = {08},
	pages        = {210},
	doi          = {10.1007/JHEP08(2023)210},
	eprint       = {2306.02026},
	archiveprefix = {arXiv},
	primaryclass = {hep-th}
}

@article{Palti:2025ydz,
    author = "Palti, Eran and Petri, Nicol\`o",
    title = "{Metrics over multi-parameter AdS vacua}",
    eprint = "2504.01316",
    archivePrefix = "arXiv",
    primaryClass = "hep-th",
    month = "4",
    year = "2025"
}

@article{Shiu:2023bay,
	title        = {{Connecting flux vacua through scalar field excursions}},
	author       = {Shiu, Gary and Tonioni, Flavio and Van Hemelryck, Vincent and Van Riet, Thomas},
	year         = {2024},
	journal      = {Phys. Rev. D},
	volume       = {109},
	number       = {6},
	pages        = {066017},
	doi          = {10.1103/PhysRevD.109.066017},
	eprint       = {2311.10828},
	archiveprefix = {arXiv},
	primaryclass = {hep-th},
	reportnumber = {UUITP-32/23}
}

@article{Basile:2023rvm,
	title        = {{Domain walls and distances in discrete landscapes}},
	author       = {Basile, Ivano and Montella, Carmine},
	year         = {2023},
	month        = {9},
	eprint       = {2309.04519},
	archiveprefix = {arXiv},
	primaryclass = {hep-th}
}

@article{Lust:2019zwm,
	title        = {{AdS and the Swampland}},
	author       = {L\"ust, Dieter and Palti, Eran and Vafa, Cumrun},
	year         = {2019},
	journal      = {Phys. Lett. B},
	volume       = {797},
	pages        = {134867},
	doi          = {10.1016/j.physletb.2019.134867},
	eprint       = {1906.05225},
	archiveprefix = {arXiv},
	primaryclass = {hep-th}
}

@article{topping2009ricci,
  title={Ricci Flow: The Foundations via Optimal Transportation},
  author={Topping, Peter},
  year={2009}
}

@article{mccann2010ricci,
  title={Ricci flow, entropy and optimal transportation},
  author={McCann, Robert J and Topping, Peter M},
  journal={American Journal of Mathematics},
  volume={132},
  number={3},
  pages={711--730},
  year={2010},
  publisher={Johns Hopkins University Press}
}

@article{kopfer2024optimal,
  title={Optimal transport and generalized Ricci flow},
  author={Kopfer, Eva and Streets, Jeffrey and others},
  journal={SIGMA. Symmetry, Integrability and Geometry: Methods and Applications},
  volume={20},
  pages={003},
  year={2024},
  publisher={SIGMA. Symmetry, Integrability and Geometry: Methods and Applications}
}

@article{PhysRev.116.1322,
  title = {Dynamical Structure and Definition of Energy in General Relativity},
  author = {Arnowitt, R. and Deser, S. and Misner, C. W.},
  journal = {Phys. Rev.},
  volume = {116},
  issue = {5},
  pages = {1322--1330},
  numpages = {0},
  year = {1959},
  month = {Dec},
  publisher = {American Physical Society},
  doi = {10.1103/PhysRev.116.1322},
  url = {https://link.aps.org/doi/10.1103/PhysRev.116.1322}
}

@article{Bergmann:1972ud,
    author = "Bergmann, P. G. and Komar, A.",
    title = "{The coordinate group symmetries of general relativity}",
    doi = "10.1007/BF00671650",
    journal = "Int. J. Theor. Phys.",
    volume = "5",
    pages = "15--28",
    year = "1972"
}

@book{Hanson:1976cn,
    author = "Hanson, Andrew J. and Regge, Tullio and Teitelboim, Claudio",
    title = "{Constrained Hamiltonian Systems}",
    reportNumber = "RX-748, PRINT-75-0141 (IAS,PRINCETON)",
    publisher = "Accademia Nazionale dei Lincei",
    year = "1976"
}

@article{DeWitt:1967yk,
    author = "DeWitt, Bryce S.",
    editor = "Fang, Li-Zhi and Ruffini, R.",
    title = "{Quantum Theory of Gravity. 1. The Canonical Theory}",
    doi = "10.1103/PhysRev.160.1113",
    journal = "Phys. Rev.",
    volume = "160",
    pages = "1113--1148",
    year = "1967"
}

@article{Isham:1992ms,
    author = "Isham, C. J.",
    editor = "Ibort, L. A. and Rodriguez, M. A.",
    title = "{Canonical quantum gravity and the problem of time}",
    eprint = "gr-qc/9210011",
    archivePrefix = "arXiv",
    reportNumber = "IMPERIAL-TP-91-92-25",
    journal = "NATO Sci. Ser. C",
    volume = "409",
    pages = "157--287",
    year = "1993"
}

@article{Hartle:1983ai,
    author = "Hartle, J. B. and Hawking, S. W.",
    editor = "Fang, Li-Zhi and Ruffini, R.",
    title = "{Wave Function of the Universe}",
    reportNumber = "PRINT-83-0937 (CAMBRIDGE)",
    doi = "10.1103/PhysRevD.28.2960",
    journal = "Phys. Rev. D",
    volume = "28",
    pages = "2960--2975",
    year = "1983"
}

@article{Halliwell:1988wc,
    author = "Halliwell, Jonathan J.",
    title = "{Derivation of the Wheeler-De Witt Equation from a Path Integral for Minisuperspace Models}",
    reportNumber = "NSF-ITP-88-25",
    doi = "10.1103/PhysRevD.38.2468",
    journal = "Phys. Rev. D",
    volume = "38",
    pages = "2468",
    year = "1988"
}

@article{Freedman:2003ax,
    author = "Freedman, D. Z. and Nunez, Carlos and Schnabl, M. and Skenderis, K.",
    title = "{Fake supergravity and domain wall stability}",
    eprint = "hep-th/0312055",
    archivePrefix = "arXiv",
    reportNumber = "MIT-CTP-3450, ITFA-2003-55",
    doi = "10.1103/PhysRevD.69.104027",
    journal = "Phys. Rev. D",
    volume = "69",
    pages = "104027",
    year = "2004"
}

@article{Wheeler:1968iap,
    author = "Wheeler, J. A.",
    editor = "Fang, Li-Zhi and Ruffini, R.",
    title = "{Superspace and the nature of quantum geometrodynamics}",
    journal = "Adv. Ser. Astrophys. Cosmol.",
    volume = "3",
    pages = "27--92",
    year = "1987"
}

@article{Skenderis:2006jq,
    author = "Skenderis, Kostas and Townsend, Paul K.",
    title = "{Hidden supersymmetry of domain walls and cosmologies}",
    eprint = "hep-th/0602260",
    archivePrefix = "arXiv",
    reportNumber = "DAMTP-2006-18, ITFA-2006-08",
    doi = "10.1103/PhysRevLett.96.191301",
    journal = "Phys. Rev. Lett.",
    volume = "96",
    pages = "191301",
    year = "2006"
}

@article{Celi:2004st,
    author = "Celi, Alessio and Ceresole, Anna and Dall'Agata, Gianguido and Van Proeyen, Antoine and Zagermann, Marco",
    title = "{On the fakeness of fake supergravity}",
    eprint = "hep-th/0410126",
    archivePrefix = "arXiv",
    reportNumber = "KUL-TF-04-29, SU-ITP-04-034",
    doi = "10.1103/PhysRevD.71.045009",
    journal = "Phys. Rev. D",
    volume = "71",
    pages = "045009",
    year = "2005"
}

@article{Aoufia:2025ppe,
    author = "Aoufia, Christian and Castellano, Alberto and Ib{\'a}{\~n}ez, Luis",
    title = "{Laplacians in Various Dimensions and the Swampland}",
    eprint = "2506.03253",
    archivePrefix = "arXiv",
    primaryClass = "hep-th",
    reportNumber = "IFT-UAM/CSIC-25-61, EFI-25-8",
    month = "6",
    year = "2025"
}

@article{DeBiasio:2022zuh,
    author = "De Biasio, Davide",
    title = "{On-Shell Flow}",
    eprint = "2211.04231",
    archivePrefix = "arXiv",
    primaryClass = "hep-th",
    month = "11",
    year = "2022"
}

@article{DeBiasio:2022nsd,
    author = {De Biasio, Davide and Freigang, Julian and L\"ust, Dieter and Wiseman, Toby},
    title = {{Gradient flow of Einstein-Maxwell theory and Reissner-Nordstr{\"o}m black holes}},
    eprint = "2210.14705",
    archivePrefix = "arXiv",
    primaryClass = "hep-th",
    doi = "10.1007/JHEP03(2023)074",
    journal = "JHEP",
    volume = "03",
    pages = "074",
    year = "2023"
}

@article{Velazquez:2022eco,
    author = {Vel{\'a}zquez, David Mart{\'\i}n and De Biasio, Davide and L\"ust, Dieter},
    title = "{Cobordism, singularities and the Ricci flow conjecture}",
    eprint = "2209.10297",
    archivePrefix = "arXiv",
    primaryClass = "hep-th",
    doi = "10.1007/JHEP01(2023)126",
    journal = "JHEP",
    volume = "01",
    pages = "126",
    year = "2023"
}

@article{DeBiasio:2022omq,
    author = {De Biasio, Davide and L\"ust, Dieter},
    title = "{Geometric flow of bubbles}",
    eprint = "2201.01679",
    archivePrefix = "arXiv",
    primaryClass = "hep-th",
    reportNumber = "LMU-ASC 02/22, MPP-2022-2",
    doi = "10.1016/j.nuclphysb.2022.115812",
    journal = "Nucl. Phys. B",
    volume = "980",
    pages = "115812",
    year = "2022"
}

@article{Luca:2022inb,
    author = "Luca, Giuseppe Bruno De and De Ponti, Nicol{\`o} and Mondino, Andrea and Tomasiello, Alessandro",
    title = "{Gravity from thermodynamics: Optimal transport and negative effective dimensions}",
    eprint = "2212.02511",
    archivePrefix = "arXiv",
    primaryClass = "hep-th",
    doi = "10.21468/SciPostPhys.15.2.039",
    journal = "SciPost Phys.",
    volume = "15",
    number = "2",
    pages = "039",
    year = "2023"
}

@article{DeLuca:2025klz,
    author = "De Luca, G. Bruno and De Ponti, Nicol{\`o} and Mondino, Andrea and Tomasiello, Alessandro",
    title = "{Cheng's eigenvalue comparison on metric measure spaces and applications}",
    eprint = "2507.23671",
    archivePrefix = "arXiv",
    primaryClass = "math.SP",
    month = "7",
    year = "2025"
}

@article{DeLuca:2021ojx,
    author = "De Luca, G. Bruno and De Ponti, Nicol{\`o} and Mondino, Andrea and Tomasiello, Alessandro",
    title = "{Cheeger bounds on spin-two fields}",
    eprint = "2109.11560",
    archivePrefix = "arXiv",
    primaryClass = "hep-th",
    doi = "10.1007/JHEP12(2021)217",
    journal = "JHEP",
    volume = "12",
    pages = "217",
    year = "2021"
}

@article{DeLuca:2021mcj,
    author = "De Luca, G. Bruno and Tomasiello, Alessandro",
    title = "{Leaps and bounds towards scale separation}",
    eprint = "2104.12773",
    archivePrefix = "arXiv",
    primaryClass = "hep-th",
    doi = "10.1007/JHEP12(2021)086",
    journal = "JHEP",
    volume = "12",
    pages = "086",
    year = "2021"
}

@article{Anchordoqui:2025izb,
    author = {Anchordoqui, Luis A. and L\"ust, Dieter and L{\"u}st, Severin},
    title = "{Species quantum mechanics}",
    eprint = "2510.25846",
    archivePrefix = "arXiv",
    primaryClass = "hep-th",
    reportNumber = "MPP-2025-203, LMU-ASC 24/25",
    doi = "10.1103/vwjx-lrhl",
    journal = "Phys. Rev. D",
    volume = "113",
    number = "2",
    pages = "026021",
    year = "2026"
}

@book{sakurai2020modern,
  title={Modern quantum mechanics},
  author={Sakurai, Jun John and Napolitano, Jim},
    year={2020},
  publisher={Cambridge university press}
}

@article{bertrand2013optimal,
  title={The optimal mass transport problem for relativistic costs},
  author={Bertrand, J{\'e}r{\^o}me and Puel, Marjolaine},
  journal={Calculus of Variations and Partial Differential Equations},
  volume={46},
  number={1},
  pages={353--374},
  year={2013},
  publisher={Springer}
}

@misc{mccann2023d,
      title={Displacement convexity of Boltzmann's entropy characterizes the strong energy condition from general relativity}, 
      author={Robert J McCann},
      year={2023},
      eprint={1808.01536},
      archivePrefix={arXiv},
      primaryClass={math-ph},
      url={https://arxiv.org/abs/1808.01536}, 
}

@misc{suhr2018t,
      title={Theory of optimal transport for Lorentzian cost functions}, 
      author={Stefan Suhr},
      year={2018},
      eprint={1601.04532},
      archivePrefix={arXiv},
      primaryClass={math.DG},
      url={https://arxiv.org/abs/1601.04532}, 
}

@incollection{brenier2004extended,
  title={Extended monge-kantorovich theory},
  author={Brenier, Yann},
  booktitle={Optimal Transportation and Applications: Lectures given at the CIME Summer School, held in Martina Franca, Italy, September 2-8, 2001},
  pages={91--121},
  year={2004},
  publisher={Springer}
}

@article{Hashimoto:2026kjy,
    author = "Hashimoto, Koji and Tanahashi, Norihiro",
    title = "{Holography and Optimal Transport: Emergent Wasserstein Spacetime in Harmonic Oscillator, SYK and Krylov Complexity}",
    eprint = "2604.17649",
    archivePrefix = "arXiv",
    primaryClass = "hep-th",
    reportNumber = "KUNS-3098",
    month = "4",
    year = "2026"
}

@book{griffiths2018introduction,
  title={Introduction to quantum mechanics},
  author={Griffiths, David J and Schroeter, Darrell F},
  year={2018},
  publisher={Cambridge university press}
}

@book{bender1999advanced,
  title={Advanced mathematical methods for scientists and engineers: Asymptotic methods and perturbation theory},
  author={Bender, Carl M and Orszag, Steven A},
  volume={1},
  year={1999},
  publisher={Springer}
}

@article{Anchordoqui:2026nit,
    author = {Anchordoqui, Luis and Etheredge, Muldrow and L\"ust, Dieter},
    title = "{Moduli Space Quantum Mechanics}",
    eprint = "2603.06795",
    archivePrefix = "arXiv",
    primaryClass = "hep-th",
    reportNumber = "MPP-2026-27",
    month = "3",
    year = "2026"
}

@article{Das:2026hbw,
    author = "Das, Rathindra Nath and Demulder, Saskia",
    title = "{Integrability breaking in semiclassical strings in Koopman-Krylov space}",
    eprint = "2602.23421",
    archivePrefix = "arXiv",
    primaryClass = "hep-th",
    month = "2",
    year = "2026"
}

@article{Herraez:2025clp,
    author = {Herr{\'a}ez, Alvaro and L{\"u}st, Dieter and Masias, Joaquin and Montella, Carmine},
    title = "{A short overview on the Black Hole-Tower Correspondence and Species Thermodynamics}",
    eprint = "2506.02335",
    archivePrefix = "arXiv",
    primaryClass = "hep-th",
    reportNumber = "MPP-2025-114",
    doi = "10.22323/1.490.0161",
    journal = "PoS",
    volume = "CORFU2024",
    pages = "161",
    year = "2025"
}

@article{Calderon-Infante:2026ymy,
    author = "Calder{\'o}n-Infante, Jos{\'e} and Cheng, Gongrui and Herr{\'a}ez, Alvaro and Van Riet, Thomas",
    title = "{End-of-the-World Singularities: The Good, the Bad, and the Heated-up}",
    eprint = "2603.18133",
    archivePrefix = "arXiv",
    primaryClass = "hep-th",
    month = "3",
    year = "2026"
}
\bibliographystyle{JHEP}

\end{document}